\documentclass[lettersize,journal]{IEEEtran}
\usepackage[dvipsnames, svgnames, x11names]{xcolor}
\usepackage{amsmath,amssymb,amsfonts}
\usepackage{algorithmic}
\usepackage{algorithm}
\usepackage{array}
\usepackage[caption=false,font=normalsize,labelfont=sf,textfont=sf]{subfig}
\usepackage{textcomp}
\usepackage{stfloats}
\usepackage{hyperref}
\usepackage{url}

\usepackage{verbatim}
\usepackage{graphicx}
\usepackage{cite}
\usepackage{enumitem}
\usepackage{multirow}
\usepackage{tcolorbox}
\usepackage{booktabs}
\usepackage{tabularx}
\usepackage{pifont}
\usepackage{threeparttable}
\usepackage{cite}
\usepackage{color}
\usepackage{bbding}
\usepackage[T1]{fontenc}
\usepackage[scaled=0.88]{beramono}

\usepackage{fp}
\newcommand{\ShowAbsoluteNumber}[1]{%
\ifnum #1<10%
{\hspace*{0pt}#1}%
\else%
\ifnum #1<100%
{\hspace*{0pt}#1}%
\else%
\ifnum #1<1000%
{\hspace*{0pt}#1}%
\else%
{\numprint{#1}}%
\fi%
\fi%
\fi%
}

\newcommand{\ShowPercentage}[2]{%
\FPeval\percentage{round(#1/#2*100,0)}%
\ifnum \percentage=0%
{\scriptsize(<1\%)}%
\else%
\ifnum \percentage<10%
{\small(\hspace*{0pt}\FPprint{percentage}\%)}%
\else%
{\small(\FPprint{percentage}\%)}%
\fi%
\fi%
\hspace*{0.3ex}%
}
\newlength\BARSIZE  
\newcommand{\inlinechart}[2]{%
\FPeval{\BLACKBARSIZE}{#1/#2}\textcolor{black!80}{\rule{\BLACKBARSIZE\BARSIZE}{1.6ex}}%
\FPeval{\BLACKBARSIZE}{1 - (#1/#2)}\textcolor{black!10}{\rule{\BLACKBARSIZE\BARSIZE}{1.6ex}}%
}

\begin{document}

\begin{sloppypar}

\title{\huge Empirical Review of Smart Contract and DeFi Security: \\ Vulnerability Detection and Automated Repair}

\author{Peng Qian, Rui Cao, Zhenguang Liu, Wenqing Li, Ming Li, Lun Zhang, Yufeng Xu, \\ Jianhai Chen, and Qinming He

%\thanks{This work was supported in part by the National Key R\&D Program of China (2021YFB2700500); in part by the Key R\&D Program of Zhejiang Province (2022C01086). } }           
\thanks{Peng Qian is with College of Computer Science and Technology, Zhejiang University, Hangzhou 310058, China and Goplus Security  (e-mail: pqian@zju.edu.cn).}
\thanks{Rui Cao is with School of Computer Science and Technology, Zhejiang Gongshang University, Hangzhou, China (e-mail: primercrui@gmail.com).}
\thanks{Zhenguang Liu, Wenqing Li, Jianhai Chen, and Qinming He are with College of Computer Science and Technology, Zhejiang University, Hangzhou 310058, China (e-mail: liuzhenguang2008@gmail.com, lwqfighting@zju.edu.cn; chenjh919@zju.edu.cn; hqm@zju.edu.cn). }
\thanks{Ming Li, Lun Zhang, and Yufeng Xu are with Goplus Security, Hangzhou, China (e-mail: Mike@gopluslabs.io; Allen@gopluslabs.io; eskil@gopluslabs.io). }
\thanks{(Corresponding author: Jianhai Chen)}
}

% The paper headers
\markboth{}%
{Qian \MakeLowercase{\textit{et al.}}: Comprehensive Understanding of DeFi Attacks and Empirical Review of State-of-the-Art DeFi Bug Detection Techniques}

%\IEEEpubid{0000--0000/00\$00.00~\copyright~2021 IEEE}
% Remember, if you use this you must call \IEEEpubidadjcol in the second
% column for its text to clear the IEEEpubid mark.

\maketitle

\begin{abstract}
Decentralized Finance (DeFi) is emerging as a peer-to-peer financial ecosystem, enabling participants to trade products on a permissionless blockchain. Built on blockchain and smart contracts, the DeFi ecosystem has experienced explosive growth in recent years. Unfortunately, smart contracts hold a massive amount of value, making them an attractive target for attacks. So far, attacks against smart contracts and DeFi protocols have resulted in billions of dollars in financial losses, severely threatening the security of the entire DeFi ecosystem. Researchers have proposed various security tools for smart contracts and DeFi protocols as countermeasures. However, a comprehensive investigation of these efforts is still lacking, leaving a crucial gap in our understanding of how to enhance the security posture of the smart contract and DeFi landscape.

To fill the gap, this paper reviews the progress made in the field of smart contract and DeFi security from the perspective of both vulnerability detection and automated repair. First, we analyze the DeFi smart contract security issues and challenges. Specifically, we lucubrate various DeFi attack incidents and summarize the attacks into six categories. Then, we present an empirical study of 42 state-of-the-art techniques that can detect smart contract and DeFi vulnerabilities. In particular, we evaluate the effectiveness of traditional smart contract bug detection tools in analyzing complex DeFi protocols. Additionally, we investigate 8 existing automated repair tools for smart contracts and DeFi protocols, providing insight into their advantages and disadvantages. To make this work useful for as wide of an audience as possible, we also identify several open issues and challenges in the DeFi ecosystem that should be addressed in the future. As a side contribution, we release an annotated dataset that consists of $99$ DeFi protocols ($7,340$ DeFi smart contracts) and concerns six types of DeFi attacks, hoping to facilitate the DeFi community.
\end{abstract}

\begin{IEEEkeywords}
Smart Contract; Decentralized Finance (DeFi); Blockchain; Vulnerability Detection; Automated Repair
\end{IEEEkeywords}

\section{Introduction}
\label{introduction}
Blockchain and its killer applications, {e.g.,} Bitcoin~\cite{nakamoto2008bitcoin} and smart contract~\cite{zou2019smart}, are taking the world by storm, giving rise to a variety of interesting and compelling decentralized applications~\cite{rosa2018blockchain,gao2019towards,benisi2020blockchain}. A blockchain is essentially a replicated and distributed ledger that is shared among all bookkeeping nodes in a peer-to-peer network following a consensus protocol~\cite{yu2019proof,liu2021smart}. Each block in the blockchain consists of a number of transactions. Every time a new transaction occurs, a record of that transaction is added to the ledger of every bookkeeping node~\cite{benvcic2018distributed}. The duplicate ledgers stored in the worldwide participating nodes ensure that transactions are immutable once recorded, endowing the blockchain with tamper-proof and decentralized nature~\cite{liu2021combining}.

Smart contracts are programs running on a blockchain system. They encode predefined terms into executable contract code. Once a smart contract is deployed on the blockchain, its defined rules will be strictly followed during execution. Smart contracts make the automatic execution of contract terms possible, facilitating complex decentralized applications~\cite{chen2018survey,qian2019digital}. So far, tens of millions of contracts have been deployed on Ethereum, one of the most prominent blockchain platforms, enabling a variety of applications, such as wallet~\cite{chen2019cryptoar}, gambling game~\cite{hu2021transaction}, supply chain~\cite{moosavi2021blockchain}, healthcare~\cite{agbo2019blockchain}, and cross-industry finance~\cite{williams2020cross}.

Recently, we have witnessed a dramatic rise in the popularity of {decentralized finance} (DeFi)  applications~\cite{chen2019decentralized,zhou2021just,werner2021sok,jensen2021introduction}. 
The total value locked (TVL) in DeFi has increased to as high as \$248.84 billion due to the rapid growth of these applications.
DeFi has shown its potential to expand the use of blockchain from simple value transfer to complex financial services~\cite{amler2021defi}. Popular DeFi protocols now enable a variety of decentralized services, including lending and borrowing~\cite{bartoletti2021sok}, portfolio management~\cite{raheman2021architecture,bartoletti2021theory}, asset exchanges~\cite{bartoletti2021maximizing}, and derivatives~\cite{bartoletti2021towards}, all without the need for trusted parties.

\begin{figure}
\centering
\includegraphics[width=8.5cm]{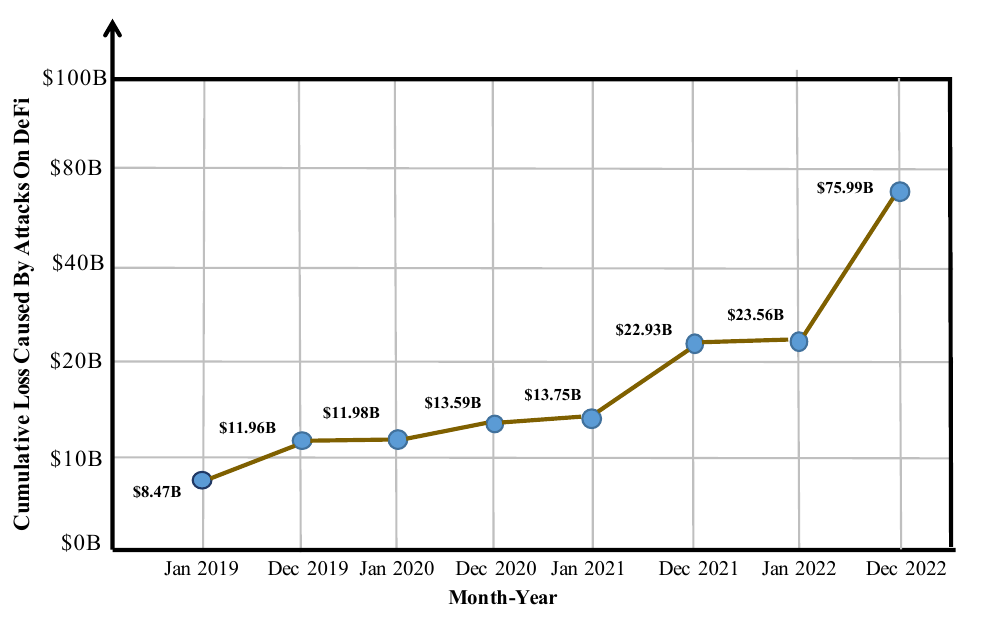}
\caption{The cumulative losses caused by attacks targeting DeFi protocols.} 
\label{curve}
\end{figure}

\textbf{Motivation I: Understanding DeFi Attacks.}\quad 
While DeFi has been constantly improving, it is still in its infancy and the new DeFi ecosystem is crude, leaving  a large room for security attacks~\cite{gudgeon2020decentralized,qin2021attacking}. 
%Akin to the traditional finance, DeFi is being plagued by predatory traders, showcasing a sea of creative market manipulation techniques, such as high-frequency attacks~\cite{zhou2021high}, pump and dump schemes~\cite{xu2019anatomy}, and wash trading~\cite{victor2021detecting}. In particular, a DeFi protocols is usually composed of multiple interactive smart contracts. The composability of DeFi protocols leads to the emergence of attack opportunities for adversaries throughout the tightly intertwined DeFi space~\cite{wang2021blockeye}. 
For example, in February 2020, the well-known DeFi protocol \texttt{bZx} suffered from two consecutive attacks~\cite{oosthoek2021flash}. Attackers made use of the logic flaws in \texttt{bZx} to achieve arbitrage (i.e., stealing over \$8 million ETH at that time) at a low cost. 
This case is not isolated, and attacks on DeFi protocols happen every few months~\cite{wang2021promutator}. In April 2020, cybercriminals hacked into the DeFi protocol \texttt{UniswapV1} and exploited the {reentrancy} vulnerability to steal 1,278 \texttt{ETH}~\cite{Group12}. In October 2021, the DeFi protocol \texttt{Cream.Finance} suffered from an external attack, yielding losses of more than \$130 million~\cite{Inspex}. Recently, DeFi cross-chain project \texttt{Poly Network}~\cite{Poly} and NFT game \texttt{Ronin}~\cite{ronin} were attacked by hackers, losing \$611 million and \$622 million respectively, which are the two most severe attacks in DeFi at the time of writing. 
According to the statistics from the REKT Database~\cite{rekt}, DeFi protocols have lost a total of \$77.1 billion due to scams, hacks, and exploits, out of which only \$6.5 billion has been returned. Fig.~\ref{curve} shows the cumulative financial losses caused by attacks on DeFi over the past three years. 
Obviously, frequent attacks have caused huge losses to the DeFi ecosystem and severely hindered its development.

Notwithstanding recent efforts have been directed towards investigating DeFi attacks~\cite{amler2021defi,wang2021towards,tolmach2021formal}, a broad and well-structured overview of this important research topic is still missing. In particular, current studies mostly revolve around financial applications and services of DeFi~\cite{qin2021cefi,chohan2021decentralized}, while lacking an in-depth review on DeFi attacks. \emph{Motivated by this, we strive to provide a comprehensive understanding of various DeFi attacks, hoping to facilitate the DeFi community.}

\textbf{Motivation II: Investigating State-of-the-Art Smart Contract and DeFi Security Tools.}\quad 
Like traditional computer programs, smart contracts contain vulnerabilities. Distinct from the traditional programs, exposed bugs in the on-chain smart contracts cannot be patched due to the immutability of the blockchain~\cite{marino2016setting}. Since DeFi protocols typically consist of multiple smart contracts, they inevitably carry the risk of external attacks. Therefore, it is important to identify potential vulnerabilities in smart contracts. A number of works~\cite{xue2020cross,perez2021smart,brent2020ethainter} have been developed to automatically detect flaws in smart contracts. They mainly focus on static analysis~\cite{ghaleb2020effective,grishchenko2018foundations,schneidewind2020ethor} and fuzzing techniques~\cite{wustholz2020targeted,zong2020fuzzguard,grieco2020echidna} to discover vulnerabilities. For example, Slither~\cite{feist2019slither} converts smart contracts into an intermediate representation and performs taint analysis to detect vulnerabilities. ContractFuzzer~\cite{contractfuzzer} applies fuzzing techniques to smart contracts and reveals vulnerabilities by monitoring runtime behavior during fuzzing. 
\emph{However, most of these approaches are tailored for traditional smart contracts that usually comprise a single contract. Their ability to perform vulnerability analysis on complex DeFi protocols has yet to be investigated.}

Several recent efforts~\cite{cao2021flashot,wang2021blockeye,wang2022impact} are devoted to detecting DeFi attacks by analyzing the transactions in DeFi protocols, from which they extract transaction patterns to expose potential attacks. For example, \textsc{ProMutator}~\cite{wang2021promutator} detects the price manipulation attack by reconstructing probable DeFi use patterns from historical transactions. \textsc{DeFiRanger}~\cite{wu2021defiranger} identifies DeFi price oracle attacks according to recovered high-level DeFi semantics. 
\emph{Unfortunately, although the interest in DeFi attack detection is increasing, there is no systematic overview to probe the state-of-the-art DeFi attack detection approaches and gauge their effectiveness.}

%While a number of tools have been developed to detect vulnerabilities in smart contracts, they exhibit certain limitations in the capacity for fixing smart contracts automatically. 

Another line of work focuses on identifying and patching vulnerabilities in smart contracts and DeFi protocols. For example, \textsc{sGuard}~\cite{SGUARD} compiles a smart contract into the bytecode, source map, and abstract syntax tree (AST). The bytecode is used to find bugs, while the source map and AST help fix the contract at the source code level.~\cite{10.1145/3551349.3559560} develop an automated repair approach for DeFi protocols that can fix violations of functional specifications expressed as a property while providing solid correctness guarantees. \emph{To our knowledge, the review of existing automated repair tools for smart contracts and DeFi protocols is yet to be explored.}

%EVMPatch~\cite{} uses binary rewriting to replace a vulnerable basic block with a trampoline, which transfers the control flow to a free-of-vulnerability basic block appending to the end of the bytecode. 

{\textbf{Our Work.}}\quad
In this paper, we aim to provide a comprehensive understanding of various DeFi attacks and an empirical review of state-of-the-art smart contract and DeFi security tools. 
\emph{On the one hand}, we study 57 reported DeFi attack incidents, from which we summarize six common categories of DeFi attacks. In particular, we present concrete examples to illustrate the attack flow of each type of DeFi attack. 
Notably, we construct and release an annotated dataset consisting of $99$ DeFi protocols (a total of $7,340$ smart contracts). 
We ensure the traceability of each project by providing its address or URL. For each DeFi protocol, we annotate the type of attack or vulnerability it suffered. 
\emph{On the other hand}, we investigate 50 security analysis tools that can either detect or repair vulnerabilities in smart contracts and DeFi protocols. 
Specifically, we first review the principles of bug detection tools for traditional smart contracts, and then conduct extensive experiments to evaluate their performance when applied to complex DeFi protocols. 
Then, we illustrate the design of existing DeFi attack hunting techniques and DeFi smart contract automated repair approaches, as well as their workflow.
Finally, we analyze the limitations of existing security tools when dealing with DeFi protocols, and provide overall novel insights. Our study also identifies several open issues and challenges in the DeFi ecosystem that need to be addressed by future work.

{\textbf{Contributions.}}\quad The key contributions of this paper are:
\begin{itemize}[itemsep=1pt, topsep=1pt, leftmargin=\dimexpr\labelwidth + 2 \labelsep\relax]
\item This work, to the best of our knowledge, is the first to present a comprehensive review of various DeFi attacks and a systematic evaluation of the state-of-the-art smart contract and DeFi security tools, empirically revealing both the capabilities and limitations of existing approaches.
\item In light of 57 reported DeFi attack incidents, we summarize six categories of DeFi attacks or vulnerabilities. Further, we also construct a benchmark dataset, which contains a total of $7,340$ DeFi smart contracts. To facilitate the community, we have released this dataset for public use at \url{https://github.com/Messi-Q/DeFi-Protocol}. 
\item We empirically study 42 bug detection tools and 8 automated repair techniques for smart contracts and DeFi protocols, with the analysis of their performance. We also discuss several open issues and challenges in the DeFi ecosystem that need to be tackled by future research.
\end{itemize}

\textbf{Paper Organizations.}\quad The remainder of the paper is organized as follows. In Section~\ref{background}, we give a brief introduction to the  background of blockchain, smart contract, and decentralized finance (DeFi). Thereafter, we summarize six common types of DeFi attacks and present concrete examples in Section~\ref{attacks}. In Section~\ref{techniques}, we investigate state-of-the-art bug detection and repair techniques for smart contracts and DeFi protocols. Section~\ref{evaluation} describes the dataset construction, and presents the evaluation results of available automated vulnerability detection tools on DeFi protocols. In Section~\ref{challenges}, we discuss several open issues and challenges. Finally, we conclude the paper in Section~\ref{conclusion}.

\section{Background}
\label{background}
Before diving into the key components, please allow us to introduce the required background on blockchain, smart contract, and decentralized finance.

\subsection{Blockchain In A Nutshell}
\label{blockchain}
Blockchain, a distributed ledger maintained in a decentralized manner, was first popularized as the technology behind Bitcoin~\cite{bohme2015bitcoin}. The ledger is shared among all participating nodes in the blockchain network. A public blockchain network has an open and transparent infrastructure, which is not owned or controlled by any centralized organization~\cite{chatterjee2017overview}. Put differently, blockchain can guarantee the fidelity and security of the recorded data without the need for a trusted intermediary.

More specifically, the distributed ledger contains a sequence of blocks, each of which records a set of transactions. Every block links to its previous block, forming a chain. When a user wants to add a new transaction to the ledger, the transaction data is verified by the so-called miners (bookkeeping nodes) and put into a new block~\cite{dos2019efficient}.  Once the new block is confirmed by miners that follow a consensus protocol, it is added to the chain~\cite{bamakan2020survey}. Each miner stores a duplicate transaction ledger, which endows the blockchain with decentralized and immutable nature, that is, nobody can alter and delete the data recorded in the ledger. 
%Unfortunately, miners herein retain the privilege to control single-handedly the transaction order of their mined blocks, an information asymmetry that is being exploited for profit~\cite{qin2021attacking}.

\textbf{Evolution of Blockchain.}\quad The first implementation of a blockchain system was {Bitcoin}, {i.e.,} decentralized digital currency verified using cryptography~\cite{wang2020designated}. The second wave of decentralized blockchain applications started with the advent of Ethereum~\cite{tacs2019building}, which introduced smart contracts and enabled multiple types of dapps. In recent years, a new subfield of blockchain, decentralized finance~\cite{liu2021first}, which specializes in advancing financial technologies and services on top of a permissionless blockchain (typically Ethereum), has attracted considerable attention worldwide. For a thorough background on blockchain, readers can refer to SoKs, such as~\cite{bonneau2015sok,atzei2017survey,bano2019sok}. %In what follows, we proceed to introduce smart contract, and decentralized finance. 

\subsection{Ethereum and Smart Contract}
\label{smart_contract}
Ethereum is one of the most prevalent blockchain platforms, which is the first to introduce the functionality of smart contracts~\cite{buterin2019next}. Ethereum smart contracts are developed in high-level programming languages such as Solidity~\cite{solidity} and Vyper~\cite{vyper}. Smart contracts are essentially a set of digital agreements with encoded rules that can be enforced without the involvement of a trusted third party. They are compiled to bytecode and executed within a virtual machine called the Ethereum Virtual Machine (EVM). Once deployed, smart contracts are uploaded to the blockchain and broadcast to all ledger nodes for backup. 

A smart contract can implement arbitrary rules for manipulating digital assets~\cite{zou2019smart}. The rules defined in a smart contract are strictly and automatically followed during execution, effectuating the `code is law' logic~\cite{liu2021combining}. 
%Transactions of a smart contract are executed by miners. Miners will receive a fee (called \emph{gas}) for running transactions, and the fee is paid by the users who issue the execution requests. Any transaction that runs out of gas will be aborted~\cite{albert2021don}. Detailed descriptions of Ethereum can be found in \emph{Ethereum Yellow Paper}~\cite{wood2014ethereum}.
While Ethereum is becoming one of the most influential blockchains, it has been exposed to a large number of smart contract security vulnerabilities~\cite{perez2019smart}. Since DeFi protocols are usually composed of multiple smart contracts, it inevitably becomes an attractive attack target, which has caused numerous financial losses~\cite{huang2019smart}. 
%Therefore, in this paper, we embrace to investigate the DeFi attacks and vulnerabilities that occurred on Ethereum, promoting readers to have a deep understanding of various DeFi attacks and facilitate the DeFi security community.

\begin{figure*}
\centering
\includegraphics[width=15.6cm]{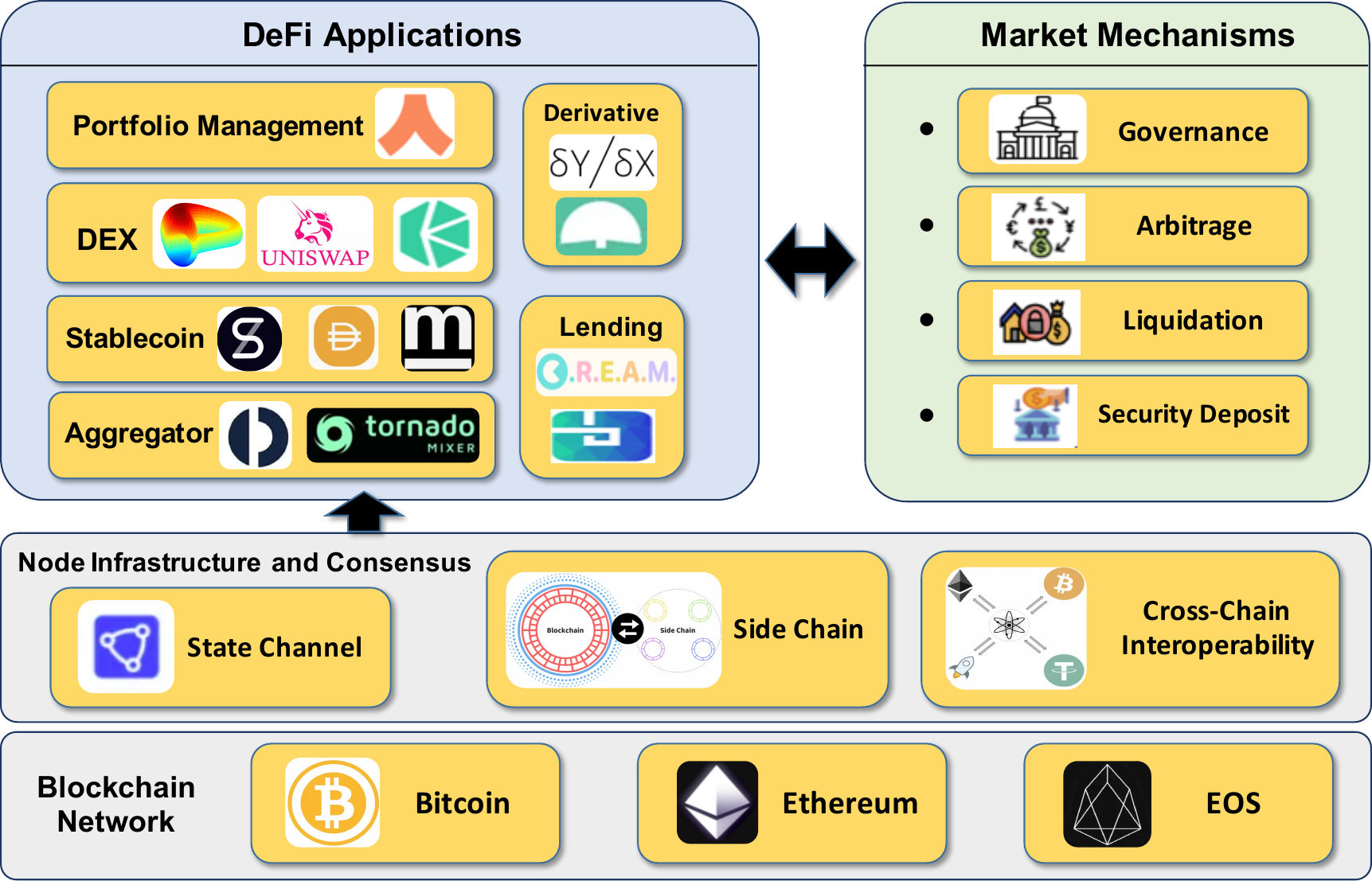}
\caption{A high-level overview of DeFi ecosystem. DeFi builds upon a distributed blockchain network, enabling various DeFi applications, such as portfolio management, decentralized exchanges, derivatives, lending and borrowing, stablecoin, and aggregator.} 
\label{fig}
\end{figure*}

\subsection{DeFi: Decentralized Finance}
\label{decentralized_finance}
Decentralized finance, a peer-to-peer financial system powered by the blockchain, is flourishing~\cite{schar2021decentralized}. With the widespread adoption of smart contracts, the concept of decentralized finance truly came to fruition, even to the point of hosting an economy exceeding \$200 billion~\cite{qin2021cefi,statista}. 
%Smart contract-enabled blockchains allow developers to construct flexible financial products on top of a blockchain system. 
DeFi builds upon the permissionless foundation provided by blockchains. Anyone is free to propose a novel financial contract, and anyone can interact with, transfer assets to, and withdraw assets from it, as long as the participants abide by the immutable contract rules. A DeFi protocol typically consists of multiple interactive smart contracts that run on top of the blockchain's state machine. 
%The state transition of a DeFi application will be reflected on its underlying blockchain, and the corresponding transactions issued by users in the peer-to-peer blockchain network will be broadcast to every node in the blockchain. 
Currently, a variety of financial services are migrating to the DeFi ecosystem, including lending and borrowing, decentralized exchanges, portfolio management, derivatives, and many others~\cite{moncada2020next,abdulhakeem2021powered,amler2021defi,makerdao}. To provide an intuition of what DeFi is and what DeFi can do, we present a high-level overview of DeFi in Fig.~\ref{fig}. Explicitly, we summarize the applications of DeFi into six categories.

\subsubsection{Lending and Borrowing}
Lending and borrowing of assets in a DeFi application are realized through protocols for funds loaning, which are referred to as the DeFi lending protocols~\cite{saengchote2021decentralized}. Decentralized lending services constitute the largest class of DeFi applications, with a cumulative TVL of over \$40 billion~\cite{qin2021quantifying}. They offer loans to individuals or businesses using smart contracts as negotiators or intermediaries, as smart contracts automate the lending and borrowing process~\cite{gudgeon2020defi}. 

In DeFi applications, loans are generally of two forms. One is the over-collateralized loan, where the borrower is required to post collateral, {i.e.,} provide something of value as security to cover the debt~\cite{qin2021empirical}. In this way, collateralization ensures that the lender can recover the loaned value and provides the borrower with an incentive to repay the loan. Another interesting form is the non-collateralized loan, {i.e.,} flash loan~\cite{wang2021towards}. Flash loans allow users to borrow any available amount of assets from a liquidity pool without upfront collateral, as long as the loan is repaid at the end of the transaction. Flash loans leverage the atomicity of the blockchain transaction ({i.e.,} a transaction reverts to its previous state if the loan is not repaid within the current transaction) to facilitate several use cases, {e.g.,} decentralized exchange arbitrage~\cite{daian2020flash} and collateral swaps~\cite{schar2021decentralized}. 

In general, a transaction reverts due to three possible reasons: 1) the transaction sender does not specify sufficient transaction fees, 2) the transaction does not satisfy a condition set forth by the interacting smart contract, or 3) the transaction conflicts ({e.g.,} double spending) with another transaction~\cite{zakhary2019atomic}. %This concept of a state reversion enables flash loans to be executed atomically within only one blockchain transaction.
Flash loans, therefore, entail two interesting properties. First, the lender is guaranteed that the borrower will repay the loan. If the repayment is not performed, the loan will not be given. Second, the borrower can technically request any amount of capital, up to the number of funds available in a flash loan pool. %, given a constant payment that corresponds to the blockchain transaction fees. 
While providing convenience, flash loans allow adversaries to launch malicious attacks with a large number of assets that they do not actually own. Undesirable trades using the flash loan have resulted in numerous notorious attacks on DeFi applications, such as~\cite{aave,balancer,compound}.

\subsubsection{Decentralized Exchange}
Decentralized exchange (DEX) is essentially a kind of DeFi protocol that enables the on-chain exchange of digital assets~\cite{wang2022cyclic}. Users can trade different tokens in a decentralized manner by interacting with smart contracts. Decentralized exchanges have accumulated over \$25 billion locked funds. 
For example, one of the largest DEXs is \texttt{Uniswap}~\cite{uniswap}, whose users have locked up around \$8 billion in token value. %\texttt{Uniswap} contract is publicly available on Ethereum, and any user can directly interact with it. 
Compared to traditional centralized exchanges, DEXs have advantages in privacy protection and asset management. There are two modes in DEXs, {i.e.,} list of booking (LOB) and automated market maker (AMM). DEXs in LOB mode maintain an off-chain order book to record users’ bids and asks, {namely}, the order matching is performed off-chain. DEXs in AMM mode achieve a fully decentralized exchange. AMM allows liquidity providers, the traders who are willing to provide liquidity to the market, to deposit assets into a liquidity pool. The market maker can deposit two or more tokens into a liquidity pool with a self-defined weight. The trade rate between cryptocurrencies in the pool is automatically calculated based on the pricing mechanism~\cite{zhou2021high}. AMM has become the most popular mode in DEXs due to its flexible liquidity~\cite{hertzog2017bancor,kyber}.

\subsubsection{Portfolio Management}
As more and more DeFi protocols motivate clients to provide liquidity, a new type of project, known as portfolio management, debuts to help users ({i.e.,} liquidity providers) invest their assets~\cite{heimbach2021behavior,chitra2022defi}. They automatically find the DeFi protocols that provide the highest annual percentage yield (APY). However, liquidity allocation is an arduous task for liquidity providers who seek to maximize their profits due to the complex and expansive space of yield-generating options. Therefore, the management of on-chain assets that serve as decentralized investment funds can be automated through DeFi protocols. An investment strategy that entails transacting with other DeFi protocols is encoded in the smart contracts, and the invested assets are deposited into the contract~\cite{mohan2020automated}.

\subsubsection{Derivative}
DeFi derivatives are built upon the smart contracts that derive value from the performance of an underlying entity, such as currencies, bonds, and interest rates~\cite{alao2021towards}. Tokenized derivatives can be created without trusted third parties and are able to prevent the influence of malicious attacks. While approximately 99\% of the derivative trading volume is created on centralized exchanges, a number of DeFi protocols have emerged that provide similar functionality, with a particular focus on futures, perpetual swaps, and options~\cite{wachter2021measuring}. Popular examples of DeFi derivatives include \texttt{CompliFi}~\cite{compliFi}, \texttt{dYdX}~\cite{dYdX}, and \texttt{BarnBridge}~\cite{barnBridge}, etc.

\subsubsection{Stablecoin}
Stablecoins are a class of cryptocurrencies designed to ensure price stability~\cite{klages2020stablecoins,mita2019stablecoin}. Typically, stablecoins are stabilized by either being directly/indirectly backed or intervened through various stabilization mechanisms. The popular stablecoins, such as \texttt{USDC} or \texttt{USDT}, are custodial and fall outside the scope of DeFi, as they primarily rely on a trusted third party. In decentralized settings, the challenge for the protocol designer is to construct a stablecoin that achieves price stability in an economically secure and stable manner and where all necessary parties can continue to participate profitably~\cite{saengchote2021defi}. Price stability is achieved through on-chain collateral that provides a foundation of secured loans from which the stablecoin derives its economic value. Non-custodial stablecoins aim to be independent of the societal institutions that custodial designs rely on. Popular stablecoins in DeFi include \texttt{Tether}~\cite{tether}, \texttt{DAI}~\cite{DAI}, \texttt{Diem}~\cite{Diem}, etc.

\subsubsection{Aggregator}
DeFi aggregator is essentially a platform that consolidates trades from various decentralized platforms into one place, increasing the efficiency of cryptocurrency trading~\cite{cousaert2021sok}. Typically, a DeFi aggregator leverages multiple DEXs and implements various buying and selling strategies to help users maximize profits, while mitigating gas fees and DEXs trading commissions~\cite{liu2021first}. Not only do aggregators pull the best prices, but some DeFi aggregators even offer a unique, user-friendly way to analyze and combine other users' trading strategies via a convenient drag-and-drop mechanism~\cite{li2021defi}. In particular, with the advent of DeFi aggregators, new entrants to the industry can benefit from DeFi without understanding the complex technologies such as trading, decentralized services, and blockchain. Overall, an aggregator helps users make better trading decisions. Popular DeFi aggregators include \texttt{1inch}~\cite{1inch}, \texttt{Matcha}~\cite{Matcha}, and \texttt{Plasma.Finance}~\cite{Plasma.Finance}.

\section{Overview of DeFi Attack}
\label{attacks}
DeFi has taken an incredible wave of popularity and enabled growing penetration in various industries. Nevertheless, due to the trillion dollars wealth it holds, DeFi attracts numerous external attacks, which severely threaten the security of the entire DeFi ecosystem~\cite{carter2021defi,wang2022speculative}. While multifarious DeFi attacks have attracted considerable attention, there is still a lack of a well-structured overview for providing an in-depth review and analysis of DeFi attacks. Motivated by this, we present a comprehensive review of various DeFi attacks, hoping to inspire the readers. 
Specifically, we examine the well-known DeFi attack incidents over the past three years, which are listed in Table~\ref{table1}. 
We summarize these attacks into six categories, i.e., \emph{flash loan attack (pump and arbitrage, price manipulation, reentrancy), deflation token attack, sandwich attack}, and \emph{rug pull attack}. It is worth noting that many existing works have investigated a variety of vulnerabilities in traditional smart contracts, such as \emph{timestamp dependency} and \emph{unhandled exception}~\cite{atzei2017survey,qian2022smart,chu2023survey}. Instead, in this work, we mainly take an insight into attacks and vulnerabilities targeting DeFi protocols. Next, we will go through the details of the DeFi attacks one by one.

\subsection{Flash Loan Attack}
\label{flash_loan_attack}
As an emerging service in the DeFi ecosystem, the flash loan allows users to apply for a non-collateral loan. While providing convenience, it enables adversaries to launch malicious attacks with a large number of assets that they do not have~\cite{gudgeon2020decentralized}. After scrutinizing existing DeFi attacks, we found that around $50\%$ of DeFi attack incidents involve flash loans~\cite{wang2021towards}. Flash loan attack has become the most prominent form among various DeFi attacks. In particular, we summarize three major types of flash loan attacks, namely, \emph{pump and arbitrage}~\cite{qin2021attacking}, \emph{price manipulation}~\cite{wu2021defiranger}, and \emph{reentrancy}~\cite{Inspex}. For each type of flash loan attack, we present the intuition to identify the essence of the attack. We then deliver a concrete example to clarify the attack.

\begin{figure}
\centering
\includegraphics[width=8.5cm]{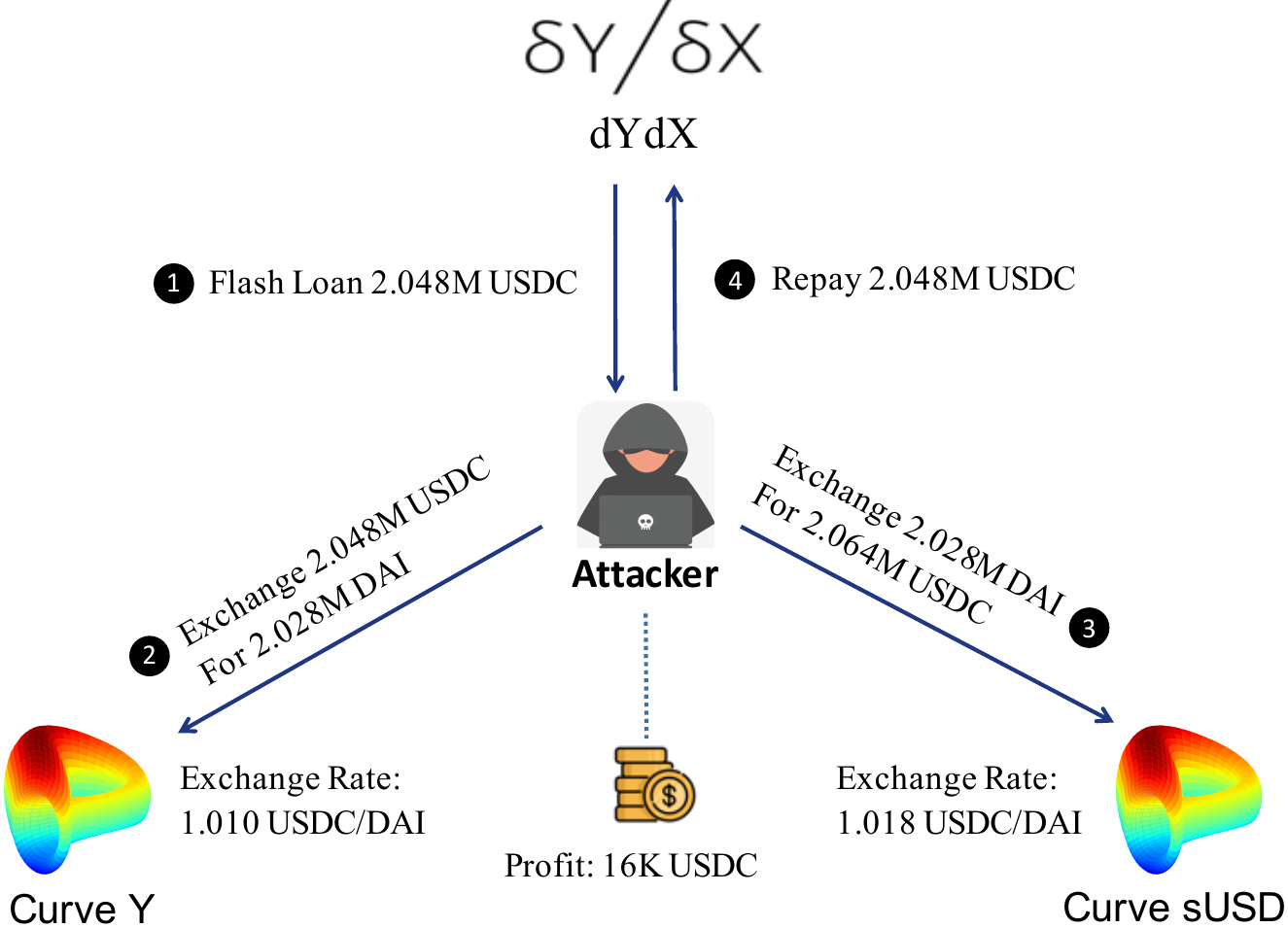}
\caption{A pump and arbitrage attack on the DeFi protocol — \texttt{dYdX}.} 
\label{fig1}
\end{figure}

\subsubsection{Pump and Arbitrage}
\label{pump_arbitrage}
Generally speaking, arbitrage is the practice of making a profit by trading on different exchanges that provide different prices for the same asset. Since the DeFi market reacts more slowly to events on the blockchain network than the real-world market, attackers can take advantage of market inefficiencies to buy and sell an asset at different prices, gaining financial benefits. With the flash loan, attackers can achieve arbitrage without any pre-owned asset. Particularly, once the price difference is found, attackers can instantly borrow a considerable amount of assets with a flash loan service and further achieve arbitrage.

\textbf{Intuition.}\quad The nucleus of the pump and arbitrage attack is that the attacker leverages the different prices of the same asset on different exchanges. For example, the attacker pumps the \texttt{ETH/WBTC} exchange rate on a constant product AMM ({e.g.,} \texttt{Uniswap}) with the leveraged funds of \texttt{ETH} in a margin trade. Afterwards, the attacker purchases \texttt{ETH} at a cheaper price on the distorted price market with the borrowed \texttt{WBTC} from a lending platform. 

\textbf{Example.}\quad We present, as an example, the specific details of the pump and arbitrage attack on the DeFi protocol \texttt{dYdX}\footnote{\texttt{dYdX}: \url{https://dydx.exchange}}, which is shown in Fig.~\ref{fig1}. Specifically, the attacker first borrows a flash loan of 2.048 million \texttt{USDC} from the exchange \texttt{dYdX} in step \ding{182}. Then, the attacker exchanges all holding \texttt{USDC} for 2.028 million \texttt{DAI} in another exchange \texttt{Curve Y} with an exchange rate of 1.010 \texttt{USDC/DAI} in step \ding{183}. Thereafter, he/she exchanges all \texttt{DAI} for 2.064 million \texttt{USDC} in the third exchange \texttt{Curve sUSD} with an exchange rate of 1.018 \texttt{USDC/DAI} in step \ding{184}. Finally, the attacker repays the original borrowed 2.048 million \texttt{USDC} in step \ding{185}. With the two exchanges, the attacker finally gains an arbitrage of 16 thousand \texttt{USDC} ({i.e.,} 16 thousand = 2.064 million - 2.048 million).

\subsubsection{Price Manipulation}
\label{price_attack}
Price manipulation is one of the most common flash loan attacks. In such attacks, the adversary distorts the reported price and propagates the distortion using a vulnerable customized function, causing the victim to receive inaccurate price information. As a result, the victim is tricked into overvaluing or undervaluing the target asset. Typically, the process of a price manipulation attack consists of the following four steps, all of which are executed within the same transaction to avoid being interrupted by other operations~\cite{wang2021promutator}.
\begin{itemize}[itemsep=1pt, topsep=1pt, leftmargin=\dimexpr\labelwidth + 2 \labelsep\relax]
\item \textbf{Prepare Target Asset.} First, an attacker prepares a target asset, whose price the attacker intends to inflate. Furthermore, the attacker borrows a large number of funds with a flash loan for the next phase.
\item \textbf{Inflate Asset Price.} Then, the attacker manipulates the price of the target asset by significantly reducing its reserves in the corresponding AMM liquidity pools. This is typically done by swapping a large amount of the target asset for another token. The attacker reports the manipulated price of the target asset to the victim.
\item \textbf{Profit From Victim.} Since the victim may overvalue the target asset, the attacker makes a profit by exchanging the target asset for another asset through the victim's services ({e.g.,} collateralized borrowing).
\item \textbf{Recover Asset Price.} Finally, the attacker performs the reverse actions to restore the unbalanced AMM liquidity pools to their original state. As a result, the attacker avoids the losses caused by the second-step price slippage and repays all his original assets, paying only for the swap fees.
\end{itemize} 

\textbf{Intuition.}\quad The core of the price manipulation attack is that the attacker lowers the token exchange rate by using a flash loan. In the second step, the adversary benefits from the decreased exchange rate. As an example, the attacker can gain a profit from the decreased \texttt{ETH/sUSD} exchange rate by borrowing \texttt{ETH} against \texttt{sUSD} as collateral.

\begin{figure}
\centering
\includegraphics[width=8.4cm]{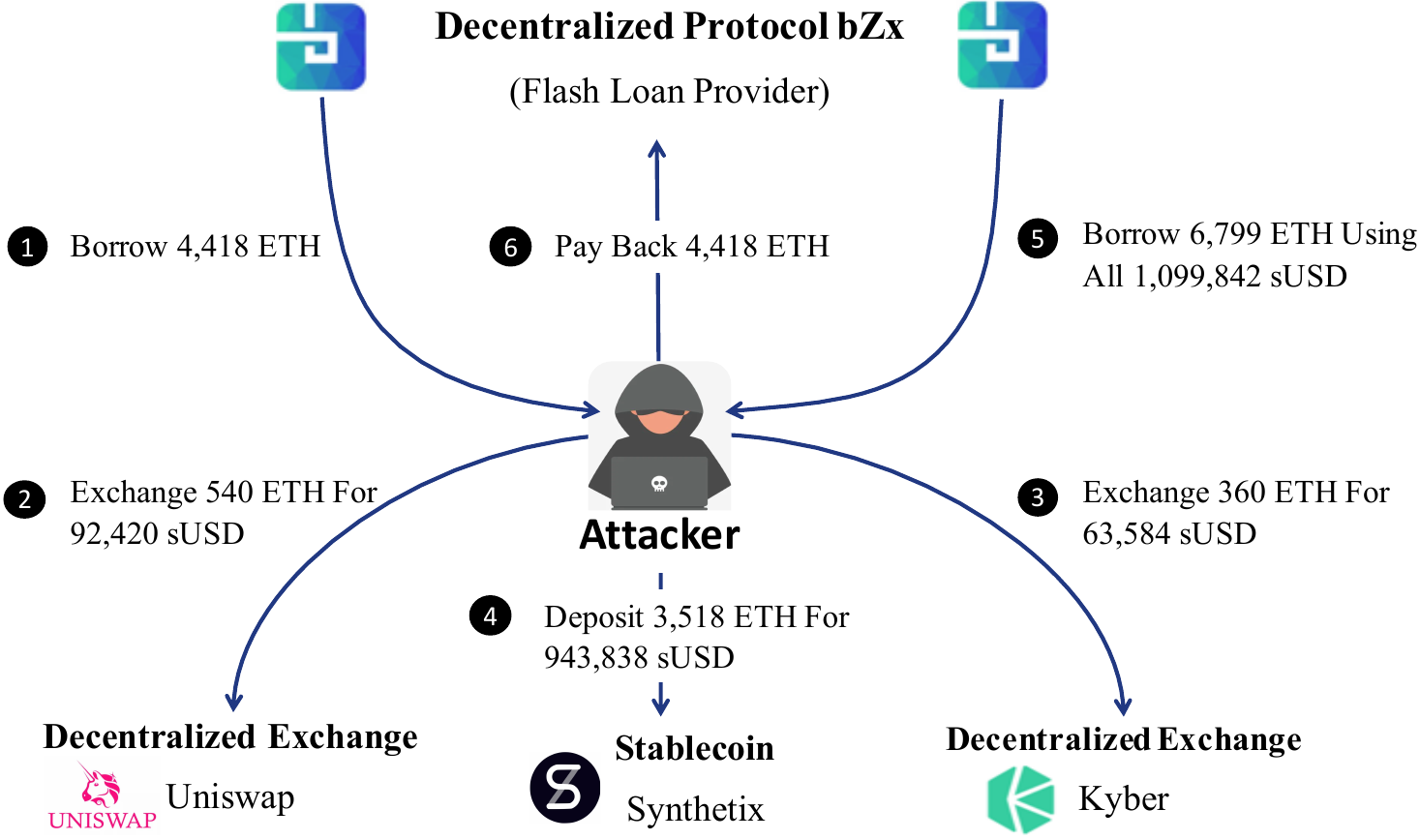}
\caption{A price manipulation attack on the DeFi protocol — \texttt{bZx}.} 
\label{fig2}
\end{figure}

\textbf{Example.}\quad To further elaborate on the price manipulation attack, we present a real-world attack example on the DeFi protocol \texttt{bZx}\footnote{\texttt{bZx}: \url{https://bzx.network} }, as shown in Fig.~\ref{fig2}. In this case, the attacker exploits the dependency of \texttt{bZx} on other DeFi protocols ({i.e.,} \texttt{Uniswap} and \texttt{Kyber}) to manipulate the asset exchange rates, making profits within a single atomic transaction. In particular, the attacker takes a sequence of six transactions, consisting of borrowing, exchanging, and repaying assets ({e.g.,} \texttt{ETH} and \texttt{sUSD}). Note that these transactions are packed into the same transaction in the exact order. 

Specifically, the attacker first borrows 4,418 \texttt{ETH} from \texttt{bZx} (step \ding{182} in Fig.~\ref{fig2}), then the attacker uses the 4,418 borrowed \texttt{ETH} to exchange for \texttt{sUSD} using other DeFi protocols, {i.e.,} \texttt{Uniswap}, \texttt{Synthetix}, and \texttt{Kyber}, respectively (steps \ding{183}--\ding{185} in Fig.~\ref{fig2}). Since \texttt{bZx} relies on \texttt{Uniswap} and \texttt{Kyber} for price oracles, which are susceptible to large amounts of transactions, the attacker can thus heavily skew the exchange rate of \texttt{ETH/sUSD} in \texttt{bZx} in his/her favor. After that, the attacker triggers step \ding{186}, namely, borrowing 6,799 \texttt{ETH} using all holding 1,099,842 \texttt{sUSD}. Finally, the attacker pays back 4,418 \texttt{ETH} in step \ding{187}, which is borrowed at the very beginning. The outcome of steps \ding{182}--\ding{187} is that the attacker gains a net profit of 2,381 \texttt{ETH} ({i.e.,} 2,381 = 6,799 - 4,418), with only a small amount of \texttt{ETH} to pay a gas fee.

\subsubsection{Reentrancy} 
\label{reentrancy_attack}
In traditional smart contracts, reentrancy is a well-known vulnerability that caused the notorious DAO attack \cite{ye2020clairvoyance}. 
When a function $F$ of a \emph{victim} contract transfers money to a malicious \emph{attack} contract $C$, due to the default settings of smart contracts, the \emph{fallback} function of $C$ is automatically triggered. The \emph{attack} contract $C$ can set a malicious operation in its \emph{fallback} function, {i.e.,} calling $F$ again to perform an illegal second-time transfer. As the current execution of $F$ is waiting for the first-time transfer to finish, the balance of $C$ may not be reduced yet, making $F$ wrongly believe that $C$ still has enough balance and transfer to $C$ again. Therefore, the \emph{attack} contract $C$ can exploit the reentrancy vulnerability to successfully steal additional money from the \emph{victim} contract. Similarly, DeFi protocols have also suffered from such a reentrancy attack~\cite{coinCodeCap}. However, unlike a traditional reentrancy attack on a single smart contract, the flash loan reentrancy attack on DeFi protocols usually involves multiple smart contracts. 
%In the smart contracts, a reentrancy attack occurs when a contract $\mathcal{S}$ makes an external call to another untrusted contract. Then, the untrusted contract makes a recursive call back to $\mathcal{S}$ in an attempt to steal money~\cite{ye2020clairvoyance}. 

\begin{figure}
\centering
\includegraphics[width=8.4cm]{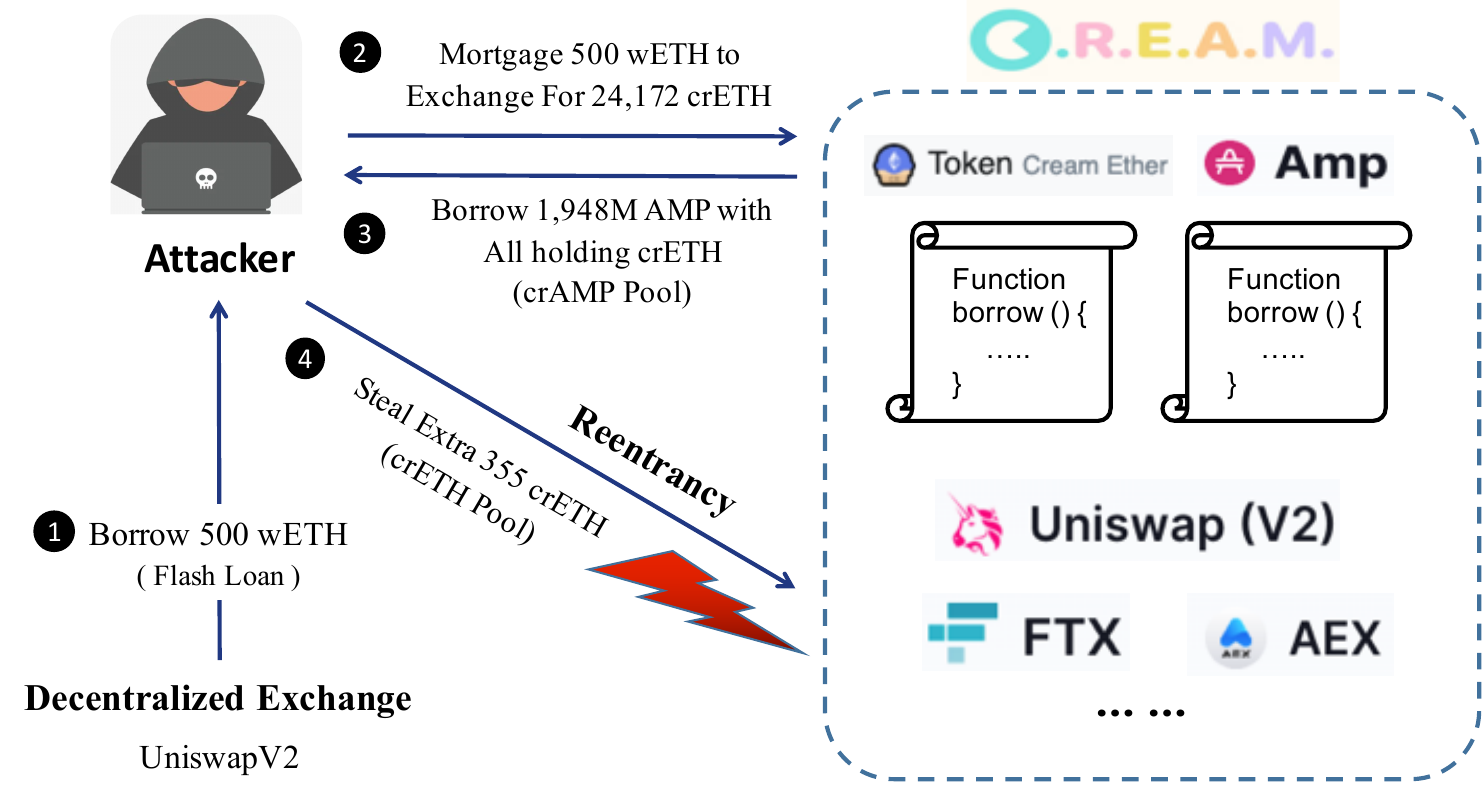}
\caption{A reentrancy attack on the DeFi protocol — \texttt{Cream.Finance}.} 
\label{fig3}
\end{figure}

\textbf{Intuition.}\quad In essence, a reentrancy attack is caused by some untrusted external calls that interrupt some atomic transactions ({e.g.,} transferring and accounting), resulting in inconsistent variable states. The core of the flash loan reentrancy attack is that the attacker exploits the inconsistent state of different loan pools before and after a contract's lending ({i.e.,} {transferring}) to achieve multiple unregistered loans.

\textbf{Example.}\quad To further illustrate the flash loan reentrancy attack, we present a typical example of the DeFi protocol \texttt{Cream.Finance}\footnote{\texttt{Cream.Finance}: \url{https://cream.finance}} in Fig.~\ref{fig3}. Specifically, attacker $\mathcal{A}$ borrows 500 \texttt{wETH} from \texttt{UniSwapV2} with a flash loan in step \ding{182}. Then, $\mathcal{A}$ collateralizes the borrowed 500 \texttt{wETH} to \texttt{Cream.Finance} in exchange for 24,172 \texttt{crETH} in step \ding{183}. Note that \texttt{Cream.Finance} consists of multiple loan pools, each of which is essentially a smart contract. Thereafter, $\mathcal{A}$ invokes the function \emph{borrow} in the \texttt{crAMP} pool and exchanges for 1,948 million \texttt{AMP} with all holding \texttt{crETH} in step \ding{184}. However, the transfer of an \texttt{AMP} automatically triggers the \emph{fallback} function in the attacker's contract. Since the borrowing status of $\mathcal{A}$ in the \texttt{crAMP} pool has not been updated, attacker $\mathcal{A}$ calls the function \emph{borrow} in another pool \texttt{crETH} in step \ding{185}, thus gaining an extra \texttt{crETH} from the pool.

\subsection{Deflation Token Attack}
\label{deflation_token}
Deflation is a term used in crypto finance to describe a drop in the value of an asset due to certain factors such as over-minting~\cite{lu2021freeswap}. Token deflation refers to the phenomenon that the total supply of tokens decreases each time a token transfer happens. With the rise of decentralized finance, various deflation tokens are emerging. Some types of deflation tokens in DeFi are implemented in the form of deducting a certain percentage of tokens for destruction and redistribution each time a user performs a transaction transfer~\cite{amler2021defi}. This may result in the actual number of tokens received by the recipient being less than the amount paid by the sender. The principle of the deflation token attack is that the attacker exploits such characteristics to reduce the number of a target token by performing multiple in and out transfers, in order to profit from the difference in the amount of the deflation token.

\begin{figure}
\centering
\includegraphics[width=8.3cm]{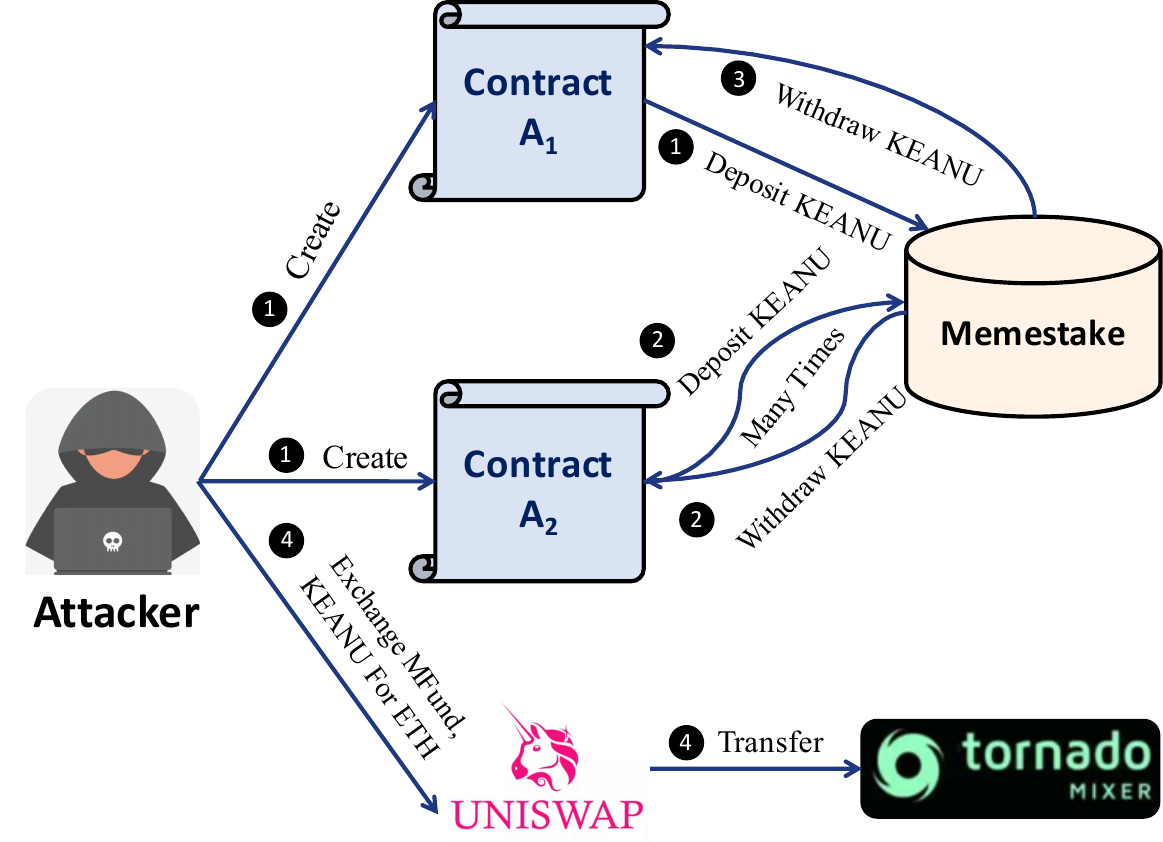}
\caption{A real-world deflation token attack on DeFi protocol — \texttt{Sanshu Inu}.} 
\label{fig4}
\end{figure}

\textbf{Example.}\quad Fig.~\ref{fig4} presents a real-world example to illustrate the deflation token attack on the \texttt{Memestake} contract of the DeFi protocol \texttt{Sanshu Inu}\footnote{\texttt{Sanshu Inu}: \url{https://sanshuinufinance.com}}. The overall attack process can be summarized into the following four steps.
\begin{itemize}[itemsep=1pt, topsep=1pt, leftmargin=\dimexpr\labelwidth + 2 \labelsep\relax]
\item First, the attacker creates two contracts and initializes them accordingly. Specifically, contract $A_1$ is an investment contract that deposits 2,049 billion \texttt{KEANU} tokens into the \texttt{Memestake} pool, and $A_2$ is the attack contract that is used to manipulate the reward calculation of \texttt{Memestake}.
\item Then, the attacker calls $A_2$ to deposit/withdraw a large number of \texttt{KEANU} tokens to/from \texttt{Memestake} many times, forcing \texttt{Memestake} to trade \texttt{KEANU} in large quantities. Since \texttt{KEANU} is a deflation token, each transaction costs $2\%$ of the transaction amount, resulting in the real number of tokens that users deposit to \texttt{Memestake} being smaller than the registered number in \emph{user.amount} maintained by \texttt{Memestake}. However, when $A_2$ performs the withdrawal operation, \texttt{Memestake} still transfers the recorded amount of tokens to $A_2$, which leads to the continuous reduction of the token holdings of \texttt{KEANU} in the \texttt{Memestake} pool.
\item Next, the attacker uses $A_2$ to modify the value of \texttt{accMfundPerShare} in the \texttt{Memestake}. This value depends on the number of \texttt{KEANU} tokens  in the pool that the attacker manipulated in {Step} \ding{183}. That is, the real number of \texttt{KEANU} tokens in the \texttt{Memestake} has been reduced to a small one. Thus, when $A_1$ withdraws the \texttt{KEANU} tokens deposited in {Step} \ding{182}, it can receive a reward of \texttt{MFund} tokens far in excess of the normal value.
\item Finally, the attacker swaps the extracted \texttt{MFund} and \texttt{KEANU} tokens to \texttt{ETH}, and transfers them away through \texttt{Tornado}, making a net profit of 56 \texttt{ETH}.
\end{itemize}

\subsection{Sandwich Attack}
\label{sandwich_attack}
A sandwich attack is a typical predatory trading strategy in which a trader wraps a victim transaction with two malicious transactions, one before the victim transaction and one after the victim transaction~\cite{wang2022impact,zhou2021high,yuksel2021mitigating,qin2021quantifying}. In such attacks, the malicious attacker first scans the mempool for pending transactions and finds that a user (i.e., the victim) is trying to trade an asset X for another asset Y. Then, the attacker buys asset Y at a low price. This transaction may be earlier than the victim's transaction. Once this transaction is completed before the victim's transaction, the price of asset Y will increase accordingly\footnote{ The exchange rate of each transaction is determined by preset algorithms and market liquidity reserves~\cite{adams2021uniswap}. A buy order will increase the price of an asset, while a sell order decreases the price of the asset. }. When the victim's transaction is carried out, he/she will receive a smaller quantity of asset Y than he/she was supposed to receive due to the increase in the price of asset Y. Finally, the attacker sandwiches the victim's transaction by selling off asset Y at a higher price than he/she bought it, thereby profiting from the manipulated price. This attack sequence allows the attacker to pocket a profit by front-running and back-running a trader, creating an artificial price rise.

It is worth mentioning that, in the blockchain, the miners sort the transactions in the memory pool according to the \emph{gas price}, and then select the transactions in the order of the gas price from high to low when creating a block~\cite{weber2017availability,zarir2021developing}. Therefore, an attacker can achieve the sandwich attack as long as the gas price of the front-running transaction is set slightly higher than that of the victim transaction, and the gas price of the back-running transaction is lower than that of the victim transaction.

\textbf{Intuition.}\quad The core of the sandwich attack is that the attacker hunts for a victim transaction. Then, he/she quickly buys the asset at a low price and ensures that this transaction is scheduled just before the victim transaction ({i.e.,} front-running). Finally, the attacker sells the asset shortly after the victim's transaction ({i.e.,} back-running) to make a profit.

\begin{figure}
\centering
\includegraphics[width=8.5cm]{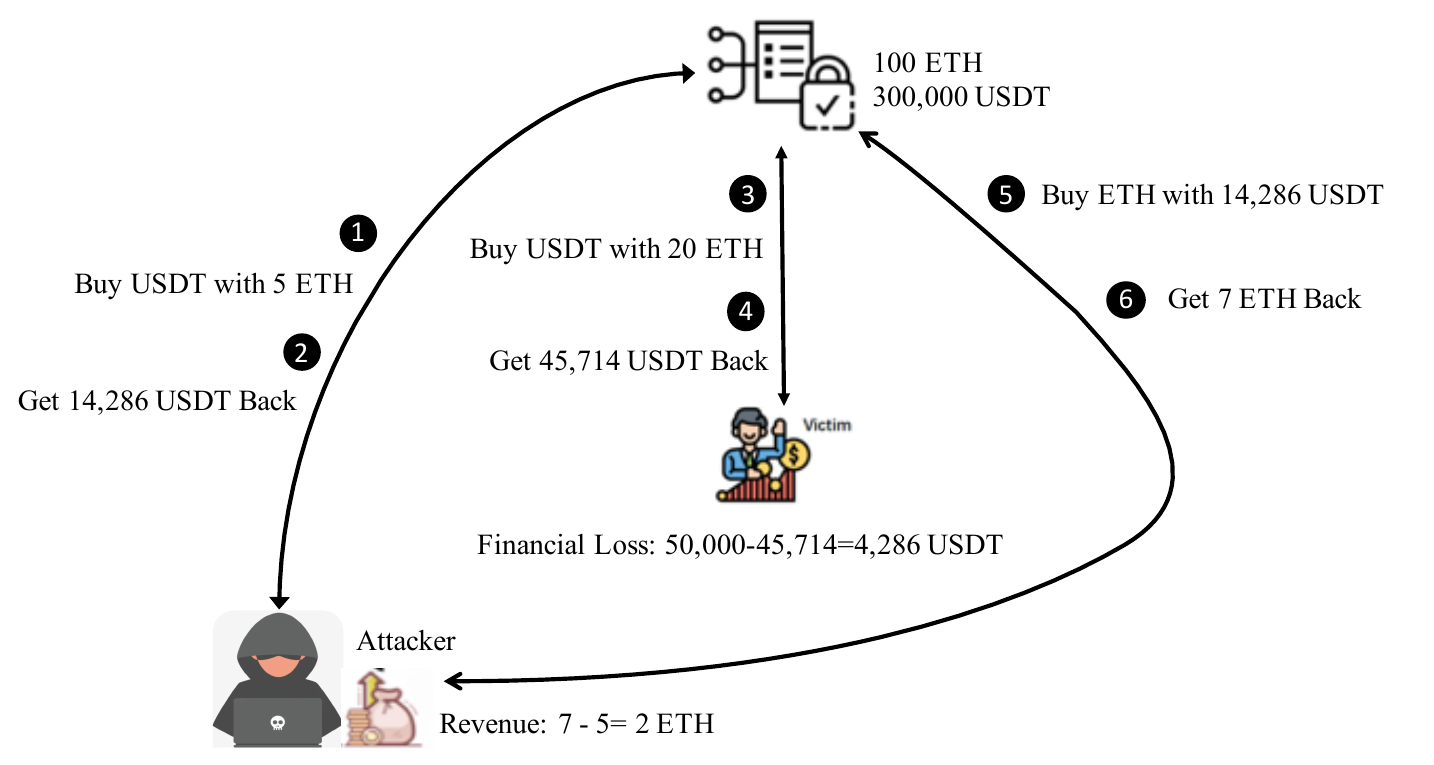}
\caption{A simplified example of the sandwich attack on AMM DEXs.} 
\label{sandwich}
\end{figure}

\textbf{Example.}\quad As shown in Fig.~\ref{sandwich}, the attacker front-runs the victim transaction with a buy order of \texttt{USDT} and back-runs the victim transaction with a sell order. Specifically, the victim transaction aims to buy \texttt{USDT} with 20 \texttt{ETH}. If there are no attackers, the trader will get back 500,000 \texttt{USDT}. However, the attacker front-runs to exchange 14,286 \texttt{USDT} for 5 \texttt{ETH} in steps \ding{182}--\ding{183} and sandwiches the victim transaction. Since the reserves in the liquidity pool change after each transaction. The original state of \texttt{ETH/USDT} is 1/3,000. After the attacker's first transaction, the state of \texttt{ETH/USDT} becomes 1/2,721. Therefore, in steps \ding{184}--\ding{185}, the victim can only exchange 45,714 \texttt{USDT} for 20 \texttt{ETH}, resulting in a loss of 4,286 \texttt{USDT}. After the victim's transaction, the state of \texttt{ETH/USDT} in the liquidity pool changes to 1/1,920. Finally, the attacker uses all holding 14,286 \texttt{USDT} to exchange for 7 \texttt{ETH} in steps \ding{186}--\ding{187}, gaining a profit of 2 \texttt{ETH}.

% shows a simplified process example of the sandwich attack. We can observe that $T_V$ is the transaction issued by a victim.  $T_{A1}$ and $T_{A2}$ denote the first and second transactions of the attacker, respectively. $T_V$ is initially placed before $T_{A1}$, but $T_{A1}$ front-runs $T_V$. Then, after $T_V$ is carried out, $T_{A2}$ back-runs it, thereby completing the sandwich attack. 

\subsection{Rug Pull Attack}
\label{rug_pull}
Rug pull refers to the draining of investments from DEX liquidity pools or the sudden dereliction of a project, sweeping away investors' funds without any warning signs~\cite{xia2021demystifying,scharfman2022decentralized}. The rug pull attack occurs mostly in decentralized exchanges (DEXs) and is a typical scam in DeFi. Scammers invest a lot of money in liquidity pools and publish attractive advertisements on social media to attract investors. Once investors deposit tokens into these liquidity pools, the scammers will ``sweep the carpet'', namely, withdraw all the tokens from the pools.

A typical type of rug pull attack involves manipulating the token price with the attacker's reserves, such as provoking the token value to drop precipitously to zero. The rug pull attacks in DeFi have caused more than \$2.8 billion in losses in 2021, making it the biggest scam in the DeFi ecosystem (accounting for $37\%$ of all scam revenue in 2021)~\cite{rugfull}.

\textbf{Example.}\quad In July 2021, the digital collectible platform \texttt{Bondly Finance} was compromised~\cite{bondly}. According to the disclosure of the project, the attacker gained access to the password account of the CEO of \texttt{Bondly Finance} through a carefully planned strategy. The password account contains the {mnemonic recovery phrase} of his hardware wallet, which allows the attacker to access the \texttt{BONDLY} smart contract after copying the password. The attacker exploited this vulnerability to mint 373 million \texttt{BONDLY} on Ethereum, resulting in a significant drop in the token price and a loss of \$5.9 million.

\renewcommand{\arraystretch}{1.1}
\begin{table*}
      %\tiny
	\centering
	\caption{\footnotesize A summary of well-known DeFi protocols that have suffered from external attacks. Specifically, \textbf{Category} means the category of DeFi financial services. \textbf{Attack Type} denotes the attack type that a DeFi protocols had endured. \textbf{Loss} indicates the corresponding financial losses caused by the attack. \textbf{Date} indicates the time when the attack occurred. } 
	\resizebox{1.005\textwidth}{!}{
		\begin{tabular}{|c|c|c|c|c|c|c|} 
		\hline
		 \textbf{\#}&\textbf{DeFi Protocols}&\textbf{Category}&\textbf{Attack Type}&\textbf{Loss}\tnote{2}&\textbf{Date}& \textbf{Ref.}  \\
			\hline
	           1&{bZx-V1}& Lending and Borrowing&\multirow{26}*{Flash Loan Attack}&{\$9.4 Million} & 2020.02.17  &~\cite{bZx-V1} \\
			 \cline{1-3}\cline{5-7}
		    2& {MakerDAO}&{Other}&~&{\$9 Million} & 2020.03.12  &~\cite{makerdao2}  \\
			 \cline{1-3}\cline{5-7}
		 3& {UniSwap}& {Decentralized Exchange}&~&{\$0.3 Million} & 2020.04.18 &~\cite{uniswap2}  \\
			 \cline{1-3}\cline{5-7}
			 4& {bZx-V2}& Lending and Borrowing&~&{\$8.1 Million} &  2020.09.15  &~\cite{bZx-V2} \\
			   \cline{1-3}\cline{5-7}
			 5& {Eminence Finance}&{Gaming}&~&{\$15 Million} &  2020.09.29  &~\cite{eminence} \\
			 \cline{1-3}\cline{5-7}
			 6& {Harvest Finance}&{Derivative}&~&{\$33.8 Million} &  2020.10.26  &~\cite{harvest}  \\
			\cline{1-3}\cline{5-7}
			 7&  {Akropolis}&{Aggregator}&~&{\$2 Million} &  2020.11.12  & ~\cite{akropolis2} \\
			  \cline{1-3}\cline{5-7}
			 8& {Pickle Finance}& {Yield Aggregator}&~&{\$20 Million} & 2020.11.22  & ~\cite{pickle2}  \\
			 \cline{1-3}\cline{5-7}
			 9& {Warp Finance}&  Lending and Borrowing&~&{\$7.7 Million} & 2020.12.17  &~\cite{warp}  \\
			\cline{1-3}\cline{5-7}
			 10&{Alpha Homora}&{Derivative}& ~ &{\$37.5 Million} & 2021.02.13 & ~\cite{alpha} \\
			\cline{1-3}\cline{5-7}
			 11& {Spartan}& {Decentralized Exchange}&~&{\$40 Million} & 2021.05.02  &~\cite{spartan} \\
			 \cline{1-3}\cline{5-7}
			 12& {Value DeFi}& {Derivative}&~&{\$28 Million} & 2021.05.07  &~\cite{valuedefi1,valuedefi2,valuedefi3}   \\
			 \cline{1-3}\cline{5-7}
			 13& {Venus}& {Lending and Borrowing}&~&{\$300 Million} & 2021.05.19  & ~\cite{venus}   \\
			\cline{1-3}\cline{5-7}
			 14& {PancakeBunny}& {Aggregator}&~&{\$47.1 Million} & 2021.05.20  &~\cite{pancakeBunny,pancakeBunny2}  \\
			 \cline{1-3}\cline{5-7}
			 15& {Burgerswap}&{Decentralized Exchange}&~&{\$7.2 Million} & 2021.05.28  &~\cite{Burgerswap} \\
			 \cline{1-3}\cline{5-7}
			 16& {JulSwap}&{Decentralized Exchange}&~&{\$0.7 Million} &  2021.05.28 &~\cite{julswap}  \\
			  \cline{1-3}\cline{5-7}
			 17& {Belt Finance}&{Decentralized Exchange}&~&{\$50 Million} &  2021.05.30 &~\cite{belt2} \\
			 \cline{1-3}\cline{5-7}
			 18& {SushiSwap}& {Decentralized Exchange}&~&{\$0.11 Million} & 2021.07.21  & ~\cite{sushiSwap} \\
			 \cline{1-3}\cline{5-7}
			 19& {THORChain}&{Cross-Chain Bridge}&~&{\$5.8 Million} & 2021.07.22 &~\cite{THORChain1}  \\
			 \cline{1-3}\cline{5-7}
			 20& {Vee Finance}& {Lending and Borrowing}&~&{\$13.5 Million} & 2021.09.21 &~\cite{vee}  \\
			\cline{1-3}\cline{5-7}
			 21& {Indexed Finance }&{Aggregator}&~&{\$16 Million} &  2021.10.14  &~\cite{indexed}  \\
			\cline{1-3}\cline{5-7}
			 22& {Cream Finance}& Lending and Borrowing&~&{\$130 Million} &  2021.10.27 &~\cite{cream.finance2} \\
			\cline{1-3}\cline{5-7}
			 23& {Nerve}&{Portfolio Management}&~&{\$8 Million} &  2021.11.15 &~\cite{nerve}  \\
			\cline{1-3}\cline{5-7}
			 24& {Beanstalk}&{Stablecoin}&~&{\$182 Million} & 2022.04.17  &~\cite{beanstalk} \\
			\cline{1-3}\cline{5-7}
			 25& {Saddle Finance}& {Decentralized Exchange}&~&{\$10 Million} & 2022.04.30 &~\cite{saddle}  \\
			\hline
			 26& {Balancer}&{Portfolio Management}&\multirow{3}*{Deflation Token Attack}&{\$0.5 Million} &  2020.06.28 &~\cite{balancer2} \\
			 \cline{1-3}\cline{5-7}
			 27& {Sanshu Inu}&{Gaming}&~&{\$5 Million} &  2020.12.28  &~\cite{sanshu2}  \\
			  \cline{1-3}\cline{5-7}
			 28& {Ankr}&{Other}&~&{\$5 Million} &  2022.12.02  &~\cite{Ankr}  \\
			 \hline
			 29& {Lendf.Me}&{Lending and Borrowing}&\multirow{4}*{Reentrancy Attack}&{\$25 Million} &  2020.04.18 &~\cite{lendf.me}  \\
			 \cline{1-3}\cline{5-7}
			 30& {Origin}& {Other}&~&{\$7 Million} & 2020.11.17  &~\cite{origin}  \\
			 \cline{1-3}\cline{5-7}
			 31& {Grim Finance}&{Derivative}&~&{\$30 Million} &  2021.12.19  &~\cite{grim}   \\
			 \cline{1-3}\cline{5-7}
			 32& {Rari Capital}& {Lending and Borrowing}&~&{\$90 Million} & 2022.04.30 &~\cite{rari2,rari3}  \\
			 \cline{1-3}\cline{5-7}
			 33& {FEI protocol}& {Stablecoin}&~&{\$80 Million} & 2022.04.30 &~\cite{Fei}  \\
			  \hline
			 34& {Mango}&{Decentralized Exchange}&{Sandwich Attack}&{\$100 Million}  & 2022.10.12  & ~\cite{Mango} \\
			 \hline
			 35& {BONDLY}&{Stablecoin}&{Rug Pull Attack}&{\$22 Million}  & 2021.07.15  & ~\cite{BONDLY2} \\
			 \hline
			 36& {Bancor}&{Decentralized Exchange}&\multirow{7}*{Smart Contract Logic Bug}&{\$23.5 Million} & 2020.06.16  &~\cite{bancor} \\
			\cline{1-3}\cline{5-7}
			 37& {DODO Pool}&{Decentralized Exchange}&~&{\$3.8 Million} &  2021.03.09  &~\cite{dodo} \\
			 \cline{1-3}\cline{5-7}
			 38& {Uranium Finance}& {Decentralized Exchange}&~&{\$50 Million} & 2021.04.28 &~\cite{Uranium}  \\
			 \cline{1-3}\cline{5-7}
			 39& {ChainSwap}& {Cross-Chain Bridge } &~&{\$4.8 Million}  &  2021.07.11 &~\cite{ChainSwap1,ChainSwap2} \\
			 \cline{1-3}\cline{5-7}
			 40& {Popsicle}& {Derivative}&~&{\$25 Million} & 2021.08.04  &~\cite{popsicle2} \\
			\cline{1-3}\cline{5-7}
			 41& {Poly Network}& {Cross-Chain Bridge}&~&{\$611 Million} &  2021.08.10 &~\cite{poly2}   \\
			\cline{1-3}\cline{5-7}
			 42& {Compound Finance}& Lending and Borrowing&~&{\$160 Million} &  2021.10.03  & ~\cite{compound2}  \\
			  \hline
			 43& {PAID Network}& {Decentralized Exchange}&\multirow{2}*{Compromised Private Key}&{\$160 Million} &  2021.03.05 &~\cite{PAID2}  \\
			 \cline{1-3}\cline{5-7}
			 44& {Ronin}& {Gaming}&~&{\$622 Million} & 2022.03.23 &~\cite{ronin}  \\
                         \hline
			 45& {Oypn}& {Insurance}&\multirow{13}*{Other}&{\$0.37 Million} &  2020.08.04  &~\cite{oypn}  \\
			 \cline{1-3}\cline{5-7}
			 46& {Chainlink}& {Stablecoin} &~&{\$0.34 Million} &  2020.08.30  &~\cite{Chainlink} \\
			 \cline{1-3}\cline{5-7}
			 47& {Cover}&{Insurance}&~&{\$5 Million} &  2020.12.28  &~\cite{cover}  \\
			 \cline{1-3}\cline{5-7}
			 48& {yCredit Finance}& {Decentralized Exchange}&~&{\$326 Million} & 2021.01.02 &~\cite{ycredit2}  \\
			  \cline{1-3}\cline{5-7}
			 49& {Yearn Finance}& {Aggregator}&~&{\$11 Million} & 2021.02.04 &~\cite{yearn2}  \\
			  \cline{1-3}\cline{5-7}
			 50& {Furucombo}&{Aggregator}&~&{\$15 Million} &  2021.02.27  &~\cite{furucombo2}   \\
			 \cline{1-3}\cline{5-7}
			 51& {EasyFi}&{Lending and Borrowing}&~&{\$80 Million} &  2021.04.20 &~\cite{easyfi2} \\
			 \cline{1-3}\cline{5-7}
			 52& {VaultSX}& {Other}&~&{\$13.5 Million} & 2021.05.16 &~\cite{vaultSX}  \\
			  \cline{1-3}\cline{5-7}
			 53& {AnySwap}&{Decentralized Exchange}&~&{\$7.9 Million}  & 2021.07.10  &~\cite{anyswap} \\
			 \cline{1-3}\cline{5-7}
			 54& {SafeDollar}& {Stablecoin}&~&{\$0.25 Million} & 2021.07.28  &~\cite{safeDollar}  \\	
                          \cline{1-3}\cline{5-7}
			 55& {BadgerDAO}&{Other}&~&{\$120 Million} &  2021.12.02 &~\cite{badgerdao} \\
			 \cline{1-3}\cline{5-7}
			 56& {Wormhole}&  Message Passing & ~ &{\$7.7 Million} &  2022.02.02 &~\cite{Wormhole}  \\
			  \cline{1-3}\cline{5-7}
			 57& {Nomad}&  Cross-chain Bridge & ~ &{\$190 Million} &  2022.08.02 &~\cite{Nomad}  \\
			\hline
	\end{tabular}
}
\label{table1}
\end{table*}

Due to the complexity and diversity of the DeFi ecosystem, we are unable to describe each type of DeFi attack and vulnerability in detail. We have listed the well-known DeFi protocols that have suffered from a DeFi attack over the past three years in Table~\ref{table1}. %n particular, we also provide an anchored address or URL to ensure the traceability of each DeFi protocol in our dataset, expecting to facilitate future research. 
Additionally, we refer the readers to~\cite{quadriga} for more hacks, frauds, and scams in DeFi.

\begin{tcolorbox}[arc = 1mm, colback = black!2!white, colframe = black!65!white, boxrule=0.6mm]
\textbf{Summary: Why are various DeFi attack incidents emerging?} 

  \textbf{(1)} DeFi is open-source, suggesting that its code is available to everyone, which inevitably introduces security risks. 

  \textbf{(2)} A DeFi protocol is vulnerable to external exploitation due to its complexity and composability. 

  \textbf{(3)} DeFi protocols tend to be launched in a rush. In order to earn the new financial market share promptly, some developers did not perform adequate security checks of the DeFi protocol, leaving a potential risk of security vulnerabilities.
\end{tcolorbox}

\section{Review of Smart Contract and DeFi Security Tools} 
\label{techniques} 
To safeguard the security of participant funds and privacy in smart contracts and DeFi protocols, recent advancements put forward corresponding solutions by incorporating vulnerability detection and automated repair techniques.
\begin{itemize}[itemsep=1pt, topsep=1pt, leftmargin=\dimexpr\labelwidth + 2 \labelsep\relax]
\item \textbf{Vulnerability Detection.} A plethora of work has been designed to automatically identify vulnerabilities in smart contracts. Existing surveys have provided a relatively comprehensive overview of current bug-finding tools for smart contracts. However, a fine-grained classification and empirical comparison of existing methods is still lacking. Moreover, many of these approaches are tailored for traditional smart contracts, which are typically single contracts. It is necessary and critical to verify whether these methods can effectively handle complex DeFi protocols. Towards these, we not only perform an empirical analysis of existing smart contract bug detection tools but also evaluate their effectiveness in detecting vulnerabilities in DeFi protocols. 

Furthermore, several recent efforts have explored detecting attacks against DeFi protocols by analyzing the DeFi transactions and restoring the high-level DeFi semantics. Given the lack of review of attack detection tools on DeFi protocols, we illustrate the design of existing DeFi attack hunting techniques, as well as their workflow.
\item \textbf{Automated Repair.} While existing tools are able to identify vulnerabilities and highlight the affected code lines in smart contracts, they exhibit certain limitations in automatically providing patches. In this context, smart contract automated repair is emerged as a countermeasure. Unfortunately, due to the immutability of the underlying blockchain, the smart contract cannot be updated once deployed. Therefore, traditional program repair techniques cannot be applied to smart contracts. This makes it challenging to fix vulnerabilities in smart contracts.  Recently, several efforts have been proposed to automatically patch vulnerabilities in smart contracts and DeFi protocols, leading to a new branch of smart contract security. However, a systematic review of existing automated repair methods for smart contracts is still missing. Towards this, we conduct a comprehensive review of current automated repair approaches, expecting to provide overall insights and facilitate future work.
\end{itemize}

In what follows, we will introduce the existing bug detection and automated repair techniques for smart contracts and DeFi protocols, respectively.

\subsection{Vulnerability Detection}
\label{bug_detection}
In this subsection, we investigate 42 well-known approaches that are able to detect smart contract vulnerabilities and DeFi attacks, which are listed in Table~\ref{table2}. More specifically, we first demonstrate the principles of the smart contract vulnerability detection tools, organized by the specific analysis techniques on which they are mainly based on. Furthermore, we summarize the vulnerability types supported by the smart contract vulnerability detection tools and list them in Table~\ref{table3}. Finally, we elaborate on the design of existing DeFi attack hunting approaches as well as their workflow, hoping to push forward the  boundaries of this research direction.

\renewcommand{\arraystretch}{1.07}
\begin{table*}[h]
	\centering
	\caption{\footnotesize Overview of state-of-the-art methods for detecting vulnerabilities and attacks targeting smart contracts and DeFi protocols.}
	\resizebox{1.00\textwidth}{!}{
		\begin{tabular}{|c|c|c|c|c|c|}
			\hline
			\textbf{\#} & \textbf{Tool} & \textbf{Type}& \textbf{Analysis Level} & \textbf{Public Available} &\textbf{Reference}\\
			\hline
			1 & {\textsc{Securify}} & \multirow{4}{*}{{Formal Verification}} & Bytecode & https://github.com/eth-sri/securify2 &~\cite{securify,securify2} \\
			\cline{1-2}\cline{4-6}  
			2 & {\textsc{VeriSmart}} &  & Source Code & https://github.com/kupl/VeriSmart-public &~\cite{so2020verismart} \\
			\cline{1-2}\cline{4-6}  
			3 & {\textsc{VeriSol}} & & Source Code & https://github.com/microsoft/verisol &~\cite{wang2018formal} \\
			\cline{1-2}\cline{4-6}  
			4 & {\textsc{Zeus}} &  & Source Code & Not Available &~\cite{kalra2018zeus} \\
			\hline 
			5 & {\textsc{DefectChecker}} & \multirow{10}{*}{{Symbolic Execution}} & Bytecode & https://github.com/Jiachi-Chen/DefectChecker &~\cite{chen2021defectchecker} \\
			\cline{1-2}\cline{4-6}
			6 & {\textsc{HoneyBadger}} &  & Bytecode & https://github.com/christoftorres/HoneyBadger &~\cite{torres2019art} \\
			\cline{1-2}\cline{4-6}
			7 & {\textsc{Maian}} &  & Bytecode & https://github.com/MAIAN-tool/MAIAN &~\cite{nikolic2018finding} \\
			\cline{1-2}\cline{4-6}
			8 & {\textsc{Manticore}} &  & Bytecode & https://github.com/trailofbits/manticore &~\cite{mossberg2019manticore} \\
			\cline{1-2}\cline{4-6}
			9 & {\textsc{Mythril}} &  &  Bytecode & https://github.com/ConsenSys/mythril &~\cite{mythril} \\
			\cline{1-2}\cline{4-6}
			10 & {\textsc{Oyente}} &  & Bytecode & https://github.com/melonproject/oyente &~\cite{oyente} \\ 
			\cline{1-2}\cline{4-6}
			11& {\textsc{Osiris}} &  &  Bytecode & https://github.com/christoftorres/Osiris &~\cite{torres2018osiris} \\
			\cline{1-2}\cline{4-6}
			12 &  {\textsc{Sereum}} &  & Bytecode & https://github.com/uni-due-syssec/eth-reentrancy-attack-patterns &~\cite{rodler2018sereum} \\
			\cline{1-2}\cline{4-6}
			13 & {\textsc{TeEther}} &  & Bytecode & https://github.com/nescio007/teether &~\cite{krupp2018teether} \\
			\cline{1-2}\cline{4-6}
			14 &  {\textsc{VerX}} &  & Source Code & https://github.com/eth-sri/verx-benchmarks &~\cite{permenev2020verx} \\
			\hline
			%15 & \textsc{ContractGuard} & \multirow{5}{*}{{Intermediate Representation}} & Bytecode & https://github.com/contractguard/experiments &~\cite{wang2019contractguard} \\
			\cline{1-2}\cline{4-6}
			15 & {\textsc{Ethir}} & \multirow{4}{*}{{Intermediate Representation}} & Bytecode & https://github.com/costa-group/ethIR &~\cite{albert2018ethir} \\
			\cline{1-2}\cline{4-6}
			16 & {\textsc{Smartcheck}} &  & Source Code & https://github.com/smartdec/smartcheck &~\cite{tikhomirov2018smartcheck} \\
			\cline{1-2}\cline{4-6}
			17 & {\textsc{Slither}} &  & Source Code & https://github.com/crytic/slither &~\cite{feist2019slither} \\  
			\cline{1-2}\cline{4-6}
			18 & {\textsc{Vandal}} &  & Bytecode & https://github.com/usyd-blockchain/vandal &~\cite{brent2018vandal} \\
			\hline
			19 & {\textsc{ContractFuzzer}} & \multirow{12}{*}{{Fuzzing Test}} & Bytecode & https://github.com/gongbell/ContractFuzzer &~\cite{contractfuzzer} \\   
			\cline{1-2}\cline{4-6}
			20 & {\textsc{ContraMaster}} &  & Source Code & https://github.com/ntu-SRSLab/vultron &~\cite{wang2020oracle,wang2019vultron} \\   
			\cline{1-2}\cline{4-6}
			21 & {\textsc{Regurad}} &  & Source Code & Not Available &~\cite{liu2018reguard} \\
			\cline{1-2}\cline{4-6}
			22 & {\textsc{ILF}} &  & Source Code & https://github.com/eth-sri/ilf &~\cite{he2019learning} \\
			\cline{1-2}\cline{4-6}
			23 & {\textsc{Harvey}} &  & Source Code & Not Available &~\cite{wustholz2020harvey,wustholz2020targeted} \\   
			\cline{1-2}\cline{4-6}
			24 & {\textsc{ConFuzzius}} &  & Bytecode  & https://github.com/christoftorres/ConFuzzius &~\cite{torres2021confuzzius} \\   
			\cline{1-2}\cline{4-6}
			25 & {\textsc{sFuzz}} &  & Bytecode & https://github.com/duytai/sFuzz &~\cite{nguyen2020sfuzz} \\   
			\cline{1-2}\cline{4-6}
			26 & {\textsc{xFuzz}} &  & Source Code &https://github.com/ToolmanInside/xfuzz\_tool  &~\cite{9795233} \\   
			\cline{1-2}\cline{4-6}
			27 & {\textsc{Smartian}} &  & Bytecode & https://github.com/SoftSec-KAIST/Smartian &~\cite{choi2021smartian} \\   
			\cline{1-2}\cline{4-6}
			28 & {\textsc{RLF}} &  & Source Code & https://github.com/Demonhero0/rlf &~\cite{rlf} \\  
			\cline{1-2}\cline{4-6}
			29 & {\textsc{IR-Fuzz}} &  & Source Code & https://github.com/Messi-Q/IR-Fuzz &~\cite{10018241} \\  
			\cline{1-2}\cline{4-6}
			30 & {\textsc{ItyFuzz}} &  & Source Code & https://github.com/fuzzland/ityfuzz &~\cite{ItyFuzz} \\  
			\hline
			31 & {\textsc{SaferSC}} &  \multirow{8}{*}{{Deep Learning}} & Bytecode & https://github.com/wesleyjtann/Safe-SmartContracts &~\cite{tann2018towards} \\  
			\cline{1-2}\cline{4-6}
			32 & {\textsc{ReChecker}} &  & Source Code & https://github.com/Messi-Q/ReChecker &~\cite{qian2020towards} \\
			\cline{1-2}\cline{4-6}
			33 & {\textsc{ContractWard}} & & Bytecode & Not Available &~\cite{wang2020contractward} \\
			\cline{1-2}\cline{4-6}
			34 & {\textsc{S-gram}} &  & Source Code & https://github.com/njaliu/sgram-artifact &~\cite{liu2018s} \\  
			\cline{1-2}\cline{4-6}
			35 & {\textsc{TMP}} &   &Source Code & https://github.com/Messi-Q/GNNSCVulDetector &~\cite{zhuangsmart,liu2021combining} \\
			\cline{1-2}\cline{4-6}
			36 & {\textsc{DeeSCVHunter}} &   &Source Code & Not Available &~\cite{9534324} \\
			%\cline{1-2}\cline{4-6}
			%37 & {\textsc{VSCL}} &   &Source Code & Not Available &~\cite{9461050} \\
			\cline{1-2}\cline{4-6}
			37 & {\textsc{CodeNet}} &   &Source Code & Not Available &~\cite{9740682} \\
			\cline{1-2}\cline{4-6}
			38 & {\textsc{DL-MDF}} &   &Source Code & Not Available &~\cite{s23167246} \\
			\hline
			39 & {\textsc{BlockEye}} &  \multirow{4}{*}{{DeFi Attack Hunting}} & DeFi protocol & Not Available &~\cite{wang2021blockeye} \\
			\cline{1-2}\cline{4-6}
			40 & {\textsc{DeFiRanger}} &   & DeFi protocol & Not Available &~\cite{wu2021defiranger} \\
			\cline{1-2}\cline{4-6}
			41 & {\textsc{Flashot}} &   & DeFi protocol & Not Available &~\cite{cao2021flashot} \\
			\cline{1-2}\cline{4-6}
			42 & {\textsc{ProMutator}} &   & DeFi protocol & https://github.com/csienslab/ProMutator &~\cite{wang2021promutator} \\
			\hline
	\end{tabular} 
	} 
\label{table2}
\end{table*}

\renewcommand{\arraystretch}{1.2}
\begin{table*}
	\centering
	\caption{\footnotesize A summary of bug types supported by each tool. For each type: AC is shot for access control; AF implies assertion failure; BD indicates block dependency (block.timestamp, block.number, block.blockhash, etc); DS refers to denial of service; DD denotes dangerous delegatecall; FE represents freezing Ether; GS means gasless send; LE is leaking Ether; RE stands for reentrancy; IO denotes integer over- /under- flow; SC implies suicidal contract; SA refers to short addresses; TO is transaction order dependency; TX means transaction origin use; UC indicates unchecked low-level call; UE is unhandled exception. For more details of various smart contract vulnerabilities, we refer interested readers to~\cite{atzei2017survey,durieux2020empirical,10.1145/3460319.3464837}. }
	\resizebox{1.00\textwidth}{!}{
		\begin{tabular}{ l l l l l l l l l l l l l l l l l }
			\hline
			\multirow{2}{*}{\textbf{Tool}} & \multicolumn{16}{c}{\textbf{Vulnerability Type}}  \\
			\cline{2-17}
			~ & \textbf{AC} & \textbf{AF} & \textbf{BD} & \textbf{DS} & \textbf{DD} & \textbf{FE} & \textbf{GS} & \textbf{LE} & \textbf{RE} & \textbf{IO} & \textbf{SC} &  \textbf{SA} &  \textbf{TO} & \textbf{TX}& \textbf{UC} & \textbf{UE}      \\
			\hline
			\textsc{CodeNet}~\cite{9740682}  &  {\color{DarkRed} \ding{55} } & {\color{DarkRed} \ding{55}  } &  {\color{OliveGreen} \ding{51} } &  {\color{DarkRed} \ding{55} } & {\color{DarkRed} \ding{55}  }  & {\color{DarkRed} \ding{55} } & {\color{DarkRed} \ding{55} } & {\color{DarkRed} \ding{55}  } & {\color{OliveGreen} \ding{51} } & {\color{DarkRed} \ding{55} } & {\color{DarkRed} \ding{55}  } &  {\color{DarkRed} \ding{55} } & {\color{DarkRed} \ding{55}} &  {\color{OliveGreen} \ding{51} } &  {\color{OliveGreen} \ding{51}  } & {\color{DarkRed} \ding{55}  }  \\
			{\textsc{ConFuzzius}}~\cite{torres2021confuzzius} &  {\color{DarkRed} \ding{55} } & {\color{OliveGreen} \ding{51} } &  {\color{OliveGreen} \ding{51} } &  {\color{DarkRed} \ding{55} } & {\color{OliveGreen} \ding{51} }  & {\color{OliveGreen} \ding{51} } & {\color{DarkRed} \ding{55} } & {\color{OliveGreen} \ding{51} } & {\color{OliveGreen} \ding{51} } & {\color{OliveGreen} \ding{51} } & {\color{OliveGreen} \ding{51} } &  {\color{DarkRed} \ding{55} } & {\color{OliveGreen} \ding{51} } &  {\color{DarkRed} \ding{55} } &  {\color{DarkRed} \ding{55} } & {\color{OliveGreen} \ding{51} } \\
			 {\textsc{ContractFuzzer}}~\cite{contractfuzzer} & {\color{OliveGreen} \ding{51} } &  {\color{DarkRed} \ding{55} } & {\color{OliveGreen} \ding{51} } &  {\color{DarkRed} \ding{55} } & {\color{OliveGreen} \ding{51} } & {\color{OliveGreen} \ding{51} } & {\color{OliveGreen} \ding{51} } &  {\color{DarkRed} \ding{55} } & {\color{OliveGreen} \ding{51} } &  {\color{DarkRed} \ding{55} } &  {\color{DarkRed} \ding{55} } &  {\color{DarkRed} \ding{55} } &  {\color{DarkRed} \ding{55} } &  {\color{DarkRed} \ding{55} } &  {\color{DarkRed} \ding{55} } & {\color{OliveGreen} \ding{51} } \\
			{\textsc{ContraMaster}}~\cite{wang2020oracle,wang2019vultron} &  {\color{DarkRed} \ding{55} }  &  {\color{DarkRed} \ding{55} } &  {\color{DarkRed} \ding{55} } &  {\color{DarkRed} \ding{55} } &  {\color{DarkRed} \ding{55} } &  {\color{DarkRed} \ding{55} } & {\color{OliveGreen} \ding{51} } &  {\color{DarkRed} \ding{55} } & {\color{OliveGreen} \ding{51} } & {\color{OliveGreen} \ding{51} } &  {\color{DarkRed} \ding{55} } &  {\color{DarkRed} \ding{55} } &  {\color{DarkRed} \ding{55} } &  {\color{DarkRed} \ding{55} } &  {\color{DarkRed} \ding{55} } & {\color{OliveGreen} \ding{51} } \\
			%\textsc{ContractGuard}~\cite{wang2019contractguard} & {\color{DarkRed} \ding{55} } & {\color{DarkRed} \ding{55} } & {\color{DarkRed} \ding{55} } & {\color{OliveGreen} \ding{51} } & {\color{OliveGreen} \ding{51} } & {\color{DarkRed} \ding{55} } & {\color{DarkRed} \ding{55} } &  {\color{DarkRed} \ding{55} } & {\color{OliveGreen} \ding{51} } & {\color{OliveGreen} \ding{51} } & {\color{DarkRed} \ding{55} } & {\color{DarkRed} \ding{55} } & {\color{DarkRed} \ding{55} } & {\color{OliveGreen} \ding{51} } &  {\color{OliveGreen} \ding{51} } & {\color{DarkRed} \ding{55} } \\
			{\textsc{ContractWard}}~\cite{wang2020contractward} &  {\color{DarkRed} \ding{55} } &  {\color{DarkRed} \ding{55} } & {\color{OliveGreen} \ding{51} }  &  {\color{DarkRed} \ding{55} } &  {\color{DarkRed} \ding{55} } &  {\color{DarkRed} \ding{55} } & {\color{DarkRed} \ding{55} } &  {\color{DarkRed} \ding{55} } & {\color{OliveGreen} \ding{51} }  & {\color{OliveGreen} \ding{51} }  &  {\color{DarkRed} \ding{55} } &  {\color{DarkRed} \ding{55} } & {\color{OliveGreen} \ding{51} }  &  {\color{DarkRed} \ding{55} } &  {\color{DarkRed} \ding{55} } &  {\color{DarkRed} \ding{55} } \\
			\textsc{DeeSCVHunter}~\cite{9534324} & {\color{DarkRed} \ding{55} } & {\color{DarkRed} \ding{55} } & {\color{OliveGreen} \ding{51} }  & {\color{DarkRed} \ding{55} }  & {\color{DarkRed} \ding{55} } &  {\color{DarkRed} \ding{55} } & {\color{DarkRed} \ding{55} } &  {\color{DarkRed} \ding{55} } & {\color{OliveGreen} \ding{51} }  & {\color{DarkRed} \ding{55} } & {\color{DarkRed} \ding{55} }  & {\color{DarkRed} \ding{55} } &  {\color{DarkRed} \ding{55} } & {\color{DarkRed} \ding{55} } & {\color{DarkRed} \ding{55} }  & {\color{DarkRed} \ding{55} }    \\
			\textsc{DefectChecker}~\cite{chen2021defectchecker} & {\color{DarkRed} \ding{55} } & {\color{DarkRed} \ding{55} } & {\color{OliveGreen} \ding{51} }  & {\color{OliveGreen} \ding{51} }  & {\color{DarkRed} \ding{55} } &  {\color{OliveGreen} \ding{51} } & {\color{DarkRed} \ding{55} } &  {\color{OliveGreen} \ding{51} } & {\color{OliveGreen} \ding{51} }  & {\color{DarkRed} \ding{55} } & {\color{DarkRed} \ding{55} }  & {\color{DarkRed} \ding{55} } &  {\color{OliveGreen} \ding{51} } & {\color{DarkRed} \ding{55} } & {\color{OliveGreen} \ding{51} }  & {\color{DarkRed} \ding{55} }  \\
			\textsc{DL-MDF}~\cite{s23167246} & {\color{DarkRed} \ding{55} } & {\color{DarkRed} \ding{55} } & {\color{OliveGreen} \ding{51} }  & {\color{DarkRed} \ding{55} }  & {\color{DarkRed} \ding{55} } &  {\color{OliveGreen} \ding{51} } & {\color{DarkRed} \ding{55} } &  {\color{DarkRed} \ding{55}  } & {\color{OliveGreen} \ding{51} }  & {\color{OliveGreen} \ding{51}} & {\color{DarkRed} \ding{55} }  & {\color{DarkRed} \ding{55} } &  {\color{DarkRed} \ding{55} } & {\color{DarkRed} \ding{55} } & {\color{DarkRed} \ding{55}}  & {\color{DarkRed} \ding{55} }  \\
			{\textsc{Ethir}}~\cite{albert2018ethir} & {\color{OliveGreen} \ding{51} }  & {\color{OliveGreen} \ding{51} } & {\color{OliveGreen} \ding{51} } & {\color{DarkRed} \ding{55} } & {\color{OliveGreen} \ding{51} } &  {\color{DarkRed} \ding{55} } & {\color{DarkRed} \ding{55} } &  {\color{DarkRed} \ding{55} } & {\color{OliveGreen} \ding{51} } & {\color{DarkRed} \ding{55} } &  {\color{DarkRed} \ding{55} } &  {\color{DarkRed} \ding{55} } & {\color{OliveGreen} \ding{51} } &  {\color{DarkRed} \ding{55} } &  {\color{DarkRed} \ding{55} } &  {\color{DarkRed} \ding{55} } \\
			{\textsc{Harvey}}~\cite{wustholz2020harvey,wustholz2020targeted} &  {\color{DarkRed} \ding{55} } & {\color{OliveGreen} \ding{51} } &  {\color{DarkRed} \ding{55} } &  {\color{DarkRed} \ding{55} } &  {\color{DarkRed} \ding{55} } &  {\color{DarkRed} \ding{55} } & {\color{DarkRed} \ding{55} } &  {\color{DarkRed} \ding{55} } & {\color{OliveGreen} \ding{51} } & {\color{OliveGreen} \ding{51} } &  {\color{DarkRed} \ding{55} } &  {\color{DarkRed} \ding{55} } &  {\color{DarkRed} \ding{55} } &  {\color{DarkRed} \ding{55} } & {\color{OliveGreen} \ding{51} } & {\color{OliveGreen} \ding{51} } \\
			{\textsc{HoneyBadger}}~\cite{torres2019art} &  {\color{DarkRed} \ding{55} } &   {\color{DarkRed} \ding{55} } & {\color{DarkRed} \ding{55} } &  {\color{DarkRed} \ding{55} } & {\color{DarkRed} \ding{55} }  & {\color{OliveGreen} \ding{51} } & {\color{DarkRed} \ding{55} } &  {\color{OliveGreen} \ding{51} } &  {\color{DarkRed} \ding{55} } & {\color{OliveGreen} \ding{51} } & {\color{DarkRed} \ding{55} } &  {\color{DarkRed} \ding{55} } &  {\color{DarkRed} \ding{55} } &  {\color{DarkRed} \ding{55} } & {\color{DarkRed} \ding{55} } &  {\color{DarkRed} \ding{55} }  \\
			{\textsc{ILF}}~\cite{he2019learning} & {\color{OliveGreen} \ding{51} }  &  {\color{DarkRed} \ding{55} } & {\color{OliveGreen} \ding{51} } &  {\color{DarkRed} \ding{55} } & {\color{OliveGreen} \ding{51} } & {\color{OliveGreen} \ding{51} } & {\color{DarkRed} \ding{55} }  &  {\color{OliveGreen} \ding{51} } & {\color{OliveGreen} \ding{51} } &  {\color{DarkRed} \ding{55} } & {\color{OliveGreen} \ding{51} } &  {\color{DarkRed} \ding{55} } &  {\color{DarkRed} \ding{55} } &  {\color{DarkRed} \ding{55} } &  {\color{DarkRed} \ding{55} } & {\color{OliveGreen} \ding{51} } \\
			{\textsc{IR-Fuzz}}~\cite{10018241} & {\color{OliveGreen} \ding{51} }  &  {\color{DarkRed} \ding{55} } & {\color{OliveGreen} \ding{51} } &  {\color{DarkRed} \ding{55} } & {\color{OliveGreen} \ding{51} } & {\color{OliveGreen} \ding{51} } & {\color{OliveGreen} \ding{51} }  &  {\color{DarkRed} \ding{55} } & {\color{OliveGreen} \ding{51} } &  {\color{OliveGreen} \ding{51} } & {\color{OliveGreen} \ding{51} } &  {\color{DarkRed} \ding{55} } &  {\color{DarkRed} \ding{55} } &  {\color{OliveGreen} \ding{51} } &  {\color{OliveGreen} \ding{51} } & {\color{OliveGreen} \ding{51} } \\
			{\textsc{ItyFuzz}}~\cite{ItyFuzz} & {\color{OliveGreen} \ding{51} }  &  {\color{DarkRed} \ding{55} } & {\color{DarkRed} \ding{55} } &  {\color{OliveGreen} \ding{51} } & {\color{DarkRed} \ding{55} } & {\color{DarkRed} \ding{55} } & {\color{DarkRed} \ding{55} }  &  {\color{DarkRed} \ding{55}  } & {\color{OliveGreen} \ding{51} } &  {\color{DarkRed} \ding{55} } & {\color{DarkRed} \ding{55} } &  {\color{DarkRed} \ding{55} } &  {\color{DarkRed} \ding{55} } &  {\color{DarkRed} \ding{55} } &  {\color{OliveGreen} \ding{51} } & {\color{DarkRed} \ding{55} } \\
			{\textsc{Maian}}~\cite{nikolic2018finding} &   {\color{OliveGreen} \ding{51} } &  {\color{DarkRed} \ding{55} } &  {\color{DarkRed} \ding{55} } &  {\color{DarkRed} \ding{55} } &  {\color{DarkRed} \ding{55} } & {\color{OliveGreen} \ding{51} } & {\color{DarkRed} \ding{55} }  &  {\color{OliveGreen} \ding{51} } &  {\color{DarkRed} \ding{55} } &  {\color{DarkRed} \ding{55} } & {\color{OliveGreen} \ding{51} }  &  {\color{DarkRed} \ding{55} } &  {\color{DarkRed} \ding{55} } &  {\color{DarkRed} \ding{55} } &  {\color{DarkRed} \ding{55} } &  {\color{DarkRed} \ding{55} } \\
			{\textsc{Manticore}}~\cite{mossberg2019manticore} & {\color{OliveGreen} \ding{51} } & {\color{OliveGreen} \ding{51} }  & {\color{OliveGreen} \ding{51} }  & {\color{DarkRed} \ding{55} } & {\color{OliveGreen} \ding{51} }  & {\color{DarkRed} \ding{55} } & {\color{DarkRed} \ding{55} }  &  {\color{OliveGreen} \ding{51} }  & {\color{OliveGreen} \ding{51} }  & {\color{OliveGreen} \ding{51} }  & {\color{OliveGreen} \ding{51} } &  {\color{DarkRed} \ding{55} } & {\color{DarkRed} \ding{55} } & {\color{OliveGreen} \ding{51} }  & {\color{OliveGreen} \ding{51} } & {\color{OliveGreen} \ding{51} }  \\
			{\textsc{Mythril}}~\cite{mythril} & {\color{OliveGreen} \ding{51} } & {\color{OliveGreen} \ding{51} } &  {\color{OliveGreen} \ding{51} } & {\color{DarkRed} \ding{55} } & {\color{OliveGreen} \ding{51} }  & {\color{DarkRed} \ding{55} }  & {\color{DarkRed} \ding{55} }  &  {\color{OliveGreen} \ding{51} } & {\color{OliveGreen} \ding{51} } & {\color{OliveGreen} \ding{51} } & {\color{OliveGreen} \ding{51} } & {\color{DarkRed} \ding{55} } & {\color{DarkRed} \ding{55} } & {\color{OliveGreen} \ding{51} }  & {\color{OliveGreen} \ding{51} } & {\color{OliveGreen} \ding{51} } \\
			{\textsc{Oyente}}~\cite{oyente} & {\color{OliveGreen} \ding{51} }  & {\color{OliveGreen} \ding{51} } & {\color{OliveGreen} \ding{51} } & {\color{OliveGreen} \ding{51} } & {\color{DarkRed} \ding{55} } & {\color{DarkRed} \ding{55} } & {\color{DarkRed} \ding{55} } &  {\color{DarkRed} \ding{55} } & {\color{OliveGreen} \ding{51} } & {\color{DarkRed} \ding{55} } & {\color{DarkRed} \ding{55} } & {\color{DarkRed} \ding{55} } & {\color{OliveGreen} \ding{51} } & {\color{DarkRed} \ding{55} } & {\color{DarkRed} \ding{55} } & {\color{DarkRed} \ding{55} } \\
			 {\textsc{Osiris}}~\cite{torres2018osiris} & {\color{DarkRed} \ding{55} }   &  {\color{OliveGreen} \ding{51} } & {\color{OliveGreen} \ding{51} } & {\color{OliveGreen} \ding{51} } & {\color{DarkRed} \ding{55} }  & {\color{DarkRed} \ding{55} }  & {\color{DarkRed} \ding{55} } &  {\color{DarkRed} \ding{55} }  & {\color{OliveGreen} \ding{51} } & {\color{OliveGreen} \ding{51} } & {\color{DarkRed} \ding{55} }  & {\color{DarkRed} \ding{55} }  & {\color{DarkRed} \ding{55} }  & {\color{DarkRed} \ding{55} }  & {\color{DarkRed} \ding{55} }  & {\color{DarkRed} \ding{55} }  \\
			 {\textsc{ReChecker}}~\cite{qian2020towards} & {\color{DarkRed} \ding{55} }  & {\color{DarkRed} \ding{55} } & {\color{DarkRed} \ding{55} } & {\color{DarkRed} \ding{55} } & {\color{DarkRed} \ding{55} } & {\color{DarkRed} \ding{55} } & {\color{DarkRed} \ding{55} } &  {\color{DarkRed} \ding{55} } & {\color{OliveGreen} \ding{51} } & {\color{DarkRed} \ding{55} } & {\color{DarkRed} \ding{55} } & {\color{DarkRed} \ding{55} } & {\color{DarkRed} \ding{55} } & {\color{DarkRed} \ding{55} } & {\color{DarkRed} \ding{55} } & {\color{DarkRed} \ding{55} } \\
			 {\textsc{Regurad}}~\cite{liu2018reguard} & {\color{DarkRed} \ding{55} }  & {\color{DarkRed} \ding{55} } & {\color{DarkRed} \ding{55} } & {\color{DarkRed} \ding{55} } & {\color{DarkRed} \ding{55} } & {\color{DarkRed} \ding{55} } & {\color{DarkRed} \ding{55} } &   {\color{DarkRed} \ding{55} } & {\color{OliveGreen} \ding{51} } & {\color{DarkRed} \ding{55} } & {\color{DarkRed} \ding{55} } & {\color{DarkRed} \ding{55} } & {\color{DarkRed} \ding{55} } & {\color{DarkRed} \ding{55} } & {\color{DarkRed} \ding{55} } & {\color{DarkRed} \ding{55} } \\
			  {\textsc{RLF}}~\cite{rlf} & {\color{DarkRed} \ding{55} }  & {\color{DarkRed} \ding{55} } & {\color{OliveGreen} \ding{51} } & {\color{DarkRed} \ding{55} } & {\color{OliveGreen} \ding{51} } & {\color{OliveGreen} \ding{51} } & {\color{DarkRed} \ding{55} } &   {\color{OliveGreen} \ding{51} } & {\color{DarkRed} \ding{55} } & {\color{DarkRed} \ding{55} } & {\color{OliveGreen} \ding{51} } & {\color{DarkRed} \ding{55} } & {\color{DarkRed} \ding{55} } & {\color{DarkRed} \ding{55} } & {\color{DarkRed} \ding{55} } & {\color{OliveGreen} \ding{51} } \\
			 {\textsc{SaferSC}}~\cite{tann2018towards} & {\color{DarkRed} \ding{55} } & {\color{DarkRed} \ding{55} } & {\color{DarkRed} \ding{55} } & {\color{DarkRed} \ding{55} } & {\color{DarkRed} \ding{55} }  & {\color{OliveGreen} \ding{51} } & {\color{DarkRed} \ding{55} } &  {\color{OliveGreen} \ding{51} } & {\color{DarkRed} \ding{55} } & {\color{DarkRed} \ding{55} } &  {\color{OliveGreen} \ding{51} }  & {\color{DarkRed} \ding{55} } & {\color{DarkRed} \ding{55} } & {\color{DarkRed} \ding{55} } & {\color{DarkRed} \ding{55} } & {\color{DarkRed} \ding{55} } \\
			{\textsc{Securify}}~\cite{securify,securify2} &  {\color{OliveGreen} \ding{51} } & {\color{DarkRed} \ding{55} } & {\color{OliveGreen} \ding{51} } & {\color{DarkRed} \ding{55} } &  {\color{OliveGreen} \ding{51} } &  {\color{OliveGreen} \ding{51} }  &  {\color{DarkRed} \ding{55} } &  {\color{OliveGreen} \ding{51} } &  {\color{OliveGreen} \ding{51} } & {\color{OliveGreen} \ding{51} } & {\color{OliveGreen} \ding{51} } & {\color{DarkRed} \ding{55} } & {\color{OliveGreen} \ding{51} } & {\color{OliveGreen} \ding{51} } & {\color{OliveGreen} \ding{51} } & {\color{OliveGreen} \ding{51} } \\
			{\textsc{Sereum}}~\cite{rodler2018sereum} & {\color{DarkRed} \ding{55} } & {\color{DarkRed} \ding{55} } & {\color{DarkRed} \ding{55} }  & {\color{DarkRed} \ding{55} } & {\color{DarkRed} \ding{55} } & {\color{DarkRed} \ding{55} } & {\color{DarkRed} \ding{55} } &  {\color{DarkRed} \ding{55} } & {\color{OliveGreen} \ding{51} } & {\color{DarkRed} \ding{55} } & {\color{DarkRed} \ding{55} } & {\color{DarkRed} \ding{55} } & {\color{DarkRed} \ding{55} } & {\color{DarkRed} \ding{55} } & {\color{DarkRed} \ding{55} } & {\color{DarkRed} \ding{55} } \\
			{\textsc{S-gram}}~\cite{liu2018s} & {\color{DarkRed} \ding{55} } & {\color{OliveGreen} \ding{51} } & {\color{OliveGreen} \ding{51} }  & {\color{DarkRed} \ding{55} } & {\color{DarkRed} \ding{55} } & {\color{OliveGreen} \ding{51} } &  {\color{DarkRed} \ding{55} } &  {\color{DarkRed} \ding{55} } & {\color{OliveGreen} \ding{51} } & {\color{OliveGreen} \ding{51} } & {\color{DarkRed} \ding{55} } & {\color{DarkRed} \ding{55} } & {\color{DarkRed} \ding{55} } & {\color{DarkRed} \ding{55} } & {\color{DarkRed} \ding{55} } & {\color{DarkRed} \ding{55} } \\
			{\textsc{sFuzz}}~\cite{nguyen2020sfuzz} & {\color{OliveGreen} \ding{51} } & {\color{DarkRed} \ding{55} } &  {\color{OliveGreen} \ding{51} } & {\color{DarkRed} \ding{55} } & {\color{OliveGreen} \ding{51} } & {\color{OliveGreen} \ding{51} } & {\color{OliveGreen} \ding{51} } &  {\color{DarkRed} \ding{55} }  & {\color{OliveGreen} \ding{51} } & {\color{OliveGreen} \ding{51} } & {\color{DarkRed} \ding{55} } & {\color{DarkRed} \ding{55} } & {\color{DarkRed} \ding{55} } & {\color{DarkRed} \ding{55} } & {\color{DarkRed} \ding{55} } & {\color{OliveGreen} \ding{51} } \\
			{\textsc{Smartcheck}}~\cite{tikhomirov2018smartcheck} & {\color{OliveGreen} \ding{51} } & {\color{DarkRed} \ding{55} } & {\color{OliveGreen} \ding{51} } & {\color{OliveGreen} \ding{51} } & {\color{DarkRed} \ding{55} } & {\color{OliveGreen} \ding{51} } &  {\color{DarkRed} \ding{55} } &  {\color{DarkRed} \ding{55} }  & {\color{OliveGreen} \ding{51} } & {\color{OliveGreen} \ding{51} } & {\color{DarkRed} \ding{55} }  & {\color{DarkRed} \ding{55} } & {\color{DarkRed} \ding{55} } & {\color{OliveGreen} \ding{51} } & {\color{DarkRed} \ding{55} } & {\color{OliveGreen} \ding{51} } \\
			{\textsc{Smartian}}~\cite{choi2021smartian} & {\color{DarkRed} \ding{55} } & {\color{OliveGreen} \ding{51} } & {\color{OliveGreen} \ding{51} } & {\color{DarkRed} \ding{55} } & {\color{OliveGreen} \ding{51} } & {\color{OliveGreen} \ding{51} } & {\color{DarkRed} \ding{55} }  &  {\color{OliveGreen} \ding{51} } & {\color{OliveGreen} \ding{51} } & {\color{OliveGreen} \ding{51} } & {\color{OliveGreen} \ding{51} }  & {\color{DarkRed} \ding{55} } & {\color{DarkRed} \ding{55} } & {\color{OliveGreen} \ding{51} } & {\color{OliveGreen} \ding{51} } & {\color{OliveGreen} \ding{51} } \\
			{\textsc{Slither}}~\cite{feist2019slither} & {\color{OliveGreen} \ding{51} } & {\color{DarkRed} \ding{55} } & {\color{OliveGreen} \ding{51} } & {\color{OliveGreen} \ding{51} } & {\color{OliveGreen} \ding{51} } & {\color{OliveGreen} \ding{51} } & {\color{DarkRed} \ding{55} }  &  {\color{OliveGreen} \ding{51} } & {\color{OliveGreen} \ding{51} } & {\color{DarkRed} \ding{55} }  & {\color{OliveGreen} \ding{51} } & {\color{DarkRed} \ding{55} } & {\color{DarkRed} \ding{55} } & {\color{OliveGreen} \ding{51} } & {\color{OliveGreen} \ding{51} } & {\color{OliveGreen} \ding{51} } \\
			{\textsc{TeEther}}~\cite{krupp2018teether} & {\color{DarkRed} \ding{55} } & {\color{DarkRed} \ding{55} } & {\color{DarkRed} \ding{55} } & {\color{DarkRed} \ding{55} } & {\color{OliveGreen} \ding{51} } &  {\color{DarkRed} \ding{55} } & {\color{DarkRed} \ding{55} }  &  {\color{OliveGreen} \ding{51} } & {\color{DarkRed} \ding{55} } & {\color{DarkRed} \ding{55} } & {\color{OliveGreen} \ding{51} } & {\color{DarkRed} \ding{55} } & {\color{DarkRed} \ding{55} } & {\color{DarkRed} \ding{55} } & {\color{DarkRed} \ding{55} } & {\color{DarkRed} \ding{55} } \\
			{\textsc{TMP}}~\cite{zhuangsmart,liu2021combining} & {\color{DarkRed} \ding{55} }  & {\color{DarkRed} \ding{55} } & {\color{OliveGreen} \ding{51} } & {\color{DarkRed} \ding{55} }  & {\color{DarkRed} \ding{55} } & {\color{DarkRed} \ding{55} } & {\color{DarkRed} \ding{55} } &   {\color{DarkRed} \ding{55} } & {\color{OliveGreen} \ding{51} } & {\color{DarkRed} \ding{55} } & {\color{DarkRed} \ding{55} } & {\color{DarkRed} \ding{55} } & {\color{DarkRed} \ding{55} } & {\color{DarkRed} \ding{55} } & {\color{DarkRed} \ding{55} } & {\color{DarkRed} \ding{55} } \\
			{\textsc{Vandal}}~\cite{brent2018vandal} & {\color{DarkRed} \ding{55} } & {\color{DarkRed} \ding{55} } & {\color{DarkRed} \ding{55} } & {\color{DarkRed} \ding{55} } & {\color{DarkRed} \ding{55} } & {\color{DarkRed} \ding{55} } & {\color{DarkRed} \ding{55} }  &  {\color{OliveGreen} \ding{51} } & {\color{OliveGreen} \ding{51} } & {\color{DarkRed} \ding{55} } & {\color{OliveGreen} \ding{51} } & {\color{DarkRed} \ding{55} } & {\color{DarkRed} \ding{55} } & {\color{OliveGreen} \ding{51} } & {\color{OliveGreen} \ding{51} } & {\color{OliveGreen} \ding{51} } \\
			{\textsc{VeriSmart}}~\cite{so2020verismart} & {\color{DarkRed} \ding{55} } & {\color{DarkRed} \ding{55} } & {\color{DarkRed} \ding{55} } & {\color{DarkRed} \ding{55} } & {\color{DarkRed} \ding{55} } & {\color{DarkRed} \ding{55} } & {\color{DarkRed} \ding{55} } &  {\color{OliveGreen} \ding{51} }  & {\color{DarkRed} \ding{55} } & {\color{OliveGreen} \ding{51} } & {\color{DarkRed} \ding{55} } & {\color{DarkRed} \ding{55} } & {\color{DarkRed} \ding{55} } & {\color{DarkRed} \ding{55} } & {\color{DarkRed} \ding{55} } & {\color{DarkRed} \ding{55} } \\
			{\textsc{VeriSol}}~\cite{wang2018formal} & {\color{DarkRed} \ding{55} } & {\color{OliveGreen} \ding{51} }  & {\color{DarkRed} \ding{55} } & {\color{DarkRed} \ding{55} } & {\color{DarkRed} \ding{55} } & {\color{DarkRed} \ding{55} } & {\color{DarkRed} \ding{55} } &   {\color{DarkRed} \ding{55} } & {\color{DarkRed} \ding{55} } & {\color{DarkRed} \ding{55} } & {\color{DarkRed} \ding{55} } & {\color{DarkRed} \ding{55} } & {\color{DarkRed} \ding{55} } & {\color{DarkRed} \ding{55} } & {\color{DarkRed} \ding{55} } & {\color{DarkRed} \ding{55} } \\
			{\textsc{VerX}}~\cite{permenev2020verx} & {\color{OliveGreen} \ding{51} } & {\color{OliveGreen} \ding{51} } & {\color{DarkRed} \ding{55} } & {\color{DarkRed} \ding{55} } & {\color{DarkRed} \ding{55} } & {\color{DarkRed} \ding{55} } & {\color{DarkRed} \ding{55} }  &   {\color{DarkRed} \ding{55} } & {\color{DarkRed} \ding{55} } & {\color{OliveGreen} \ding{51} } & {\color{DarkRed} \ding{55} } & {\color{DarkRed} \ding{55} } & {\color{DarkRed} \ding{55} } & {\color{DarkRed} \ding{55} } & {\color{DarkRed} \ding{55} } & {\color{DarkRed} \ding{55} } \\
			%\textsc{VSCL}~\cite{9461050} &  {\color{DarkRed} \ding{55} } & {\color{DarkRed} \ding{55} } & {\color{OliveGreen} \ding{51} } &{\color{DarkRed} \ding{55} }  & {\color{OliveGreen} \ding{51} } & {\color{DarkRed} \ding{55} } & {\color{DarkRed} \ding{55} } &  {\color{DarkRed} \ding{55} } & {\color{OliveGreen} \ding{51} } & {\color{DarkRed} \ding{55} } & {\color{DarkRed} \ding{55} } & {\color{DarkRed} \ding{55} } & {\color{DarkRed} \ding{55}} & {\color{DarkRed} \ding{55}} & {\color{DarkRed} \ding{55}} & {\color{DarkRed} \ding{55} } \\
			{\textsc{xFuzz}}~\cite{9795233}  &  {\color{DarkRed} \ding{55} } & {\color{DarkRed} \ding{55} } & {\color{OliveGreen} \ding{51} } &{\color{DarkRed} \ding{55} }  & {\color{OliveGreen} \ding{51} } & {\color{DarkRed} \ding{55} } & {\color{DarkRed} \ding{55} } &  {\color{DarkRed} \ding{55} } & {\color{OliveGreen} \ding{51} } & {\color{DarkRed} \ding{55} } & {\color{DarkRed} \ding{55} } & {\color{DarkRed} \ding{55} } & {\color{DarkRed} \ding{55}} & {\color{OliveGreen} \ding{51} } & {\color{DarkRed} \ding{55}} & {\color{DarkRed} \ding{55} } \\
			{\textsc{Zeus}}~\cite{kalra2018zeus}  &  {\color{DarkRed} \ding{55} } & {\color{DarkRed} \ding{55} } & {\color{OliveGreen} \ding{51} } &{\color{DarkRed} \ding{55} }  & {\color{DarkRed} \ding{55} } & {\color{DarkRed} \ding{55} } & {\color{DarkRed} \ding{55} } &  {\color{DarkRed} \ding{55} } & {\color{OliveGreen} \ding{51} } & {\color{OliveGreen} \ding{51} } & {\color{DarkRed} \ding{55} } & {\color{DarkRed} \ding{55} } & {\color{OliveGreen} \ding{51} } & {\color{OliveGreen} \ding{51} } & {\color{OliveGreen} \ding{51} } & {\color{OliveGreen} \ding{51} } \\
			\hline
        \end{tabular}
	}
\label{table3}
\end{table*}

\subsubsection{Static Analysis}
\label{static_analysis}
Static analysis refers to the technique of analyzing programs without actually executing them. It examines code in the absence of real input data and is capable of detecting potential security violations, runtime errors, and logical inconsistencies. Static analysis techniques have been widely applied to smart contract vulnerability detection, which can be broadly divided into two categories, i.e., formal verification~\cite{Bhargavan} and symbolic execution~\cite{yang2020seraph}. In particular, when performing static analysis on smart contracts, researchers often utilize intermediate representation-based analysis methods~\cite{reis2020tezla}, which can store and preserve the rich semantics of the Source Code.

\paragraph{Formal Verification} 
Formal verification eliminates the ambiguity and incompatibility in a smart contract by transforming its concepts, judgments, and logic into a formal model~\cite{murray2019survey}. It cooperates with rigorous proofs to verify the correctness and security of functions in smart contracts. Common formal verification methods include {model checking}~\cite{gu2018formal} and {deductive verification}~\cite{liu2019formal}. Specifically, model checking enumerates all possible states of a smart contract through state-space searching and then checks whether the contract has corresponding security properties. Deductive verification uses logical formulas to describe a verification system and then proves whether the system has certain security properties through predefined rules. In general, formal verification techniques mainly propose a formal model and define the formal semantics of contracts to verify the security properties in smart contracts. We review existing formal verification methods for smart contract bug detection and give their brief principles as follows.
\begin{itemize}[itemsep=1pt, topsep=1pt, leftmargin=\dimexpr\labelwidth + 2 \labelsep\relax]
%\item \textbf{KEVM} is a formal analysis tool, which utilizes the $\mathbb{K}$ framework to construct the executable formal specifications based on the EVM bytecode stack. 
\item \textbf{\textsc{Securify}}~\cite{securify} extracts the semantic information from the bytecode of smart contracts and then takes advantage of a set of compliance and violation security patterns that capture sufficient conditions to prove the presence or absence of vulnerabilities. 
\item \textbf{\textsc{VeriSmart}}~\cite{so2020verismart} is an accurate verifier for ensuring arithmetic safety of smart contracts. Particularly, it takes advantage of a domain-specific algorithm to automatically discover and exploit transaction invariants, which are essential for precisely analyzing smart contracts.
\item \textbf{\textsc{VeriSol}}~\cite{wang2018formal} is a highly automated formal verifier that checks the semantic compliance of smart contracts against a state machine model with an access control policy.
\item \textbf{\textsc{Zeus}}~\cite{kalra2018zeus} employs abstract interpretation and symbolic model checking to ascertain verification conditions, and verifies the correctness of smart contracts during the formal verification process.
\end{itemize}

\textbf{Symbolic Execution.}\quad Symbolic execution methods analyze software programs using symbolic values as inputs rather than specific values during execution~\cite{baldoni2018survey}. Once a program branch is reached, the analyzer collects the corresponding path constraints following a constraint solver to obtain specific values that can trigger each branch. Notably, symbolic execution is capable of exploring multiple paths simultaneously. However, it also faces unavoidable problems such as path explosion. In most cases, the symbolic executor first constructs a control flow graph (CFG) from the Bytecode. It then designs appropriate constraints based on the characteristics of the vulnerabilities. Finally, the executor uses the constraint solver to analyze the control flow graph and generate the bug report. We present well-known symbolic execution-based vulnerability detection methods for smart contracts as follows.
\begin{itemize}[itemsep=1pt, topsep=1pt, leftmargin=\dimexpr\labelwidth + 2 \labelsep\relax]
\item \textbf{\textsc{DefectChecker}}~\cite{chen2021defectchecker} is a symbolic execution-based bug checker that analyzes the smart Bytecode to generate the CFG, stack event, and three code features (i.e., \emph{Money Call}, \emph{Loop Block}, and \emph{Payable Function}), and combines them to detect vulnerabilities.
\item \textbf{\textsc{HoneyBadger}}~\cite{torres2019art} is built on a taxonomy of honeypot techniques. It adopts symbolic execution and well-defined heuristics to pinpoint various types of honeypots in smart contracts.
\item \textbf{\textsc{Maian}}~\cite{nikolic2018finding} keeps track of contracts that can self-destruct or drain Ether from arbitrary addresses, or those that accept Ether but have no payout functionality. It performs inter-procedural symbolic analysis and uses a concrete validator to expose vulnerabilities.
\item \textbf{\textsc{Manticore}}~\cite{mossberg2019manticore} is a dynamic symbolic execution framework for analyzing bytecode of smart contracts, providing an effective way to maximize code coverage and find execution paths that lead to vulnerabilities and reachable self-destruct operations.
\item \textbf{\textsc{Mythril}}~\cite{mythril}, developed by \emph{ConsenSys}~\cite{consensys}, relies on concolic analysis, taint analysis, and control flow checking of the bytecode to prune the search space and expose vulnerabilities in smart contracts.
\item \textbf{\textsc{Oyente}}~\cite{oyente} is one of the pioneer smart contract analysis tools. It uses symbolic execution to identify smart contract vulnerabilities based on the control flow graph. \textsc{Oyente} is also used as a basis for several other tools, such as \textsc{HoneyBadger}, \textsc{Maian}, and \textsc{Osiris}.
\item \textbf{\textsc{Osiris}}~\cite{torres2018osiris} extends \textsc{Oyente} and supports to detect the integer overflow vulnerability of smart contracts.
\item \textbf{\textsc{Sereum}}~\cite{rodler2018sereum} employs dynamic taint tracking to monitor the data flow during the contract execution, avoiding inconsistent states and effectively preventing reentrancy attacks.
\item \textbf{\textsc{teEther}}~\cite{krupp2018teether} searches for vulnerable execution traces in a contract’s control flow graph and automatically creates an exploit for a contract given only its bytecode.
\item \textbf{\textsc{VerX}}~\cite{permenev2020verx} is a automated verifier prove functional specifications of smart contracts. It incorporates the delayed predicate abstraction approach, which combines symbolic execution during transaction execution with abstraction at transaction boundaries, facilitating the automatic verification of the security properties of smart contracts.
\end{itemize}

\textbf{Intermediate Representation.}\quad  
As the logic of smart Source Code becomes more complex, it becomes increasingly difficult to deal with the source code directly~\cite{pierro2021analysis}. A family of works, instead, can deal with pure bytecode~\cite{li2020stan,bistarelli2020ethereum}. However, they still suffer from low accuracy due to the inherent difficulty of interpreting bytecode and restoring comprehensive data and control flow dependencies in the missing source code. To facilitate the analysis of smart contracts, researchers probe to convert the source code or bytecode of a smart contract into an abstract code structure that is conducive for further processing, referred to as an intermediate representation (IR), and then analyze the IR to discover security problems. An intermediate representation of the smart contract is an arbitrary representation of a program between the source code and the bytecode~\cite{reis2020tezla,feist2019slither,lu2021neucheck,grech2019gigahorse,perez2020cost}. Since the intermediate representation abandons the complex logic in the source code while preserving rich data and control flow semantics, it can be favorable to further processing and analysis. We conclude intermediate representation-based analysis techniques for smart contracts as follows.
\begin{itemize}[itemsep=1pt, topsep=1pt, leftmargin=\dimexpr\labelwidth + 2 \labelsep\relax]
%\item \textbf{\textsc{ContractGuard}}~\cite{wang2019contractguard} is an intrusion detection tool for Ethereum smart contracts. It detects abnormal control flow caused by potential attacks and achieves intrusion detection through an effective context-tagged loop-free path.
\item \textbf{\textsc{Ethir}}~\cite{albert2018ethir} converts the control flow graph of smart contracts into the rule-based intermediate representation (RBP), and then infers the security properties of smart contracts based on the RBP.
\item \textbf{\textsc{SmartCheck}}~\cite{tikhomirov2018smartcheck} is an extensible static analysis tool for smart contracts, which converts the source code of a smart contract into an XML-based intermediate representation. It uses the lexical and syntactic analysis of the Solidity source code, looking for vulnerability patterns and bad coding practices.
\item \textbf{\textsc{Slither}}~\cite{feist2019slither} converts the smart Source Code into an intermediate representation of SlithIR. SlithIR uses a static single allocation (SSA) form and a reduced instruction set to simplify the contract analysis process while preserving the semantic information of the source code.
\item \textbf{\textsc{Vandal}}~\cite{brent2018vandal} is composed of an analysis pipeline and a decompiler. The decompiler performs an abstract interpretation to convert the bytecode into an intermediate representation in the form of logical relations, and then uses novel logic-driven methods to detect vulnerabilities in smart contracts.
\end{itemize}

\subsubsection{Fuzzing Test}
\label{dynamic_fuzzing}
Fuzzing has proven to be a successful technique for discovering software bugs over the past decades~\cite{manes2019art}. A wide variety of fuzzing methods have emerged, such as \emph{whitebox}, \emph{blackbox}, and \emph{greybox}~\cite{zheng2019firm,bohme2017directed,xu2019fuzzing}. The main idea of fuzzing techniques is to feed a mass of test inputs ({i.e.,} test cases) into the program under test and expose vulnerabilities by monitoring the reported abnormal results or exceptions~\cite{contractfuzzer}. When applied to smart contracts, a fuzzing engine first attempts to generate initial seeds and form executable transactions. With the assistance of the feedback of test results, it will dynamically adjust the generated test cases to explore as much contract state space as possible. This process is repeated until a stopping criterion is satisfied. Finally, the fuzzer will analyze the results generated during fuzzing and report to users. We review current smart contract fuzzing methods and summarize their principles as follows.
\begin{itemize}[itemsep=1pt, topsep=1pt, leftmargin=\dimexpr\labelwidth + 2 \labelsep\relax]
\item \textbf{\textsc{ContractFuzzer}}~\cite{contractfuzzer} is one of the earliest fuzzing frameworks for smart contracts, which identifies vulnerabilities by monitoring runtime behavior during a fuzzing campaign.
\item \textbf{\textsc{ContraMaster}}~\cite{wang2020oracle} is an oracle-supported dynamic exploit generation framework for detecting vulnerabilities in smart contracts. It combines the analysis of data-flows, control-flows, and contract state information to guide seed mutation.
\item \textbf{\textsc{ReGuard}}~\cite{liu2018reguard} is a fuzzing-based analyzer to automatically detect reentrancy vulnerabilities in Ethereum smart contracts. It dynamically identifies reentrancy bugs by iteratively generating random but diverse transactions.
\item \textbf{\textsc{ILF}}~\cite{he2019learning} proposes a new approach for learning an effective yet fast fuzzer from symbolic execution by phrasing the learning task in the framework of imitation learning. 
\item \textbf{\textsc{Harvey}}~\cite{wustholz2020harvey} extends standard greybox fuzzing for predicting new test inputs that are more likely to cover new paths or reveal vulnerabilities in smart contracts.
\item \textbf{\textsc{ConFuzzius}}~\cite{torres2021confuzzius} is a hybrid fuzzer for smart contracts that uses evolutionary fuzzing to exercise shallow parts of a smart contract. Specifically, it employs constraint solving to generate test cases, enforcing the fuzzer to reach complex and deep branches.
\item \textbf{\textsc{sFuzz}}~\cite{nguyen2020sfuzz} presents an efficient and lightweight multi-objective adaptive strategy to dig out potential vulnerabilities hidden in the hard-to-cover branches.
\item \textbf{\textsc{xFuzz}}~\cite{9795233} is a machine learning-guided smart contract fuzzing framework that uses machine learning predictions to guide fuzzers for vulnerability detection. 
\item \textbf{\textsc{Smartian}}~\cite{choi2021smartian} employs a novel feedback mechanism, i.e., data-flow-based feedback, to guide the greybox fuzzer to systematically generate critical transaction sequences.
\item \textbf{\textsc{RLF}}~\cite{rlf} introduces a reinforcement learning-guided fuzzing framework for smart contracts, which drives the fuzzer to generate vulnerable transaction sequences.
\item \textbf{\textsc{IR-Fuzz}}~\cite{10018241} extends sFuzz, which engages in a sequence generation strategy that contains invocation ordering and prolongation to trigger deeper states in a smart contract.
\item \textbf{\textsc{ItyFuzz}}~\cite{ItyFuzz} is a snapshot-based fuzzer for testing smart contracts, which incorporates  new waypoint mechanisms optimized to prioritize the exploration of interesting snapshot states, allowing for efficient program exploitation. In particular, \textsc{ItyFuzz} is able to analyze complex DeFi protocols and identify the price manipulation vulnerability.
\end{itemize}

\subsubsection{Deep Learning-based Methods}
\label{deep_learning}
In recent years, there has been a growing practice of detecting program security vulnerabilities using deep learning technologies~\cite{li2018vuldeepecker,lin2020software}. The advancement of deep learning has promoted the emergence of various vulnerability detection methods~\cite{chakraborty2021deep,zou2019mu}. Existing deep learning-based methods usually convert smart contracts into intermediate representations and then construct a deep neural network as the detection model. As deep learning technology has demonstrated its superiority in handling sophisticated data, it can also be applied to deal with the complex business logic of DeFi protocols. We provide existing deep learning-based smart contract vulnerability detection methods as follows, hoping to inspire others.
\begin{itemize}[itemsep=1pt, topsep=1pt, leftmargin=\dimexpr\labelwidth + 2 \labelsep\relax]
\item \textbf{\textsc{SaferSC}}~\cite{tann2018towards} is the first deep learning-based vulnerability detection model for smart contracts, which analyzes the operation code ({i.e.,} opcode) of smart contracts and trains a long short-term memory network (LSTM) as the detection model.
\item \textbf{\textsc{ReChecker}}~\cite{qian2020towards} constructs the bidirectional long short-term memory network based on an attention mechanism to precisely detect the reentrancy vulnerability. 
\item \textbf{\textsc{ContractWard}}~\cite{wang2020contractward} extracts bigram features from the smart contract opcode and adopts a variety of machine learning algorithms to detect bugs in smart contracts.
\item \textbf{\textsc{S-gram}}~\cite{liu2018s} introduces a novel semantic-aware security auditing technique for analyzing smart contracts. The key insight behind {S-gram} is a combination of $N$-gram language modeling and lightweight static contract analysis.
\item \textbf{\textsc{TMP}}~\cite{zhuangsmart} proposes to cast the source code of a smart contract into a \emph{contract graph} and builds a temporal-message-propagation graph model to identify vulnerabilities. 
\item \textbf{\textsc{DeeSCVHunter}}~\cite{9534324} uses a modularized and systematic deep learning-based framework to detect vulnerabilities in smart contracts. It employs the proposed vulnerability candidate slice (VCS) to guide the model to capture the critical part of the vulnerability. 
\item \textbf{\textsc{CodeNet}}~\cite{9740682} is a code-targeted convolutional neural network (CNN) architecture that detects vulnerable smart contracts while preserving their semantics and context. 
\item \textbf{\textsc{DL-MDF}}~\cite{s23167246} identifies smart contract vulnerabilities based on deep learning and multimodal decision fusion. Specifically, it extracts the features of the source code, operation code, and control flow of a smart contract through multiple models respectively, and then adopts a multimodal decision fusion strategy to output the detection results.
 \end{itemize}

\subsubsection{DeFi Attack Hunting}
\label{defi_attack_hunting}
Generally speaking, detecting DeFi attacks requires the ability to recover and understand the logical semantics of DeFi protocols, which is typically missing in the aforementioned solutions that find bugs only in traditional smart contracts. Current DeFi attack hunting methods~\cite{wu2021defiranger,wang2021promutator,wang2021blockeye} consider simulating the attack process to detect potential DeFi attack behaviors. Technically, they monitor the contract transactions during the attack simulation and reveal DeFi attacks using the patterns with the recovered high-level semantics of DeFi protocols. In what follows, we will elaborate on the specific design and workflow of existing DeFi attack hunting tools.

\begin{figure}[h]
\centering
\includegraphics[width=8.8cm]{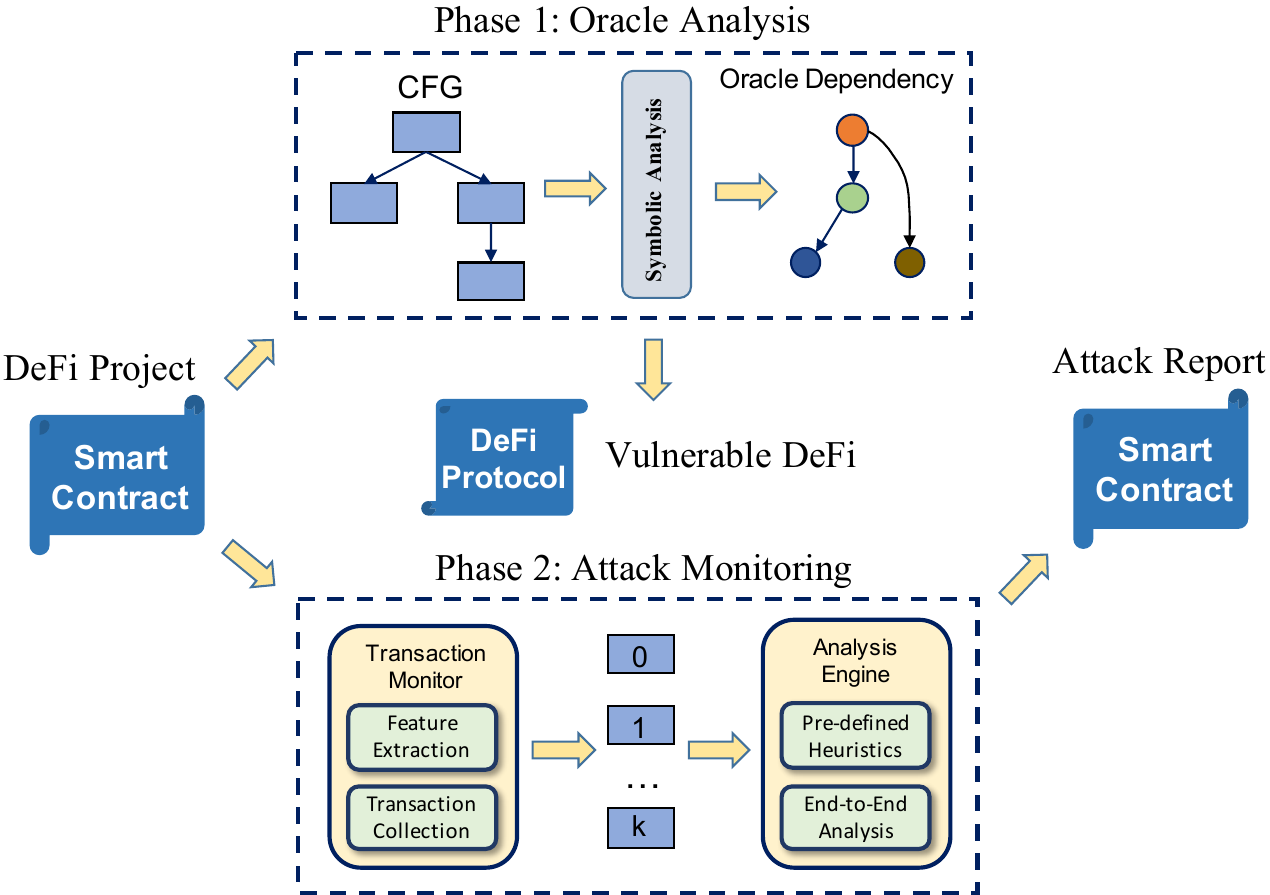}
\caption{The overall workflow of \textsc{BlockEye}, which consists of oracle analysis and attack monitoring.} 
\label{Blockeye}
\end{figure}

\paragraph{\textsc{BlockEye}}\quad 
\textsc{BlockEye}~\cite{wang2021blockeye} is a real-time attack detection system for DeFi protocols. It models interdependencies between DeFi protocols and flags potential DeFi attacks with an end-to-end analysis in real-time. The key insights behind \textsc{BlockEye} are symbolic oracle analysis and pattern-based runtime transaction validation. To illustrate the overall insights of \textsc{BlockEye}, we further describe its workflow in Fig.~\ref{Blockeye}, where \textsc{BlockEye} works in two phases. In the first phase, \textsc{BlockEye} performs symbolic analysis on smart contracts of a given DeFi protocol. Specifically, the goal of this phase is to model the inter-DeFi oracle dependency, i.e., how the oracle data provided by one DeFi protocol affects the services of another. Once oracle-dependent state updates are found, \textsc{BlockEye} identifies the DeFi protocols as potentially vulnerable. In the second phase, \textsc{BlockEye} installs a runtime monitor for vulnerable DeFi protocols to detect external attacks. Specifically, \textsc{BlockEye} employs a transaction monitor to collect related transactions based on extracted characteristics. Then, the end-to-end analysis of transactions is performed according to predefined heuristics, e.g., a large profit is made in a short period. Potential attacks are flagged by \textsc{BlockEye} if an abnormal sequence of transactions is detected. Finally, \textsc{BlockEye} generates an analysis report for attack verification.

\begin{figure}[h]
\centering
\includegraphics[width=8.8cm]{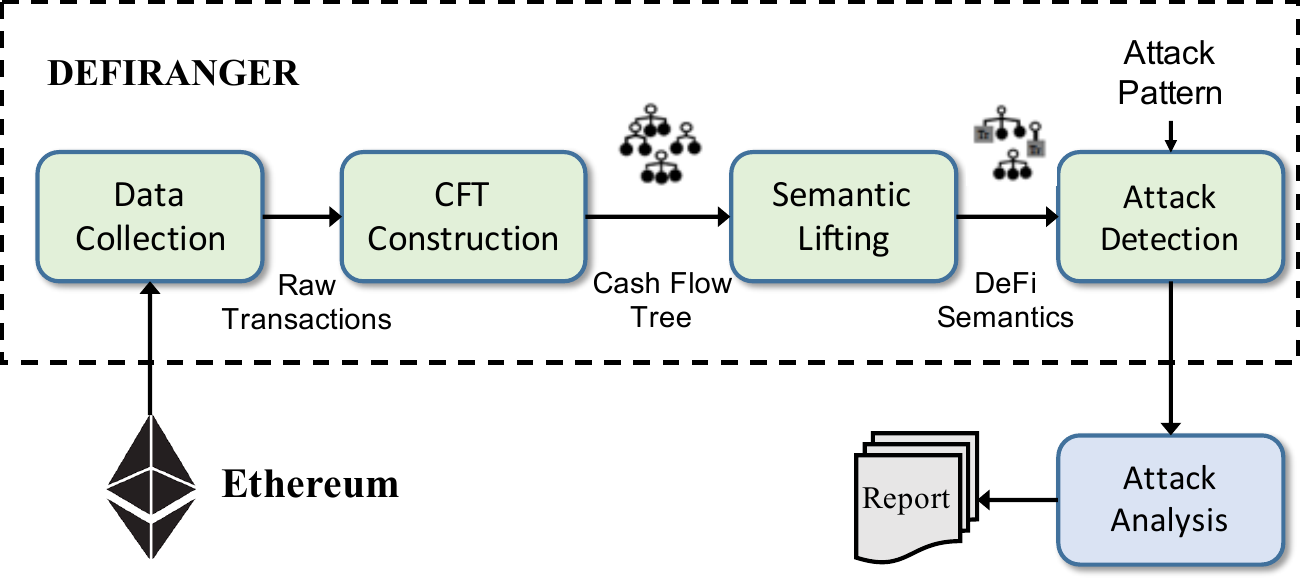}
\caption{The general workflow of \textsc{DeFiRanger}.} 
\label{DeFiRanger}
\end{figure}

\paragraph{\textsc{DeFiRanger}}\quad  
\textsc{DeFiRanger}~\cite{wu2021defiranger} is designed to detect price manipulation attacks on DeFi protocols. Specifically, \textsc{DeFiRanger} restores the high-level semantics of DeFi protocols by constructing the cash flow tree (CFT) from transactions and lifting the low-level semantics to high-level ones. After that, \textsc{DeFiRanger} detects price manipulation attacks using the patterns expressed with recovered DeFi semantics.
The workflow of \textsc{DeFiRanger} is depicted in Fig.~\ref{DeFiRanger}. \textsc{DeFiRanger} initially collects raw Ethereum transactions, and then constructs the cash flow tree that is used to convert raw transactions to token transfers. Then, \textsc{DeFiRanger} uses a lifting algorithm to recover the DeFi semantics from the CFT. Finally, \textsc{DeFiRanger} detects price manipulation attacks by matching recovered semantics with attack patterns or rules.

\begin{figure}[h]
\centering
\includegraphics[width=8.7cm]{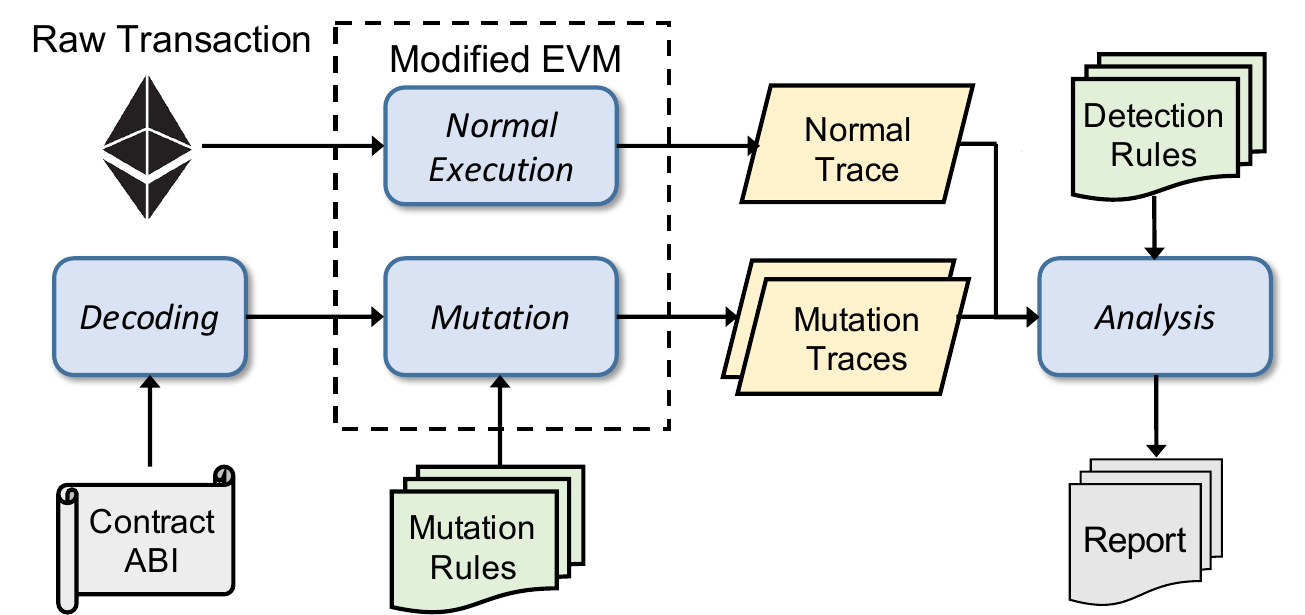}
\caption{The high-level workflow of \textsc{ProMutator}.} 
\label{ProMutator}
\end{figure}

\paragraph{\textsc{ProMutator}}\quad 
\textsc{ProMutator}~\cite{wang2021promutator} is a scalable security analysis framework that can detect price oracle vulnerabilities in DeFi protocols. \textsc{ProMutator} simulates price oracle attacks on a DeFi protocol locally and observes how a price oracle handles abnormal price data feeds.
Fig.~\ref{ProMutator} shows the high-level workflow of \textsc{ProMutator}, which processes each transaction through three phases: \emph{decoding, mutation}, and \emph{analysis}. \textsc{ProMutator} detects the price oracle bug by using built-in mutation and detection rules, which can be extended to cover other types of vulnerabilities using custom mutation and detection rules. \textsc{ProMutator} outputs whether a target DeFi protocol is vulnerable based on the detection rules and the analysis of mutated traces.

\paragraph{\textsc{Flashot}}\quad 
\textsc{Flashot}~\cite{cao2021flashot} is a framework that is able to reveal the microscopic process of the flash loan attack in a clear and precise way. Specifically, \textsc{Flashot} is designed to illustrate the precise asset flows intertwined with smart contracts in a standardized diagram. In addition, \textsc{Flashot} performs an in-depth analysis of a typical flash loan attack ({e.g.,} pump and arbitrage), with a more accurate model and solution regarding the optimization problem, as well as some economic explanations for the attacker's malicious behavior.

\begin{tcolorbox}[arc = 1mm, colback = black!2!white, colframe = black!65!white, boxrule=0.6mm]
\textbf{Summary:}

\textbf{(1)} Most state-of-the-art vulnerability detection approaches are tailored for traditional smart contracts or a single contract. 
DeFi protocols present unique challenges for analysis compared to those in traditional smart contracts due to the composability of DeFi applications. 
Since DeFi protocols are built on multiple interactive smart contracts, it is difficult for them to detect DeFi attacks and vulnerabilities. 

\textbf{(2)} Existing DeFi attack hunting techniques are still in their infancy, leaving a large room for improvement. So it is necessary to design scalable tools for identifying various attacks and vulnerabilities on DeFi protocols.
\end{tcolorbox}

\subsection{Automated Repair}
\label{automated_repair}
In this subsection, we review 8 existing automated repair tools for smart contracts and DeFi protocols, which are listed in Table~\ref{table_repair}. In particular, we summarize the principles of each automated repair tool, and then present their advantages and
disadvantages.

\paragraph{\textsc{SCRepair}}~\cite{10.1145/3402450} propose an automated smart contract repair tool \textsc{SCRepair} using a genetic programming search method. \textsc{SCRepair} performs parallel genetic repair by partitioning the large search space of candidate patches into smaller mutually exclusive search spaces that can be processed individually. Considering the gas consumption of patches, \textsc{SCRepair} integrates a gas-awareness technique that compares candidate patches in terms of gas cost, thus screening out the patches that are useful and lower gas cost. Notably, \textsc{SCRepair} suffers from the inherent problems: 1) the generated patches are incomplete, which can sometimes lead to ineffectiveness, 2) the selected patches may break the functionality of the original contract, and 3) it is unable to repair the off-chain contracts.

\renewcommand{\arraystretch}{1.15}
\begin{table*}
	\centering
	\caption{\small Overview of state-of-the-art tools for repairing vulnerabilities in smart contracts and DeFi protocols. \textbf{Vulnerability Number} represents that the number of vulnerability types can be repaired by each tool.}
	\resizebox{1.00\textwidth}{!}{
	 \begin{threeparttable}[b]
		\begin{tabular}{|c|c|c|c|c|}
			\hline
			\textbf{Tool} & \textbf{Analysis Level} & \textbf{Vulnerability Number} & \textbf{Public Available}  &\textbf{Reference}\\
			\hline
			\textsc{SCRepair} &  Source Code &  4 & https://github.com/xiaoly8/SCRepair   &  \cite{10.1145/3402450}  \\
			\hline
			\textsc{SMARTSHIELD} &  Bytecode  &  3 &  Not Available  &  \cite{zhang2020smartshield}   \\
			\hline
			\textsc{sGuard} &  Source Code  &   4  & https://github.com/duytai/sGuard    &  \cite{nguyen2021sguard}  \\
			\hline
			\textsc{EVMPatch} &  Bytecode   &  2 & https://github.com/uni-due-syssec/evmpatch-developer-study   &  \cite{rodler2021evmpatch}  \\
			\hline
			\textsc{Aroc} &  Source Code  &   4  & Not Available   &  \cite{jin2021aroc}  \\
			\hline
			\textsc{HCC} &  Source Code  &  2 &  Not Available  &  \cite{giesen2022practical}  \\
			\hline
			\textsc{Elysium} &  Bytecode   &  7  & https://github.com/christoftorres/Elysium  &  \cite{ferreira2022elysium}  \\
			\hline
			\textsc{DeFinery}\tnote{1} & Source Code   &  4  & https://sites.google.com/view/ase2022-definery  &  \cite{tolmach2022property}  \\
			\hline
	\end{tabular} 
	\begin{tablenotes}
	\item[1] \textsc{DeFinery} is built on \textsc{SCRepair}. \textsc{DeFinery} enables property-based automated repair of smart contracts while providing formal correctness guarantees.
	\end{tablenotes}
	\end{threeparttable}
} 
\label{table_repair}
\end{table*}

\paragraph{\textsc{SMARTSHIELD}}~\cite{zhang2020smartshield} develop a bytecode rectification system SMARTSHIELD to automatically fix three types of bugs in smart contracts. The overall workflow of SMARTSHIELD consists of three steps: 1) extracting the abstract syntax tree (AST) and the unfixed bytecode of a smart contract to capture its bytecode-level semantic information; 2) generating the patches to repair the insecure control flows and data operations based on the semantic information; 3) outputting the patched bytecode and a bug repair report.

\paragraph{\textsc{sGuard}}~\cite{nguyen2021sguard} propose an automatic fixing approach that can transform smart contracts to be free of 4 kinds of vulnerabilities. Given a smart contract, \textsc{sGuard} works in two phases: 1) collects a finite set of symbolic execution traces of the smart contract and then performs static analysis on the collected traces to identify potential vulnerabilities, and 2) applies specific fixing patterns for each type of bug at the source code level. Unfortunately, \textsc{sGuard} performs source code rewriting without further analysis of the interactions between different functions and the global dependencies of memory variables. The lack of such analysis in smart contract patching typically results in the destruction of the functionality of the original contract.

\paragraph{\textsc{EVMPatch}}~\cite{rodler2021evmpatch} present an automated repair framework, called \textsc{EVMPatch}, to patch faulty smart contracts. \textsc{EVMPatch} works in two main steps. In the first step, based on designed patch rules, \textsc{EVMPatch} uses a bytecode rewriting engine to fix a basic block that has a vulnerability and appends the fixed block to the end of the contract. In the second step, after fixing a contract, \textsc{EVMPatch} replays some of the historical transactions of the original contract on a local Ethereum client to verify that the fixed contract is correct by observing whether attack transactions are blocked and whether other transactions are normal. However, \textsc{EVMPatch} only works when the vulnerability is located within a single bytecode basic block, while having difficulties in handling vulnerabilities across separated basic blocks. 

\paragraph{\textsc{Aroc}}~\cite{jin2021aroc} propose a smart contract repairer named \textsc{Aroc} that can automatically fix vulnerable on-chain contracts without updating the contract code. \textsc{Aroc} consists of three main components: 1) an information extraction module, which captures the variable dependencies, path constraints, and contract metadata to synthesize patches given vulnerable contract source code and bug types; 2) a patch generation module, which produces patches according to the repair templates; 3) an enhanced EVM, which binds the generated patches to fix the vulnerable contract and block the malicious transactions. However, \textsc{Aroc} only works on its own enhanced EVM, and the binding features cannot yet be accepted by the official Ethereum.

\paragraph{\textsc{HCC}}~\cite{giesen2022practical} develop a smart contract compiler called HCC that can insert security hardening checks into the contract source code automatically. HCC designs a code property graph (CPG) to model the control and data flows of the contract and find potential bugs. HCC then mitigates the discovered bugs by inserting hardening patches. Hardening patches in HCC are expressed as graph patterns that are used as templates for CPG-level patches against bugs. However, HCC has difficulty repairing vulnerabilities in complex DeFi protocols because it is unable to recover complete CPGs from intertwined DeFi protocols.

\begin{figure*}
\centering
\includegraphics[width=16.7cm]{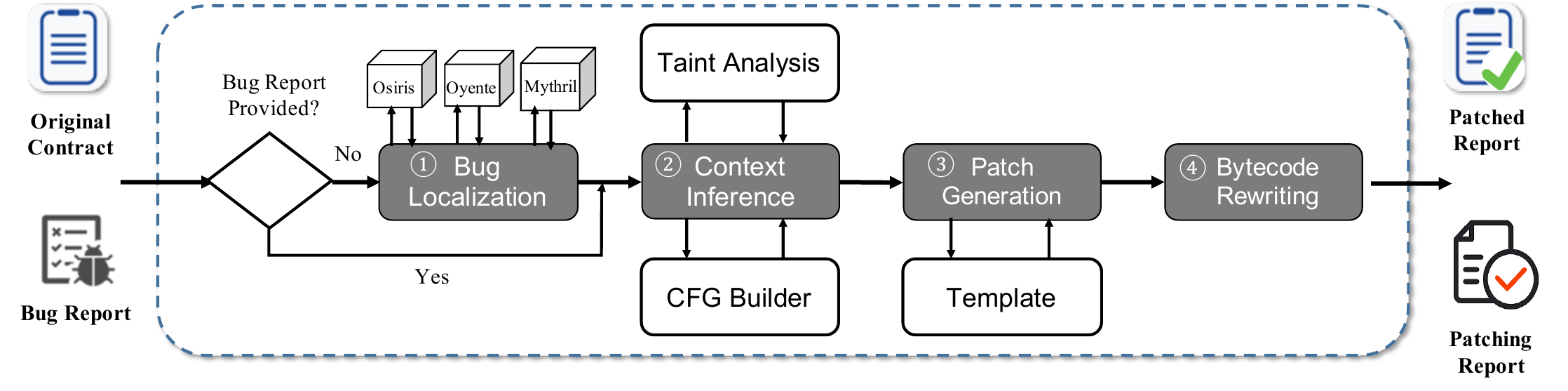}
\caption{The overall workflow and architecture of \textsc{Elysium}.} 
\label{Elysium}
\end{figure*}

\paragraph{\textsc{Elysium}}~\cite{ferreira2022elysium} propose \textsc{Elysium}, a scalable tool towards automated repair for smart contracts at the bytecode level. The overall workflow of this tool is illustrated in Fig.~\ref{Elysium}, which consists of four phases: 1) bug location, which is responsible for detecting and localizing bugs in the bytecode; 2) context inference, which extracts the control flow graph of the bytecode and infers the context information of the CFG; 3) patch generation, which produces patches by inserting previously inferred context information into given patch templates; 4) bytecode rewriting, which injects the generated patches into the CFG and transforms it back into bytecode.

\paragraph{\textsc{DeFinery}}~\cite{tolmach2022property} design an approach that enables property-based automated repair of DeFi protocols while providing formal correctness guarantees. \textsc{DeFinery} is the first attempt to repair functional specification violations in DeFi smart contracts. \textsc{DeFinery} consists of two main modules: 1) a semantic analysis part, which symbolically executes a smart contract with respect to a property and a sequence of functions that lead to its violation; 2) a patch generation part, which provides a patched version of a smart contract that satisfies the property. Notably, the patch generation module of \textsc{DeFinery} is built on \textsc{SCRepair}. \textsc{DeFinery} extends a set of statements that are defined in \textsc{SCRepair}, combines the semantic analysis module to perform the patch, and adds heuristics to select changes that are likely to repair the problem.

\begin{tcolorbox}[arc = 1mm, colback = black!2!white, colframe = black!65!white, boxrule=0.6mm]
\textbf{Summary:}

\textbf{(1)} Current automated smart contract repair techniques only address limited types of vulnerabilities and are difficult to deal with complex smart contracts, such as DeFi protocols. 

\textbf{(2)} There is still a lack of effective verifiers to prove the correctness of patches and to validate whether the intended functionality of the patched contract is equivalent to the original contract.
\end{tcolorbox}

\subsection{Other Pioneering Work for DeFi Security} 
There is a growing body of literature taking care of DeFi security from other dimensions. For example,~\cite{gudgeon2020decentralized} explore how to solve weaknesses in DeFi protocols that could lead to a DeFi security problem.~\cite{eskandari2019sok} provide an overview of the blockchain front-running attacks.~\cite{daian2020flash} study the front-running attacks in decentralized exchanges and propose the concept of Miner Extractable Value (MEV), that is, miners can extract financial income by manipulating transaction orders.~\cite{zhou2021just} present a framework called \textsc{DeFiPoser}, which can automatically create revenue-generating transactions given the blockchain state.~\cite{tolmach2021formal} develop a formal process-algebraic technique that models DeFi protocols in a compositional manner to enable efficient verification of security properties.~\cite{babel2021clockwork} introduce a formal verification framework called Clockwork Finance for mechanized reasoning about the economic security properties of composed DeFi protocols.~\cite{qin2021attacking} study the flash loan attacks and propose an optimization approach to maximize the profit of DeFi attacks.~\cite{zhou2021high} analyze sandwich attacks in decentralized exchanges.~\cite{qin2021quantifying} quantify the extracted MEV on the Ethereum, including fixed spread liquidations, and present a generalized front-running algorithm, i.e., transaction replay. They also measure various risks faced by liquidation participants and quantify the instabilities of existing lending protocols~\cite{qin2021empirical}.~\cite{wang2022speculative}, formalizing a model for undercollateralized DeFi lending platforms to assess the risks of leverage-engaging borrowers.

\paragraph{DeFi Risk Rating Tools}\quad 
Providing a credible risk assessment is conducive for developers or investors to find potential risks in a DeFi protocol timely~\cite{boussard2019steward}. Not only does this help the crypto participants make more intelligent decisions, but it also incentivizes DeFi protocols to boost their overall security and trustworthiness. Explicitly, developers offering a risk rating tool hope that the sharing of tools, standards, and development patterns will support the safe growth of the DeFi ecosystem, making DeFi more secure for current and future adopters. Here, we examine two prominent rating tools, i.e., \textsc{DeFi-Score} and \textsc{DeFi-Safety}. \textsc{DeFi-Score}~\cite{defiscore} is a framework for assessing risk in permissionless lending platforms. It is a comparable value for measuring the risk of DeFi protocols, based on factors such as smart contract risk, collateralization, and liquidity. \textsc{DeFi-Safety}~\cite{defisafety} is an independent rating organization that evaluates DeFi protocols to produce an overall safety score based on transparency and adherence to best practices.

\begin{tcolorbox}[arc = 1mm, colback = black!2!white, colframe = black!65!white, boxrule=0.6mm]
\textbf{Summary: }

\textbf{(1)} Improving the security of DeFi protocols is becoming one of the most critical and urgent tasks for Web 3.0. Current researchers have proposed various solutions for the security of DeFi protocols, such as smart contract security audits, risk assessment mechanisms, transaction replay verification, and so on.

\textbf{(2)} DeFi provides tremendous potential for innovation and growth in the financial world. Nevertheless, this potential comes with significant risks and challenges that must be addressed to ensure the long-term development of the DeFi ecosystem.

\end{tcolorbox}

\section{Performance Evaluation}
\label{evaluation}
In this section, we aim to evaluate the performance of existing bug detection tools in handling DeFi protocols. We seek to answer the following questions.
\begin{itemize}[itemsep=1pt, topsep=1pt, leftmargin=\dimexpr\labelwidth + 2 \labelsep\relax]
\item \textbf{RQ1: }How do traditional smart contract vulnerability detection tools perform when applied to DeFi protocols?
\item \textbf{RQ2: }What is the performance of state-of-the-art tools in detecting DeFi attacks or vulnerabilities?
%\item \textbf{RQ3: }\emph{What is the number of bugs reported by candidate tools when analyzing DeFi protocols?}
\end{itemize}

The following parts of this section present the setup and procedure of our experimental evaluations. {First}, we construct a benchmark dataset of DeFi protocols and select candidate vulnerability detection tools for the experiments. {Then}, we present the experimental setup, including metrics that are commonly used to evaluate vulnerability detection tools and the execution environment. {Next}, in a unified execution environment, each tool analyzes DeFi protocols and generates bug reports. {Finally}, we count the total number of bugs, manually confirm the authenticity of each bug, and calculate the value of relevant metrics, followed by the analysis of the experimental results.

\subsection{Dataset Construction}
\label{dataset}
%To the best of our knowledge, there is still a lack of dataset for DeFi attack and vulnerability detection. Indeed, it is \emph{labor-intensive} and \emph{time-consuming} to collect and label a large-scale DeFi protocol dataset. 
We construct and release a benchmark dataset, which consists of real-world DeFi protocols and concerns six types of DeFi attacks, namely, \emph{flash loan (pump and arbitrage, price manipulation, reentrancy) attack, deflation token attack, sandwich attack}, and \emph{rug pull attack}. According to the reported DeFi attack incidents, we manually label the dataset, {i.e.,} whether a DeFi protocol suffers from an external attack or has a certain type of smart contract vulnerability. Our annotations facilitate the evaluation of DeFi protocol analysis tools extensively.
%Deeasy for DeFi protocols analysis tools to be easily evaluated.

\renewcommand{\arraystretch}{1.25}
\begin{table}
	\centering
	\caption{Statistical analysis of the DeFi protocol dataset. Category represents the classification of DeFi financial services. FL indicates the flash loan attack; DT refers to deflation token attack; SW denotes DeFi sandwich attack; RP means rug pull attack; Other Attack designates  uncommon types of attacks; Normal means a DeFi protocol that has no reported attack.}
	\resizebox{0.49\textwidth}{!}{
		\begin{tabular}{| c | c | c | c | c | c | c | c | c |}
			\hline
			\textbf{Category} & \textbf{FL} & \textbf{DT} & \textbf{SW} & \textbf{RP} & \textbf{Other Attack} & \textbf{Normal} & \textbf{Total}\\
			\hline
			{Lending and Borrowing} & 9 & 2 & 3 & -- & 1 & 3 & 18 \\
			\hline
			{Decentralized Exchange} & 12 & 1 & 3 & 2 & 3 & 19 & 40 \\
			\hline
			{Portfolio Management} & 3 & -- & 1 & -- & 1 &  2 & 7 \\
			\hline
			{Derivative} & 6 & -- & 2 & 1 & 5 & 11 & 25 \\
			\hline
			{Aggregator} & 1  & 1 & -- & -- & 1 & 3 & 6 \\
			\hline
			{Other} & --  & -- & -- & -- & -- & 3 & 3 \\
			\hline
			{Total} &  31 & 4 & 9 & 3 & 11 & 41 & 99 \\
			\hline
	\end{tabular} }
\label{table4}
\end{table}

\subsubsection{Dataset Collection and Processing} 
We create the dataset by collecting DeFi protocols from three different sources: (i) Ethereum official platform (more than $98\%$), (ii) GitHub repositories, and (iii) blog posts that analyze DeFi protocols. Note that all the DeFi protocols are collected from trusted entities in the field. We also ensure the traceability of each DeFi protocol by providing the address or URL, from which they are extracted as much as possible. 

In particular, we construct the dataset from different domains of DeFi financial services, including \emph{lending and borrowing, decentralized exchange, portfolio management, derivative,} and \emph{aggregator}, which are the common types of DeFi applications. Moreover, we label the dataset by inspecting the smart source code and historical transactions of each DeFi protocol. %Then, we label whether a DeFi protocol suffers from an external attack or has a certain smart contract vulnerability. %(The description of different attacks or vulnerabilities can be found in section~\ref{attacks}).

\subsubsection{Dataset Statistic and Analysis}
We collected $99$ open-source DeFi protocols into the dataset, which consists of a total of $7,340$ smart contracts. In the dataset, $58$ DeFi protocols have experienced at least one external attack. Table~\ref{table4} presents the statistics of the dataset. Specifically, $31$ protocols are affected by a flash loan attack, of which \emph{pump and arbitrage}, \emph{price manipulation}, and \emph{reentrancy} account for $26\%$, $58\%$, and $16\%$, respectively. $4$ protocols are struck by a deflation token attack. Additionally, $9$ protocols have suffered from a DeFi sandwich attack, and $3$ protocols have suffered  a rug pull attack. There are also $11$ protocols that have been influenced by other uncommon types of DeFi attacks, such as \texttt{yCredit.Finance}~\cite{yCredit}, \texttt{Oypn}~\cite{opyn}, \texttt{Furucombo}~\cite{furucombo2}. To embrace the DeFi community, we have released the benchmark dataset at \url{https://github.com/Messi-Q/DeFi-Protocol}.

\subsection{Experimental Setup}
\label{setup}

\subsubsection{Candidate Tools} 
\label{tools}
After reviewing all of the tools in Table~\ref{table2}, we select the fully open-sourced tools for further experiments. Specifically, for static analysis tools, we focus on the most frequently compared tools, namely, \textsc{Securify}, \textsc{VeriSol}, \textsc{Maian}, \textsc{Mythril}, \textsc{Manticore}, \textsc{Oyente}, \textsc{Osiris}, \textsc{TeEther}, \textsc{VerX}, \textsc{Smartcheck}, and \textsc{Slither}. For fuzzing test methods, we choose \textsc{ContractFuzzer}, \textsc{ConFuzzius}, \textsc{ILF}, \textsc{sFuzz}, \textsc{Smartian}, and IR-Fuzz. For deep learning-based methods, \textsc{ReChecker} and \textsc{TMP} are selected for experiments. For DeFi security tools, we select \textsc{ProMutator} and \textsc{DeFi-Score}.

\subsubsection{Execution Environment}
\label{environment}
A unified execution environment is extremely important for evaluating different tools, including the unified platform and consistency of runtime parameters. Considering that each tool has an arbitrary number of self-defined settings, we choose the general settings that they all support to carry out experiments, such as the compiler type and the data dependency analyzer. All experiments are performed on the same computer equipped with an Intel Core i9 CPU at 3.3GHz, an NVIDIA RTX 2080Ti GPU, and 64GB RAM. We repeat each experiment three times and report the average results for each method.

\begin{figure}
\centering
\includegraphics[width=8.5cm]{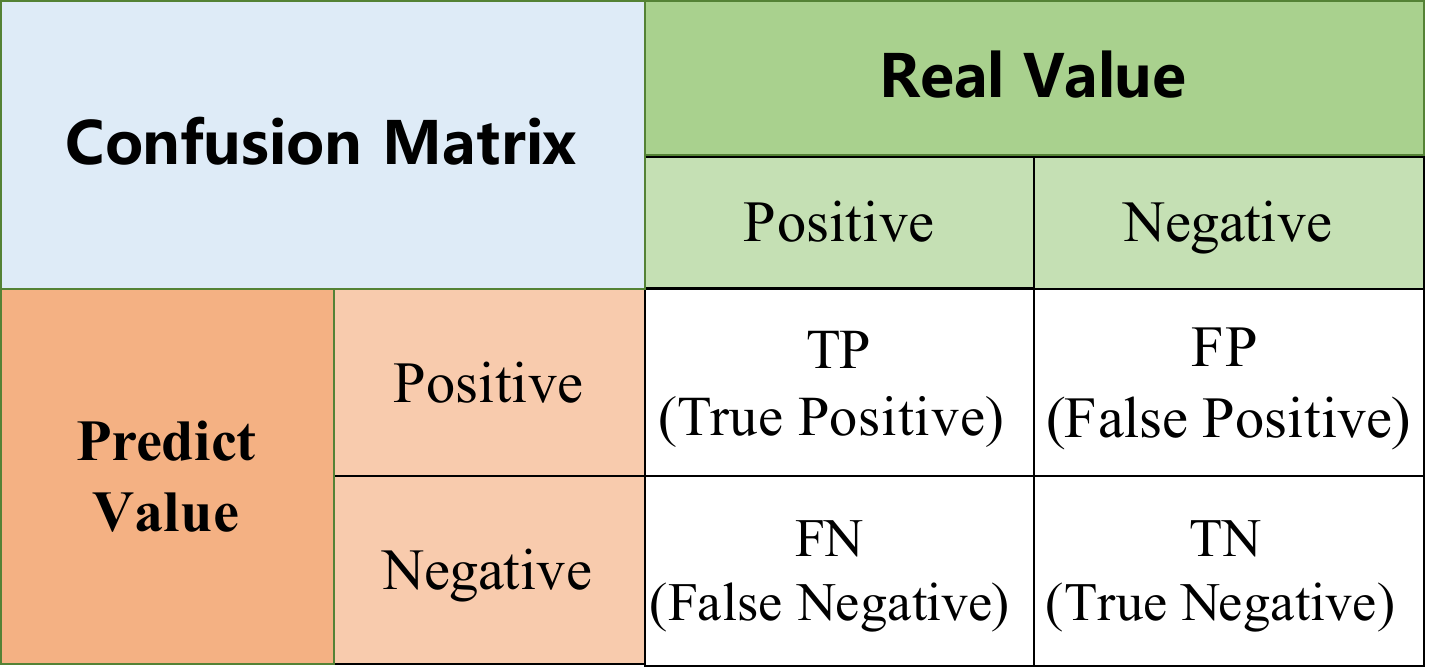}
\caption{Confusion matrix that reflects the performance of investigated tools.} 
\label{matrix}
\end{figure}

\subsection{Evaluation Metric}
\label{metric}
To measure the performance of the vulnerability detection tools, we adopt the widely used metrics, including \emph{accuracy, precision, recall}, and \emph{F1-score}. We run each tool in the dataset. When the execution is finished, each tool generates a bug report that contains the results of the vulnerability analysis. By matching it with the manually tagged labels of the DeFi protocols, we can calculate the true positive (TP), true negative (TN), false positive (FP), and false negative (FN) values of each tool. Fig.~\ref{matrix} illustrates all four possible outcomes with a 2$\times$2 confusion matrix. TP represents the number of data elements for which the detection tool correctly predicts samples with vulnerabilities. TN is the number of samples that the tool successfully detected without vulnerabilities. FP represents the number of samples that were falsely detected as having vulnerabilities. FN refers to the number of samples that were wrongly detected as protocols without vulnerabilities. Using these data, we further calculate the corresponding evaluation metrics as follows.

\emph{Accuracy.}\quad Accuracy is one of the most common metrics used to evaluate the pros and cons of the bug detection tools. Generally, it can objectively reflect the most direct ramification of a detection tool. Bug detection is essentially a binary classification problem, that is, the tool predicts whether a DeFi protocol endures an external attack or has a certain type of smart contract vulnerability.
\begin{equation}
\footnotesize Accuracy = \frac{TP + TN}{TP + FP + FN + TN}
\end{equation}

\emph{Precision.}\quad Precision is defined as the ratio of true positives and the total number of positives predicted by a tool. A greater precision value indicates a higher percentage of correct alarms and fewer false alarms, making bug verification and code modification easier.
\begin{equation}
\footnotesize Precision = \frac{TP}{TP + FP}
\end{equation}

\emph{Recall.}\quad Recall is the fraction of the total number of relevant instances that are actually detected. A higher recall value indicates that more real vulnerabilities are found, fewer hidden vulnerabilities are missed, and lower risk of unknown attacks.
\begin{equation}
\footnotesize Recall = \frac{TP}{TP + FN}
\end{equation}

\emph{F1-score.}\quad F1-score is the harmonic average of precision and recall, and it is usually treated as the critical evaluation indicator for some classification missions, {e.g.,} bug detection.
\begin{equation}
\footnotesize F1 = 2 \cdot \frac{Precision \cdot Recall}{Precision + Recall}
\end{equation}

Furthermore, we follow the settings of previous works~\cite{10.1145/3460319.3464837,durieux2020empirical} and select the \emph{unique bug} and \emph{execution time} as metrics to measure the performance of the tools.

\emph{Unique Bug.}\quad It refers to the total number of bug types reported by each tool after deduplication. When using the candidate tools to analyze DeFi protocols, we count the number of bug types reported by each tool.

\emph{Execution Time.}\quad Execution time is also a significant indicator among all methods of evaluating vulnerability detection tools. Currently, long audit time of detection tools leads to low efficiency of vulnerability analysis. In order to measure the time of execution, we record the latency that identifies the time interval between the analysis starts and ends. We present the average execution time of each DeFi protocol in the dataset.

\subsection{Results and Analysis}
\label{results_and_analysis}
In this subsection, we present the experimental results of our empirical study and the answers to the research questions. First, we verify the scalability of traditional smart contract bug detection tools in analyzing DeFi protocols. Then, we evaluate all available tools in terms of \emph{accuracy, precision, recall, F1-score, unique bug}, and \emph{execution time}. Finally, we provide an in-depth analysis of these tools and offer possible improvements for future work.

\renewcommand{\arraystretch}{1.0}
\begin{table}
	\centering
	\caption{Scalability evaluation of each tool in analyzing DeFi protocols and smart contracts with different solc versions.}
	\resizebox{0.49\textwidth}{!}{
		\begin{tabular}{ l  l  c c  c  c  c c }
			\toprule
			\# & Tools & {0.4.x} & {0.5.x} & {0.6.x} & {0.7.x} & {0.8.x} & DeFi protocols\\
			 \midrule
			1 & \textsc{Securify} & {\color{OliveGreen} \ding{51} } & {\color{DarkRed} \ding{55} } & {\color{DarkRed} \ding{55} } & {\color{DarkRed} \ding{55} } & {\color{DarkRed} \ding{55} } &  {\color{DarkRed} \ding{55} } \\
			2 & \textsc{VeriSol} & {\color{OliveGreen} \ding{51} } & {\color{DarkRed} \ding{55} } & {\color{DarkRed} \ding{55} } & {\color{DarkRed} \ding{55} } & {\color{DarkRed} \ding{55} } &  {\color{DarkRed} \ding{55} } \\
			3 & \textsc{Maian}& {\color{OliveGreen} \ding{51} } & {\color{DarkRed} \ding{55} } & {\color{DarkRed} \ding{55} } & {\color{DarkRed} \ding{55} } & {\color{DarkRed} \ding{55} }  &  {\color{DarkRed} \ding{55} } \\
			4 & \textsc{Mythril} & {\color{OliveGreen} \ding{51} } & {\color{DarkRed} \ding{55} } & {\color{DarkRed} \ding{55} } & {\color{DarkRed} \ding{55} } & {\color{DarkRed} \ding{55} } &  {\color{DarkRed} \ding{55} } \\
			5 & \textsc{Manticore }& {\color{OliveGreen} \ding{51} }  & {\color{DarkRed} \ding{55} } & {\color{DarkRed} \ding{55} } & {\color{DarkRed} \ding{55} } & {\color{DarkRed} \ding{55} } &  {\color{DarkRed} \ding{55} } \\
			6 & \textsc{Osiris} & {\color{OliveGreen} \ding{51} }  & {\color{DarkRed} \ding{55} } & {\color{DarkRed} \ding{55} } & {\color{DarkRed} \ding{55} }  & {\color{DarkRed} \ding{55} } &  {\color{DarkRed} \ding{55} } \\
			7 & \textsc{Oyente} &  {\color{OliveGreen} \ding{51} } & {\color{DarkRed} \ding{55} } & {\color{DarkRed} \ding{55} } & {\color{DarkRed} \ding{55} } & {\color{DarkRed} \ding{55} } &  {\color{DarkRed} \ding{55} } \\
			8 & \textsc{TeEther} &  {\color{OliveGreen} \ding{51} } &  {\color{OliveGreen} \ding{51} } & {\color{DarkRed} \ding{55} } & {\color{DarkRed} \ding{55} } & {\color{DarkRed} \ding{55} } & {\color{OliveGreen} \ding{51} } \\
			9 &  \textsc{VerX}  &  {\color{OliveGreen} \ding{51} } & {\color{OliveGreen} \ding{51} } & {\color{DarkRed} \ding{55} } & {\color{DarkRed} \ding{55} }  & {\color{DarkRed} \ding{55} } &  {\color{DarkRed} \ding{55} } \\
			10 & \textsc{Smartcheck} &  {\color{OliveGreen} \ding{51} } & {\color{OliveGreen} \ding{51} } & {\color{OliveGreen} \ding{51} } & {\color{OliveGreen} \ding{51} }  & {\color{OliveGreen} \ding{51} } &   {\color{OliveGreen} \ding{51} } \\
			11 & \textsc{Slither} &  {\color{OliveGreen} \ding{51} } & {\color{OliveGreen} \ding{51} } & {\color{DarkRed} \ding{55} } & {\color{DarkRed} \ding{55} } & {\color{DarkRed} \ding{55} } &  {\color{OliveGreen} \ding{51} } \\
			12 & \textsc{ContractFuzzer} &  {\color{OliveGreen} \ding{51} } & {\color{DarkRed} \ding{55} } & {\color{DarkRed} \ding{55} } & {\color{DarkRed} \ding{55} } & {\color{DarkRed} \ding{55} } &  {\color{DarkRed} \ding{55} } \\
			13 & \textsc{ConFuzzius} &  {\color{OliveGreen} \ding{51} } & {\color{DarkRed} \ding{55} } & {\color{DarkRed} \ding{55} } & {\color{DarkRed} \ding{55} }  & {\color{DarkRed} \ding{55} } &  {\color{DarkRed} \ding{55} } \\
			14 & \textsc{ILF} &  {\color{OliveGreen} \ding{51} } & {\color{OliveGreen} \ding{51} } & {\color{DarkRed} \ding{55} } & {\color{DarkRed} \ding{55} }  & {\color{DarkRed} \ding{55} } &  {\color{DarkRed} \ding{55} } \\
			15 & \textsc{sFuzz}  &  {\color{OliveGreen} \ding{51} } & {\color{DarkRed} \ding{55} } & {\color{DarkRed} \ding{55} } & {\color{DarkRed} \ding{55} }  & {\color{DarkRed} \ding{55} } &  {\color{DarkRed} \ding{55} } \\
			16 & \textsc{Smartian} &  {\color{OliveGreen} \ding{51} } & {\color{DarkRed} \ding{55} } & {\color{DarkRed} \ding{55} } & {\color{DarkRed} \ding{55} }  & {\color{DarkRed} \ding{55} } &  {\color{DarkRed} \ding{55} } \\
			17 & \textsc{IR-Fuzz}  &  {\color{OliveGreen} \ding{51} } & {\color{DarkRed} \ding{55} } & {\color{DarkRed} \ding{55} } & {\color{DarkRed} \ding{55} }  & {\color{DarkRed} \ding{55} } &  {\color{DarkRed} \ding{55} } \\
			18 & \textsc{Rechecker} &  {\color{OliveGreen} \ding{51} } & {\color{OliveGreen} \ding{51} } & {\color{OliveGreen} \ding{51} } & {\color{OliveGreen} \ding{51} }  & {\color{OliveGreen} \ding{51} } &  {\color{OliveGreen} \ding{51} } \\
			19 & \textsc{TMP} &  {\color{OliveGreen} \ding{51} } & {\color{DarkRed} \ding{55} }  & {\color{DarkRed} \ding{55} }  & {\color{DarkRed} \ding{55} }   & {\color{DarkRed} \ding{55} }  &   {\color{DarkRed} \ding{55} } \\
			20 & \textsc{DeFi-Score}  &  {\color{OliveGreen} \ding{51} } & {\color{OliveGreen} \ding{51} } & {\color{OliveGreen} \ding{51} } & {\color{OliveGreen} \ding{51} }  & {\color{DarkRed} \ding{55} } &  {\color{OliveGreen} \ding{51} } \\
			21 & \textsc{Promutator} &  {\color{OliveGreen} \ding{51} } & {\color{OliveGreen} \ding{51} } & {\color{OliveGreen} \ding{51} } & {\color{DarkRed} \ding{55} }  & {\color{DarkRed} \ding{55} } &  {\color{OliveGreen} \ding{51} } \\
			\bottomrule
	\end{tabular} }
\label{table5}
\end{table}

\subsubsection{Scalability Evaluation (Answer to RQ1)}
\label{analysis_1}
First, we run all the candidate tools on the DeFi protocol dataset to determine whether these tools can effectively analyze complex DeFi protocols. To be more specific, we evaluate the scalability of each tool by testing it on both DeFi protocols and smart contracts written in different Solidity versions. Table~\ref{table5} shows the experimental results. According to the principle behind each tool, we refer to tools 1 to 19 as traditional bug detection methods, while tools 20 to 21 are considered as DeFi analysis methods. We can observe that: {1)} most of the traditional vulnerability detection methods for smart contracts only support detecting low-version (i.e., \texttt{solc-0.4.x}) smart contracts, while only four of them are able to analyze DeFi protocols and generate detection results ({i.e.,} \textsc{TeEther}, \textsc{Smartcheck}, \textsc{Slither}, and \textsc{Rechecker}), and {2)} the DeFi analysis methods such as \textsc{Promutator} can handle not only DeFi protocols but also other versions of smart contracts.

\renewcommand{\arraystretch}{1.3}
\begin{table}
	\centering
	\caption{Performance evaluation of each tool in terms of accuracy, recall, precision, F1-score, and unique bug.}
	\resizebox{0.49\textwidth}{!}{
		\begin{tabular}{ l  l c  c  c  c  c}
			\toprule
			\# & {Tools} & {Accuracy(\%) } & {Recall(\%) } & {Precision(\%) } & {F1-score(\%) } & {Unique Bug} \\
			 \midrule
			1 & \textsc{Slither} &  58.37 & 4.50 & 83.33 & 8.54  & 25 \\
			2 & \textsc{Smartcheck} & 58.91  & 26.96 & 35.65 & 30.70  &  12 \\
			3 & \textsc{TeEther} &  63.13 & 31.32 & 75.55 & 44.28  &  1 \\
			4 & \textsc{Rechecker} &  50.48 & 86.02 & 43.28 & 57.59  &  1 \\
			5 & \textsc{DeFi-Score} &  \textbf{73.27} & \textbf{75.94} & \textbf{63.14} & \textbf{68.95}  &  1 \\
			6 & \textsc{Promutator} &  \textbf{68.45} & \textbf{72.30} & \textbf{68.69} & \textbf{70.45} &  1 \\
			\bottomrule
	\end{tabular} }
\label{table6}
\end{table}

\subsubsection{Performance Evaluation (Answer to RQ2)}
\label{analysis_2}
We then conduct experiments to evaluate the performance of available tools that can analyze DeFi protocols. Quantitative experimental results of each tool are summarized in Table~\ref{table6}. From the table, we obtain the following observations. 
\textbf{(1)} Comparing different tools, DeFi analysis methods gain a higher accuracy than other methods.For example, \textsc{DeFi-Score} and \textsc{Promutator} achieve accuracies of 73.27\% and 68.45\%, respectively, which is 10.14\% and 5.32\% higher than the state-of-the-art tool \textsc{TeEther}.
\textbf{(2)} In terms of \emph{F1-Score}, \textsc{DeFi-Score} and \textsc{Promutator} achieve 68.95\% and 70.45\% \emph{F1-score}, outperforming state-of-the-art methods by a large margin. We attribute the superior performance of DeFi analysis methods to the ability to recover high-level semantics from DeFi protocols. 
\textbf{(3)} For \emph{unique bug}, traditional vulnerability detection methods ({e.g.,} \textsc{Slither} and \textsc{Smartcheck}) generate more unique bugs, while \textsc{DeFi-Score} is used to assess risk in lending protocols and \textsc{Promutator} can only identify one specific DeFi attack, i.e., the \emph{price manipulation attack}.

\renewcommand{\arraystretch}{0.9}
\begin{table}
    \caption{Average execution time of each tool.}
    \centering
    \begin{tabularx}{0.49\textwidth}{@{}r X l r @{}}
    \toprule
\multirow{2}{*}{\#} & \multirow{2}{*}{Tools} & \multicolumn{2}{c}{Execution Time(s)}  \\
\cline{3-4}
    &            & Average & Total  \\
    \midrule
1 & \textsc{Honeybadger} & 111.68    &  11056.63 \inlinechart{11057}{261980}\\
2 & \textsc{Maian}      & 436.55    &   43218.67 \inlinechart{43219}{261980}\\
3 & \textsc{Manticore }  & 164.40    &   16275.90 \inlinechart{16276}{261980}\\
4 & \textsc{Mythril}    & 126.62    &   12535.01 \inlinechart{12535}{261980}\\
5 & \textsc{Osiris}      & 92.23    &   9130.77 \inlinechart{9131}{261980}\\
6 & \textsc{Oyente}     & 208.95    &   20686.05 \inlinechart{20686}{261980}\\
7 & \textsc{Securify}   & 208.87    &   20678.13 \inlinechart{20678}{261980}\\
8 & \textsc{Slither}    & 216.26    &   21409.97 \inlinechart{21410}{261980}\\
9 & \textsc{Smartcheck} & 528.69    &  52339.97 \inlinechart{52340}{261980}\\
10 & \textsc{TeEther} & 331.89    &  32856.88 \inlinechart{32857}{261980}\\
11 & \textsc{ILF} & 190.93    & 18902.03 \inlinechart{18902}{261980}\\
12 & \textsc{Rechecker} & 8.02    &  793.95 \inlinechart{794}{261980}\\
13 & \textsc{DeFi-Score} & 10.05    &  994.54 \inlinechart{995}{261980}\\
14 & \textsc{Promutator} & 11.11    & 1100.28 \inlinechart{1100}{261980}\\
\bottomrule
    \end{tabularx}
        \label{tab6}
\end{table}

Further, we evaluate the execution time required by the tools to analyze the DeFi protocols in the dataset. Table~\ref{tab6} presents the average and total time used by each tool. The average execution time is per project. It considers the execution of the tool on a DeFi protocol, including compilation, construction of intermediate representation, analysis, and result parsing. From the table, we can observe two different groups of execution times: the tools that take a few seconds to execute and the tools that take a few minutes. For example, all the traditional vulnerability detection tools (lines 1-11 in Table~\ref{tab6}) take a long time to analyze a DeFi protocol, ranging from 92.23s to 528.69s per protocol. In contrast, the DeFi analysis methods such as \textsc{DeFi-Score} and \textsc{Promutator} are much faster, taking only 10.05s and 11.11s per protocol, respectively. This may stem from two facts: \textbf{(1)} traditional vulnerability detection tools still have inherent difficulties in understanding the complex logic of DeFi protocols; \textbf{(2)} the specially designed DeFi analysis methods are able to recover the high-level semantics of DeFi protocols.

Interestingly, the deep learning-based method \textsc{Rechecker} is a fast tool that takes on average only 8.02s to analyze a DeFi protocol. We explain that the deep learning model spends a lot of time in the training phase, while the trained model can quickly identify the key characteristics of the DeFi protocols in the detection phase. However, we would like to point out that the execution time does not reflect the complete picture of the performance of these tools. For example, most traditional vulnerability detection tools carry out invalid runs as their output is either empty or unavailable. 

Based on the experimental results, it is not difficult to find that the performance of traditional smart contract bug detection methods on DeFi protocols is far from satisfactory. They suffer from the inherent issues in understanding the logical semantics of DeFi protocols. Particularly, a DeFi protocol typically consists of multiple interactive contracts, making it difficult for these tools to work. On the other hand, while the existing DeFi analysis methods achieve promising performance in detecting certain attacks, they are still unable to handle multiple types of attacks and discover high-level logic vulnerabilities in DeFi protocols.

\section{Discussion: Open Issues and Future Challenges}
\label{challenges}
While the interest in DeFi security is growing and considerable progress has been made in this area over the past few years~\cite{chaliasos2023smart}, there are still several open issues that need to be addressed in future research activities. In this section, we discuss the main issues that users and developers in the intertwined DeFi space face, and provide a summary of the current challenges associated with DeFi security from our perspective.

\subsection{Underlying Infrastructure}
\label{underlying_infrastructure}
One of the most critical security issues in DeFi is associated with the underlying blockchain platform. Since DeFi applications are composed of smart contracts running on the blockchain system, their executions are affected by potential blockchain vulnerabilities. For instance, eclipse attacks~\cite{heilman2015eclipse,gervais2016security} to the blockchain could separate the network, which results in churning logic among different market transactions. Feather forking~\cite{zhang2019lay} and block reorganization~\cite{hou2019squirrl} may cause consensus vulnerabilities, and thus lead to double-spending issues in DeFi markets. Furthermore, the limited throughput of the blockchain may cause network congestion~\cite{ahn2020packet}, delaying the DeFi transactions.

Owing to the open and transparent nature of the blockchain system, all market states of DeFi protocols can be observed publicly in real-time. Meanwhile, users’ transactions will be broadcast to the blockchain network and stored in a mempool. The transactions in this mempool can be accessed by anyone before miners package them in a block. Therefore, attackers can forecast the future market states based on non-executed transactions and utilize this knowledge to manipulate the market operation. For instance, attackers may observe a victim transaction in the mempool and then front-run~\cite{daian2020flash} or back-run~\cite{zhou2021a2mm} them to gain profit. Moreover, DeFi protocols are isolated from outside of the blockchain system, while external information cannot be utilized by these applications in real-time. As a result, DeFi protocols rely on each other and share information to maintain the market operation. For example, some lending borrowing protocols utilize the real-time exchange rate in DEXs to determine the collateral rate~\cite{wu2021defiranger}. Nevertheless, some attackers may target dependencies between platforms by manipulating the state of one application and profiting from another one~\cite{zhou2021just}.

Security challenges related to DeFi protocols involve interactions between traders and DeFi protocols, it is difficult to detect them effectively before launching applications. Furthermore, such security issues are associated with DeFi markets and continuously impact DeFi users in the long term~\cite{wang2022impact}. Therefore, it is important to understand how users become aware, perceive, and prevent security issues in DeFi markets.

\subsection{DeFi Smart Contract}
\label{smart_contract}
Another key factor impacting on DeFi security is related to smart contracts. Like traditional programs, smart contracts may contain vulnerabilities. Smart contracts from various fields now hold over one trillion dollars worth of virtual coins, attracting numerous attacks. Cybercriminals have mined a variety of vulnerabilities in smart contracts, leading to substantial losses~\cite{perez2021smart,su2021evil,oyente}. A large number of works are designed, both in academia and industry, to unearth smart contract bugs and vulnerabilities~\cite{schneidewind2020ethor,rodler2021evmpatch,securify}. They mainly revolve around static analysis and fuzzing techniques to discover vulnerabilities. However, they have not evolved yet to embrace the composable nature of smart contracts, which makes it impossible for such tools to reason about scenarios where the issue happens due to a change in something external to smart contracts, such as a sudden change in a price returned by an oracle (which often occurs in the DeFi protocols). Furthermore, most tools reason very little about the semantic properties of smart contracts, such as how a particular execution path can influence ERC-20 token balances. Here, we summarize the limitation of existing vulnerability detection techniques  for smart contracts from five aspects: \emph{formal verification, symbolic execution, intermediate representation, fuzzing test, and deep learning}, respectively.

\paragraph{Formal Verification} Formal verification methods rely on rigorous mathematical derivation and verification, they fail to perform path analysis of smart contracts. Therefore, they lack the detection and judgment of executable paths in the contract, resulting in a high rate of false positives and negatives.

\paragraph{Symbolic Execution} Although the symbolic execution approach effectively improves vulnerability detection, it significantly increases the computational resources and time overhead during analysis. In particular, symbolic execution suffers from the inherent problems of state space and execution path explosion.

\paragraph{Intermediate Representation} Intermediate representation methods mostly rely on predefined semantic rules, they are unable to detect the complex business logic of smart contracts and are prone to false positives. In addition, they cannot traverse the execution paths that may exist in the contract.

\paragraph{Fuzzing Test} Current fuzzing approaches for smart contracts tend to generate function invocation sequences randomly, ignoring data dependencies between functions. Moreover, most existing methods fail to steer fuzzing towards the branches that are rare or more likely to possess bugs, resulting in inefficient fuzzing where resources are wasted in normal branches. Meanwhile, it is still challenging how to intelligently adjust seed mutation such that the generated test cases could reach a target conditional branch efficiently.

\paragraph{Deep Learning} Deep learning-based methods detect vulnerabilities in smart contracts by training a large number of vulnerable contracts and constructing a learned neural network as the detection model. However, the absence of high-quality smart contract vulnerability dataset severely hinders the performance and credibility of deep learning-based approaches. Furthermore, due to the black-box nature of neural networks, deep learning-based methods endure the inherent problems of inferior interpretability.

\subsection{Maximal Extractable Value}
\label{mev}
The notion of maximal extractable value (MEV) was introduced by~\cite{daian2020flash}. It refers to the total value that miners could extract from the manipulation of transactions within a given timeframe, which may include multiple blocks’ worth of transactions. MEV is predominantly captured by DeFi traders through structural arbitrage trading strategies, and miners indirectly profit from traders’ transaction fees. 

Traders on Ethereum express their willingness to pay for inclusion in a block through their transaction’s gas price, and they indicate how much they are willing to pay miners through the transaction fee. Miners, as economically rational actors, undoubtedly pick the transactions with the highest gas price and order them by gas spend in the block they are producing. Unfortunately, the financial system that is built on Ethereum creates many pure profit opportunities such as liquidations and arbitrages of many kinds.

Obadia et al.~\cite{flashbot1,flashbot2} propose to construct a permissionless, transparent, and fair ecosystem for MEV extraction to preserve the ideals of Ethereum, mitigating the negative externalities and existential risks posed by MEV. Their approach to mitigating the MEV crisis can be broken down into three parts, {namely}, {illuminate the dark forest}, {democratize extraction}, and {distribute benefits}.

\subsection{Composability}
\label{composability}
Composability is one of the core features of decentralized finance~\cite{wachter2021measuring,saengchote2021defi,babel2021clockwork}, which allows applications and protocols to interact with one another in a permissionless way. Benefiting from the open nature of DeFi, these applications are leveraging one another to create a synergistic effect and create new forms of financial services never thought possible. For example, cryptoassets in DeFi can be easily and repeatedly tokenized and interchanged between different DeFi protocols in a fashion akin to rehypothecation. 
Unfortunately, although the composability of DeFi offers the potential to construct complex, interconnected financial systems, it also bears the danger of exposing traders to possible composability risks~\cite{gudgeon2020decentralized}. 
A typical example of composability risk is the use of flash loans for manipulating instantaneous automated market makers (AMMs) and financially exploiting protocols that use those AMMs as price feeds. This has repeatedly been exploited in past attacks (\emph{e.g.,}~\cite{Thurman,harvest}). 

To deal with composability risks, recent works~\cite{tolmach2021formal} propose a process-algebraic technique to achieve property verification by modeling DeFi protocols in a compositional manner. Nonetheless, a critical gap in DeFi research towards taxonimizing and formalizing models to quantify composability risks remains. Although regarded as a fundamental basis of DeFi, there are still considerable risks to the composability of the ecosystem. In particular, ensuring the safety of protocol composition will be close to impossible for any protocol designer and forms a major challenge for DeFi going forward.

\iffalse
\subsubsection{Interdependence}
Designing new protocols for the DeFi space requires special consideration. In particular, because of the opportunity of composing different DeFi products and creating new protocols based on existing ones. The security of a single protocol cannot be analyzed in a standalone model; influences of other protocols also need to be taken into account. We show this aspect by highlighting two specific attacks presented in prior work. The first attack, called front-running, was analyzed in~\cite{daian2020flash}. The term front-running comprises all scenarios where one party tries to get her transaction recorded before a competing transaction. Any attempt to front-run may result in a so-called priority gas auction where users alternate in increasing their transactions’ gas price to incentive miners to include their transactions first.
\fi

\subsection{Oracle}
\label{oracle}
DeFi oracles are essentially a third-party service that can enable blockchain smart contracts to access external and real-world (\emph{i.e.,} off-chain) resources, such as price information and token exchange rates~\cite{liu2021first,gu2020empirical}. While oracles play a critical role in the DeFi ecosystem, they also involve a degree of security risks. To be even more specific, the underlying mechanics of DeFi oracles are vague and unexplored. First, their deployment practices, such as how frequently the price updates and how to aggregate the price value from multiple nodes, are not transparent nor accountable, leaving room for various external malicious behaviors. Second, the level of trust placed in oracles is unclear and most likely unknown to many participants of the ecosystem. Finally, the impact of a potentially malicious oracle (\emph{or} a group of oracles) on the DeFi ecosystem is still not investigated.

While the interaction between two on-chain entities (\emph{e.g.,} smart contracts) is simple, transferring information from external sources like websites to a smart contract introduces new challenges. Many DeFi applications rely on external information, such as exchange rates and price information, which are provided by DeFi oracles. Since the data originating from these oracles impact the behavior of smart contracts and users, the challenges posed by transferring external data on-chain show a major concern. In particular, the security of these DeFi applications is based on the reliability, accuracy, and correctness of the provided information from oracles.

\subsection{Privacy}
\label{privacy}
The privacy and anonymity of DeFi protocols is at present a significantly understudied area~\cite{xu2017enabling,chu2021manta,dai2021flexible}. A common view is emerging, that is the DeFi protocols cannot hit the mainstream until the problem of privacy is addressed. Since financial data is highly sensitive for many individuals, privacy is a relevant topic. In particular, users aim for private transactions such that no unauthorized party can obtain information about users’ financial activities. Obviously, users may enjoy complete privacy when trading for cryptoassets in the DeFi market. However, a large proportion of DeFi transactions at present occur in projects built on Ethereum, wherein agents at best have pseudo-anonymity~\cite{werner2021sok}. This indicates that if a trader’s real-world identity can be linked to an on-chain address, all the actions undertaken by the trader through that address are observable. While recent advances in zero-knowledge proofs~\cite{xie2019libra,boyle2019practical} and multi-party computations \cite{reich2019privacy,bayatbabolghani2018secure} hold many promises, these technologies are yet to gain traction in the context of DeFi. One of the main challenges is the large computational cost of these technologies, which makes them very expensive to use and deploy in the context of DeFi. Therefore, a decrease in the computational cost of the underlying blockchain may be critical to how widely privacy-preserving technologies can be deployed by DeFi protocols.

\section{Conclusion and Outlook}
\label{conclusion}
With further expansion of the DeFi ecosystem and its deployment in a constantly increasing number of application domains, security concerns ({e.g.,} DeFi attacks) associated with DeFi are only expected to increase in the years to come. A considerable amount of DeFi attacks, causing billions of dollars in financial losses, have severely threatened the security of the whole DeFi ecosystem.

This paper aims to provide a comprehensive understanding of various DeFi attacks and present an empirical review of the state-of-the-art techniques that can detect or repair vulnerabilities in smart contracts and DeFi protocols. As seen from the literature review, a plethora of techniques have been proposed over the years that are able to identify smart contract vulnerabilities automatically. Unfortunately, since most of them are designed for handling traditional smart contracts or a single contract, they can only discover low-level bugs in smart contracts while failing to analyze complex DeFi protocols. Although recent efforts can capture attack patterns in DeFi protocols by recovering high-level DeFi semantics from raw transactions, they still face obstacles in distinguishing compound attacks and vulnerabilities. On the other side, the immutability of the underlying blockchain makes it difficult to fix vulnerabilities in the on-chain contracts. Current automated repair tools can fix limited types of vulnerabilities and struggle to handle complex DeFi protocols. 

There are still multiple key issues that have to be addressed in the future to make them applicable widely in various attack detection, risk assessment, and automated repair. As discussed above, understanding the dependencies between intertwined DeFi spaces is critical to figuring out the logic rules of DeFi protocols. Furthermore, enhancing scalability is important for existing analysis tools to support more types of attack detection and automated fixing. These and related challenges will undoubtedly represent research priorities related to DeFi safe ecological construction in the forthcoming year.

\footnotesize
\bibliographystyle{IEEEtran}
\bibliography{DeFi}

% Generated by IEEEtran.bst, version: 1.14 (2015/08/26)
\begin{thebibliography}{100}
\providecommand{\url}[1]{#1}
\csname url@samestyle\endcsname
\providecommand{\newblock}{\relax}
\providecommand{\bibinfo}[2]{#2}
\providecommand{\BIBentrySTDinterwordspacing}{\spaceskip=0pt\relax}
\providecommand{\BIBentryALTinterwordstretchfactor}{4}
\providecommand{\BIBentryALTinterwordspacing}{\spaceskip=\fontdimen2\font plus
\BIBentryALTinterwordstretchfactor\fontdimen3\font minus
  \fontdimen4\font\relax}
\providecommand{\BIBforeignlanguage}[2]{{%
\expandafter\ifx\csname l@#1\endcsname\relax
\typeout{** WARNING: IEEEtran.bst: No hyphenation pattern has been}%
\typeout{** loaded for the language `#1'. Using the pattern for}%
\typeout{** the default language instead.}%
\else
\language=\csname l@#1\endcsname
\fi
#2}}
\providecommand{\BIBdecl}{\relax}
\BIBdecl

\bibitem{nakamoto2008bitcoin}
S.~Nakamoto, ``Bitcoin: A peer-to-peer electronic cash system,''
  \emph{Decentralized Business Review}, p. 21260, 2008.

\bibitem{zou2019smart}
W.~Zou, D.~Lo, P.~S. Kochhar, X.-B.~D. Le, X.~Xia, Y.~Feng, Z.~Chen, and B.~Xu,
  ``Smart contract development: Challenges and opportunities,'' \emph{IEEE
  Transactions on Software Engineering}, vol.~47, no.~10, pp. 2084--2106, 2019.

\bibitem{rosa2018blockchain}
R.~V. Rosa and C.~E. Rothenberg, ``Blockchain-based decentralized applications
  meet multi-administrative domain networking,'' in \emph{Proceedings of the
  ACM SIGCOMM 2018 Conference on Posters and Demos}, 2018, pp. 114--116.

\bibitem{gao2019towards}
J.~Gao, H.~Liu, Y.~Li, C.~Liu, Z.~Yang, Q.~Li, Z.~Guan, and Z.~Chen, ``Towards
  automated testing of blockchain-based decentralized applications,'' in
  \emph{2019 IEEE/ACM 27th International Conference on Program Comprehension
  (ICPC)}.\hskip 1em plus 0.5em minus 0.4em\relax IEEE, 2019, pp. 294--299.

\bibitem{benisi2020blockchain}
N.~Z. Benisi, M.~Aminian, and B.~Javadi, ``Blockchain-based decentralized
  storage networks: A survey,'' \emph{Journal of Network and Computer
  Applications}, vol. 162, p. 102656, 2020.

\bibitem{yu2019proof}
B.~Yu, J.~Liu, S.~Nepal, J.~Yu, and P.~Rimba, ``Proof-of-qos: Qos based
  blockchain consensus protocol,'' \emph{Computers \& Security}, vol.~87, p.
  101580, 2019.

\bibitem{liu2021smart}
Z.~Liu, P.~Qian, X.~Wang, L.~Zhu, Q.~He, and S.~Ji, ``Smart contract
  vulnerability detection: From pure neural network to interpretable graph
  feature and expert pattern fusion,'' in \emph{IJCAI}, 2021, pp. 2751--2759.

\bibitem{benvcic2018distributed}
F.~M. Ben{\v{c}}i{\'c} and I.~P. {\v{Z}}arko, ``Distributed ledger technology:
  Blockchain compared to directed acyclic graph,'' in \emph{2018 IEEE 38th
  International Conference on Distributed Computing Systems (ICDCS)}.\hskip 1em
  plus 0.5em minus 0.4em\relax IEEE, 2018, pp. 1569--1570.

\bibitem{liu2021combining}
Z.~Liu, P.~Qian, X.~Wang, Y.~Zhuang, L.~Qiu, and X.~Wang, ``Combining graph
  neural networks with expert knowledge for smart contract vulnerability
  detection,'' \emph{IEEE Transactions on Knowledge and Data Engineering},
  vol.~35, no.~2, pp. 1296--1310, 2023.

\bibitem{chen2018survey}
W.~Chen, Z.~Xu, S.~Shi, Y.~Zhao, and J.~Zhao, ``A survey of blockchain
  applications in different domains,'' in \emph{Proceedings of the 2018
  International Conference on Blockchain Technology and Application}, 2018, pp.
  17--21.

\bibitem{qian2019digital}
P.~Qian, Z.~Liu, X.~Wang, J.~Chen, B.~Wang, and R.~Zimmermann, ``Digital
  resource rights confirmation and infringement tracking based on smart
  contracts,'' in \emph{2019 IEEE 6th International Conference on Cloud
  Computing and Intelligence Systems (CCIS)}.\hskip 1em plus 0.5em minus
  0.4em\relax IEEE, 2019, pp. 62--67.

\bibitem{chen2019cryptoar}
Y.-P. Chen and J.-C. Ko, ``Cryptoar wallet: A blockchain cryptocurrency wallet
  application that uses augmented reality for on-chain user data display,'' in
  \emph{Proceedings of the 21st International Conference on Human-Computer
  Interaction with Mobile Devices and Services}, 2019, pp. 1--5.

\bibitem{hu2021transaction}
T.~Hu, X.~Liu, T.~Chen, X.~Zhang, X.~Huang, W.~Niu, J.~Lu, K.~Zhou, and Y.~Liu,
  ``Transaction-based classification and detection approach for ethereum smart
  contract,'' \emph{Information Processing \& Management}, vol.~58, no.~2, p.
  102462, 2021.

\bibitem{moosavi2021blockchain}
J.~Moosavi, L.~M. Naeni, A.~M. Fathollahi-Fard, and U.~Fiore, ``Blockchain in
  supply chain management: a review, bibliometric, and network analysis,''
  \emph{Environmental Science and Pollution Research}, pp. 1--15, 2021.

\bibitem{agbo2019blockchain}
C.~C. Agbo, Q.~H. Mahmoud, and J.~M. Eklund, ``Blockchain technology in
  healthcare: a systematic review,'' in \emph{Healthcare}, vol.~7, no.~2.\hskip
  1em plus 0.5em minus 0.4em\relax Multidisciplinary Digital Publishing
  Institute, 2019, p.~56.

\bibitem{williams2020cross}
I.~Williams, \emph{Cross-industry Use of Blockchain Technology and
  Opportunities for the Future}.\hskip 1em plus 0.5em minus 0.4em\relax IGI
  global, 2020.

\bibitem{chen2019decentralized}
Y.~Chen and C.~Bellavitis, ``Decentralized finance: Blockchain technology and
  the quest for an open financial system,'' \emph{Stevens Institute of
  Technology School of Business Research Paper}, 2019.

\bibitem{zhou2021just}
L.~Zhou, K.~Qin, A.~Cully, B.~Livshits, and A.~Gervais, ``On the just-in-time
  discovery of profit-generating transactions in defi protocols,'' in
  \emph{2021 IEEE Symposium on Security and Privacy (SP)}.\hskip 1em plus 0.5em
  minus 0.4em\relax IEEE, 2021, pp. 919--936.

\bibitem{werner2021sok}
S.~M. Werner, D.~Perez, L.~Gudgeon, A.~Klages-Mundt, D.~Harz, and W.~J.
  Knottenbelt, ``Sok: Decentralized finance (defi),'' \emph{arXiv preprint
  arXiv:2101.08778}, 2021.

\bibitem{jensen2021introduction}
J.~R. Jensen, V.~von Wachter, and O.~Ross, ``An introduction to decentralized
  finance (defi),'' \emph{Complex Systems Informatics and Modeling Quarterly},
  no.~26, pp. 46--54, 2021.

\bibitem{amler2021defi}
H.~Amler, L.~Eckey, S.~Faust, M.~Kaiser, P.~Sandner, and B.~Schlosser,
  ``Defi-ning defi: Challenges \& pathway,'' in \emph{2021 3rd Conference on
  Blockchain Research \& Applications for Innovative Networks and Services
  (BRAINS)}.\hskip 1em plus 0.5em minus 0.4em\relax IEEE, 2021, pp. 181--184.

\bibitem{bartoletti2021sok}
M.~Bartoletti, J.~H.-y. Chiang, and A.~L. Lafuente, ``Sok: lending pools in
  decentralized finance,'' in \emph{International Conference on Financial
  Cryptography and Data Security}.\hskip 1em plus 0.5em minus 0.4em\relax
  Springer, 2021, pp. 553--578.

\bibitem{raheman2021architecture}
A.~Raheman, A.~Kolonin, B.~Goertzel, G.~Hegykozi, and I.~Ansari, ``Architecture
  of automated crypto-finance agent,'' \emph{arXiv preprint arXiv:2107.07769},
  2021.

\bibitem{bartoletti2021theory}
M.~Bartoletti, J.~H.-y. Chiang, and A.~Lluch-Lafuente, ``A theory of automated
  market makers in defi,'' in \emph{International Conference on Coordination
  Languages and Models}.\hskip 1em plus 0.5em minus 0.4em\relax Springer, 2021,
  pp. 168--187.

\bibitem{bartoletti2021maximizing}
M.~Bartoletti, H.-y. Chiang, James, and A.~Lluch-Lafuente, ``Maximizing
  extractable value from automated market makers,'' \emph{arXiv preprint
  arXiv:2106.01870}, 2021.

\bibitem{bartoletti2021towards}
M.~Bartoletti, J.~H.-y. Chiang, and A.~L. Lafuente, ``Towards a theory of
  decentralized finance,'' in \emph{International Conference on Financial
  Cryptography and Data Security}.\hskip 1em plus 0.5em minus 0.4em\relax
  Springer, 2021, pp. 227--232.

\bibitem{gudgeon2020decentralized}
L.~Gudgeon, D.~Perez, D.~Harz, B.~Livshits, and A.~Gervais, ``The decentralized
  financial crisis,'' in \emph{2020 Crypto Valley Conference on Blockchain
  Technology (CVCBT)}.\hskip 1em plus 0.5em minus 0.4em\relax IEEE, 2020, pp.
  1--15.

\bibitem{qin2021attacking}
K.~Qin, L.~Zhou, B.~Livshits, and A.~Gervais, ``Attacking the defi ecosystem
  with flash loans for fun and profit,'' in \emph{International Conference on
  Financial Cryptography and Data Security}.\hskip 1em plus 0.5em minus
  0.4em\relax Springer, 2021, pp. 3--32.

\bibitem{oosthoek2021flash}
K.~Oosthoek, ``Flash crash for cash: Cyber threats in decentralized finance,''
  \emph{arXiv preprint arXiv:2106.10740}, 2021.

\bibitem{wang2021promutator}
S.-H. Wang, C.-C. Wu, Y.-C. Liang, L.-H. Hsieh, and H.-C. Hsiao, ``Promutator:
  Detecting vulnerable price oracles in defi by mutated transactions,'' in
  \emph{2021 IEEE European Symposium on Security and Privacy Workshops
  (EuroS\&PW)}.\hskip 1em plus 0.5em minus 0.4em\relax IEEE, 2021, pp.
  380--385.

\bibitem{Group12}
A.~Group, ``Preventing re-entrancy attacks — lessons from history,'' Website,
  Accessed: 08-08-2023,
  \url{https://medium.com/amber-group/preventing-re-entrancy-attacks-lessons-from-history-c2d96480fac3}.

\bibitem{Inspex}
Inspex, ``Reentrancy attack on cream finance,'' Website, Accessed: 08-08-2023,
  \url{https://inspexco.medium.com/reentrancy-attack-on-cream-finance-incident-analysis-1c629686b6f5}.

\bibitem{Poly}
Wikipedia, ``Poly network exploit,'' Website, Accessed: 08-08-2023,
  \url{https://en.wikipedia.org/wiki/Poly\_Network\_exploit}.

\bibitem{ronin}
A.~Hayward, ``Hacker drains \$622m from axie infinity’s ronin ethereum
  sidechain,'' Website, Accessed: 08-08-2023,
  \url{https://decrypt.co/96322/hacker-622-million-axie-infinity-ronin-ethereum}.

\bibitem{rekt}
Rekt, ``Rekt news,'' Website, Accessed: 08-08-2023,
  \url{https://de.fi/rekt-database}.

\bibitem{wang2021towards}
D.~Wang, S.~Wu, Z.~Lin, L.~Wu, X.~Yuan, Y.~Zhou, H.~Wang, and K.~Ren, ``Towards
  a first step to understand flash loan and its applications in defi
  ecosystem,'' in \emph{Proceedings of the Ninth International Workshop on
  Security in Blockchain and Cloud Computing}, 2021, pp. 23--28.

\bibitem{tolmach2021formal}
P.~Tolmach, Y.~Li, S.-W. Lin, and Y.~Liu, ``Formal analysis of composable defi
  protocols,'' in \emph{International Conference on Financial Cryptography and
  Data Security}.\hskip 1em plus 0.5em minus 0.4em\relax Springer, 2021, pp.
  149--161.

\bibitem{qin2021cefi}
K.~Qin, L.~Zhou, Y.~Afonin, L.~Lazzaretti, and A.~Gervais, ``Cefi vs.
  defi--comparing centralized to decentralized finance,'' \emph{arXiv preprint
  arXiv:2106.08157}, 2021.

\bibitem{chohan2021decentralized}
U.~W. Chohan, ``Decentralized finance (defi): an emergent alternative financial
  architecture,'' \emph{Critical Blockchain Research Initiative (CBRI) Working
  Papers}, 2021.

\bibitem{marino2016setting}
B.~Marino and A.~Juels, ``Setting standards for altering and undoing smart
  contracts,'' in \emph{International Symposium on Rules and Rule Markup
  Languages for the Semantic Web}.\hskip 1em plus 0.5em minus 0.4em\relax
  Springer, 2016, pp. 151--166.

\bibitem{xue2020cross}
Y.~Xue, M.~Ma, Y.~Lin, Y.~Sui, J.~Ye, and T.~Peng, ``Cross-contract static
  analysis for detecting practical reentrancy vulnerabilities in smart
  contracts,'' in \emph{2020 35th IEEE/ACM International Conference on
  Automated Software Engineering (ASE)}.\hskip 1em plus 0.5em minus 0.4em\relax
  IEEE, 2020, pp. 1029--1040.

\bibitem{perez2021smart}
D.~Perez and B.~Livshits, ``Smart contract vulnerabilities: Vulnerable does not
  imply exploited,'' in \emph{30th USENIX Security Symposium (USENIX Security
  21)}, 2021, pp. 1325--1341.

\bibitem{brent2020ethainter}
L.~Brent, N.~Grech, S.~Lagouvardos, B.~Scholz, and Y.~Smaragdakis, ``Ethainter:
  a smart contract security analyzer for composite vulnerabilities,'' in
  \emph{Proceedings of the 41st ACM SIGPLAN Conference on Programming Language
  Design and Implementation}, 2020, pp. 454--469.

\bibitem{ghaleb2020effective}
A.~Ghaleb and K.~Pattabiraman, ``How effective are smart contract analysis
  tools? evaluating smart contract static analysis tools using bug injection,''
  in \emph{Proceedings of the 29th ACM SIGSOFT International Symposium on
  Software Testing and Analysis}, 2020, pp. 415--427.

\bibitem{grishchenko2018foundations}
I.~Grishchenko, M.~Maffei, and C.~Schneidewind, ``Foundations and tools for the
  static analysis of ethereum smart contracts,'' in \emph{International
  Conference on Computer Aided Verification}.\hskip 1em plus 0.5em minus
  0.4em\relax Springer, 2018, pp. 51--78.

\bibitem{schneidewind2020ethor}
C.~Schneidewind, I.~Grishchenko, M.~Scherer, and M.~Maffei, ``ethor: Practical
  and provably sound static analysis of ethereum smart contracts,'' in
  \emph{Proceedings of the 2020 ACM SIGSAC Conference on Computer and
  Communications Security}, 2020, pp. 621--640.

\bibitem{wustholz2020targeted}
V.~W{\"u}stholz and M.~Christakis, ``Targeted greybox fuzzing with static
  lookahead analysis,'' in \emph{2020 IEEE/ACM 42nd International Conference on
  Software Engineering (ICSE)}.\hskip 1em plus 0.5em minus 0.4em\relax IEEE,
  2020, pp. 789--800.

\bibitem{zong2020fuzzguard}
P.~Zong, T.~Lv, D.~Wang, Z.~Deng, R.~Liang, and K.~Chen, ``$\{$FuzzGuard$\}$:
  Filtering out unreachable inputs in directed grey-box fuzzing through deep
  learning,'' in \emph{29th USENIX Security Symposium (USENIX Security 20)},
  2020, pp. 2255--2269.

\bibitem{grieco2020echidna}
G.~Grieco, W.~Song, A.~Cygan, J.~Feist, and A.~Groce, ``Echidna: effective,
  usable, and fast fuzzing for smart contracts,'' in \emph{Proceedings of the
  29th ACM SIGSOFT International Symposium on Software Testing and Analysis},
  2020, pp. 557--560.

\bibitem{feist2019slither}
J.~Feist, G.~Grieco, and A.~Groce, ``Slither: a static analysis framework for
  smart contracts,'' in \emph{2019 IEEE/ACM 2nd International Workshop on
  Emerging Trends in Software Engineering for Blockchain (WETSEB)}.\hskip 1em
  plus 0.5em minus 0.4em\relax IEEE, 2019, pp. 8--15.

\bibitem{contractfuzzer}
B.~Jiang, Y.~Liu, and W.~Chan, ``Contractfuzzer: Fuzzing smart contracts for
  vulnerability detection,'' in \emph{Proceedings of the 33rd ACM/IEEE
  International Conference on Automated Software Engineering}.\hskip 1em plus
  0.5em minus 0.4em\relax ACM, 2018, pp. 259--269.

\bibitem{cao2021flashot}
Y.~Cao, C.~Zou, and X.~Cheng, ``Flashot: a snapshot of flash loan attack on
  defi ecosystem,'' \emph{arXiv preprint arXiv:2102.00626}, 2021.

\bibitem{wang2021blockeye}
B.~Wang, H.~Liu, C.~Liu, Z.~Yang, Q.~Ren, H.~Zheng, and H.~Lei, ``Blockeye:
  Hunting for defi attacks on blockchain,'' in \emph{2021 IEEE/ACM 43rd
  International Conference on Software Engineering: Companion Proceedings
  (ICSE-Companion)}.\hskip 1em plus 0.5em minus 0.4em\relax IEEE, 2021, pp.
  17--20.

\bibitem{wang2022impact}
Y.~Wang, Z.~Partick, Y.~Yao, Z.~Lu, and R.~Wattenhofer, ``Impact and user
  perception of sandwich attacks in the defi ecosystem,'' in \emph{ACM
  Conference on Human Factors in Computing Systems (CHI), New Orleans, LA,
  USA}, 2022.

\bibitem{wu2021defiranger}
S.~Wu, D.~Wang, J.~He, Y.~Zhou, L.~Wu, X.~Yuan, Q.~He, and K.~Ren,
  ``Defiranger: Detecting price manipulation attacks on defi applications,''
  \emph{arXiv preprint arXiv:2104.15068}, 2021.

\bibitem{SGUARD}
\BIBentryALTinterwordspacing
T.~D. Nguyen, L.~H. Pham, and J.~Sun, ``{SGUARD:} towards fixing vulnerable
  smart contracts automatically,'' in \emph{42nd {IEEE} Symposium on Security
  and Privacy, {SP} 2021, San Francisco, CA, USA, 24-27 May 2021}.\hskip 1em
  plus 0.5em minus 0.4em\relax {IEEE}, 2021, pp. 1215--1229. [Online].
  Available: \url{https://doi.org/10.1109/SP40001.2021.00057}
\BIBentrySTDinterwordspacing

\bibitem{10.1145/3551349.3559560}
P.~Tolmach, Y.~Li, and S.-W. Lin, ``Property-based automated repair of defi
  protocols,'' in \emph{Proceedings of the 37th IEEE/ACM International
  Conference on Automated Software Engineering}, ser. ASE '22, New York, NY,
  USA, 2023.

\bibitem{bohme2015bitcoin}
R.~B{\"o}hme, N.~Christin, B.~Edelman, and T.~Moore, ``Bitcoin: Economics,
  technology, and governance,'' \emph{Journal of economic Perspectives},
  vol.~29, no.~2, pp. 213--38, 2015.

\bibitem{chatterjee2017overview}
R.~Chatterjee and R.~Chatterjee, ``An overview of the emerging technology:
  Blockchain,'' in \emph{2017 3rd International Conference on Computational
  Intelligence and Networks (CINE)}.\hskip 1em plus 0.5em minus 0.4em\relax
  IEEE, 2017, pp. 126--127.

\bibitem{dos2019efficient}
S.~Dos~Santos, C.~Chukwuocha, S.~Kamali, and R.~K. Thulasiram, ``An efficient
  miner strategy for selecting cryptocurrency transactions,'' in \emph{2019
  IEEE International Conference on Blockchain (Blockchain)}.\hskip 1em plus
  0.5em minus 0.4em\relax IEEE, 2019, pp. 116--123.

\bibitem{bamakan2020survey}
S.~M.~H. Bamakan, A.~Motavali, and A.~B. Bondarti, ``A survey of blockchain
  consensus algorithms performance evaluation criteria,'' \emph{Expert Systems
  with Applications}, vol. 154, p. 113385, 2020.

\bibitem{wang2020designated}
H.~Wang, D.~He, and Y.~Ji, ``Designated-verifier proof of assets for bitcoin
  exchange using elliptic curve cryptography,'' \emph{Future Generation
  Computer Systems}, vol. 107, pp. 854--862, 2020.

\bibitem{tacs2019building}
R.~Ta{\c{s}} and {\"O}.~{\"O}. Tanr{\i}{\"o}ver, ``Building a decentralized
  application on the ethereum blockchain,'' in \emph{2019 3rd International
  Symposium on Multidisciplinary Studies and Innovative Technologies
  (ISMSIT)}.\hskip 1em plus 0.5em minus 0.4em\relax IEEE, 2019, pp. 1--4.

\bibitem{liu2021first}
B.~Liu, P.~Szalachowski, and J.~Zhou, ``A first look into defi oracles,'' in
  \emph{2021 IEEE International Conference on Decentralized Applications and
  Infrastructures (DAPPS)}.\hskip 1em plus 0.5em minus 0.4em\relax IEEE, 2021,
  pp. 39--48.

\bibitem{bonneau2015sok}
J.~Bonneau, A.~Miller, J.~Clark, A.~Narayanan, J.~A. Kroll, and E.~W. Felten,
  ``Sok: Research perspectives and challenges for bitcoin and
  cryptocurrencies,'' in \emph{2015 IEEE symposium on security and
  privacy}.\hskip 1em plus 0.5em minus 0.4em\relax IEEE, 2015, pp. 104--121.

\bibitem{atzei2017survey}
N.~Atzei, M.~Bartoletti, and T.~Cimoli, ``A survey of attacks on ethereum smart
  contracts (sok),'' in \emph{Principles of Security and Trust: 6th
  International Conference, POST 2017, Held as Part of the European Joint
  Conferences on Theory and Practice of Software, ETAPS 2017, Uppsala, Sweden,
  April 22-29, 2017, Proceedings 6}.\hskip 1em plus 0.5em minus 0.4em\relax
  Springer, 2017, pp. 164--186.

\bibitem{bano2019sok}
S.~Bano, A.~Sonnino, M.~Al-Bassam, S.~Azouvi, P.~McCorry, S.~Meiklejohn, and
  G.~Danezis, ``Sok: Consensus in the age of blockchains,'' in
  \emph{Proceedings of the 1st ACM Conference on Advances in Financial
  Technologies}, 2019, pp. 183--198.

\bibitem{buterin2019next}
V.~Buterin \emph{et~al.}, ``A next-generation smart contract and decentralized
  application platform. 2014,'' \emph{URL: \url{https://github.
  com/ethereum/wiki/wiki/White-Paper}}, 2019.

\bibitem{solidity}
Solidity, ``Solidity documentation,'' Website, Accessed: 08-08-2023,
  \url{https://vyper.readthedocs.io}.

\bibitem{vyper}
Vyper, ``Vyper documentation,'' Website, Accessed: 08-08-2023,
  \url{https://docs.soliditylang.org}.

\bibitem{perez2019smart}
D.~Perez and B.~Livshits, ``Smart contract vulnerabilities: Does anyone care?''
  \emph{arXiv preprint arXiv:1902.06710}, pp. 1--15, 2019.

\bibitem{huang2019smart}
Y.~Huang, Y.~Bian, R.~Li, J.~L. Zhao, and P.~Shi, ``Smart contract security: A
  software lifecycle perspective,'' \emph{IEEE Access}, vol.~7, pp.
  150\,184--150\,202, 2019.

\bibitem{schar2021decentralized}
F.~Sch{\"a}r, ``Decentralized finance: On blockchain-and smart contract-based
  financial markets,'' \emph{FRB of St. Louis Review}, 2021.

\bibitem{statista}
F.~I.~. Investments, ``Tvl (total value locked) across multiple decentralized
  finance (defi) blockchains,'' Website, Accessed: 08-08-2023,
  \url{https://www.statista.com/statistics/1272181/defi-tvl-in-multiple-blockchains/}.

\bibitem{moncada2020next}
R.~Moncada, E.~Ferro, A.~Favenza, and P.~Freni, ``Next generation
  blockchain-based financial services,'' in \emph{European Conference on
  Parallel Processing}.\hskip 1em plus 0.5em minus 0.4em\relax Springer, 2020,
  pp. 30--41.

\bibitem{abdulhakeem2021powered}
S.~A. Abdulhakeem, Q.~Hu \emph{et~al.}, ``Powered by blockchain technology,
  defi (decentralized finance) strives to increase financial inclusion of the
  unbanked by reshaping the world financial system,'' \emph{Modern Economy},
  vol.~12, no.~01, p.~1, 2021.

\bibitem{makerdao}
Makerdao, ``The maker foundation,'' Website, Accessed: 08-08-2023,
  \url{https://makerdao.com/en/}.

\bibitem{saengchote2021decentralized}
K.~Saengchote, ``Decentralized lending and its users: Insights from compound,''
  \emph{Available at SSRN 3925344}, 2021.

\bibitem{qin2021quantifying}
K.~Qin, L.~Zhou, and A.~Gervais, ``Quantifying blockchain extractable value:
  How dark is the forest?'' \emph{arXiv preprint arXiv:2101.05511}, 2021.

\bibitem{gudgeon2020defi}
L.~Gudgeon, S.~Werner, D.~Perez, and W.~J. Knottenbelt, ``Defi protocols for
  loanable funds: Interest rates, liquidity and market efficiency,'' in
  \emph{Proceedings of the 2nd ACM Conference on Advances in Financial
  Technologies}, 2020, pp. 92--112.

\bibitem{qin2021empirical}
K.~Qin, L.~Zhou, P.~Gamito, P.~Jovanovic, and A.~Gervais, ``An empirical study
  of defi liquidations: Incentives, risks, and instabilities,'' in
  \emph{Proceedings of the 21st ACM Internet Measurement Conference}, 2021, pp.
  336--350.

\bibitem{daian2020flash}
P.~Daian, S.~Goldfeder, T.~Kell, Y.~Li, X.~Zhao, I.~Bentov, L.~Breidenbach, and
  A.~Juels, ``Flash boys 2.0: Frontrunning in decentralized exchanges, miner
  extractable value, and consensus instability,'' in \emph{2020 IEEE Symposium
  on Security and Privacy (SP)}.\hskip 1em plus 0.5em minus 0.4em\relax IEEE,
  2020, pp. 910--927.

\bibitem{zakhary2019atomic}
V.~Zakhary, D.~Agrawal, and A.~E. Abbadi, ``Atomic commitment across
  blockchains,'' \emph{arXiv preprint arXiv:1905.02847}, 2019.

\bibitem{aave}
Aave, ``Aave protocol,'' Website, Accessed: 08-08-2023,
  \url{https://aave.com/}.

\bibitem{balancer}
Balancer, ``Balancer exchange,'' Website, Accessed: 08-08-2023,
  \url{https://balancer.fi/}.

\bibitem{compound}
Compound, ``Compound protocol,'' Website, Accessed: 08-08-2023,
  \url{https://compound.finance/}.

\bibitem{wang2022cyclic}
Y.~Wang, Y.~Chen, H.~Wu, L.~Zhou, S.~Deng, and R.~Wattenhofer, ``Cyclic
  arbitrage in decentralized exchanges,'' in \emph{The Web Conference 2022
  (WWW), Lyon, France}, 2022.

\bibitem{uniswap}
Uniswap, ``Uniswap protocol,'' Website, Accessed: 08-08-2023,
  \url{https://uniswap.org/}.

\bibitem{zhou2021high}
L.~Zhou, K.~Qin, C.~F. Torres, D.~V. Le, and A.~Gervais, ``High-frequency
  trading on decentralized on-chain exchanges,'' in \emph{2021 IEEE Symposium
  on Security and Privacy (SP)}.\hskip 1em plus 0.5em minus 0.4em\relax IEEE,
  2021, pp. 428--445.

\bibitem{hertzog2017bancor}
E.~Hertzog, G.~Benartzi, and G.~Benartzi, ``Bancor protocol,'' \emph{Continuous
  Liquidity for Cryptographic Tokens through their Smart Contracts.}, 2017.

\bibitem{kyber}
kyber, ``Kyber,'' Website, Accessed: 08-08-2023, \url{https://kyber.network}.

\bibitem{heimbach2021behavior}
L.~Heimbach, Y.~Wang, and R.~Wattenhofer, ``Behavior of liquidity providers in
  decentralized exchanges,'' \emph{arXiv preprint arXiv:2105.13822}, 2021.

\bibitem{chitra2022defi}
T.~Chitra, K.~Kulkarni, G.~Angeris, A.~Evans, and V.~Xu, ``Defi liquidity
  management via optimal control: Ohm as a case study,'' 2022.

\bibitem{mohan2020automated}
V.~Mohan, ``Automated market makers and decentralized exchanges: a defi
  primer,'' \emph{Available at SSRN 3722714}, 2020.

\bibitem{alao2021towards}
O.~Alao and P.~Cuffe, ``Towards a blockchain weather derivative financial
  instrument for hedging volumetric risks of solar power producers,'' in
  \emph{2021 IEEE Madrid PowerTech}.\hskip 1em plus 0.5em minus 0.4em\relax
  IEEE, 2021, pp. 1--6.

\bibitem{wachter2021measuring}
V.~v. Wachter, J.~R. Jensen, and O.~Ross, ``Measuring asset composability as a
  proxy for defi integration,'' in \emph{International Conference on Financial
  Cryptography and Data Security}.\hskip 1em plus 0.5em minus 0.4em\relax
  Springer, 2021, pp. 109--114.

\bibitem{compliFi}
CompliFi, ``Complifi protocol,'' Website, Accessed: 08-08-2023,
  \url{https://compli.fi/}.

\bibitem{dYdX}
dYdX, ``dydx protocol,'' Website, Accessed: 08-08-2023,
  \url{https://dydx.exchange/}.

\bibitem{barnBridge}
BarnBridge, ``Tokenized risk protocol,'' Website, Accessed: 08-08-2023,
  \url{https://barnbridge.com/}.

\bibitem{klages2020stablecoins}
A.~Klages-Mundt, D.~Harz, L.~Gudgeon, J.-Y. Liu, and A.~Minca, ``Stablecoins
  2.0: Economic foundations and risk-based models,'' in \emph{Proceedings of
  the 2nd ACM Conference on Advances in Financial Technologies}, 2020, pp.
  59--79.

\bibitem{mita2019stablecoin}
M.~Mita, K.~Ito, S.~Ohsawa, and H.~Tanaka, ``What is stablecoin?: A survey on
  price stabilization mechanisms for decentralized payment systems,'' in
  \emph{2019 8th International Congress on Advanced Applied Informatics
  (IIAI-AAI)}.\hskip 1em plus 0.5em minus 0.4em\relax IEEE, 2019, pp. 60--66.

\bibitem{saengchote2021defi}
K.~Saengchote, ``Where do defi stablecoins go? a closer look at what defi
  composability really means.'' \emph{A closer look at what DeFi composability
  really means.(July 26, 2021)}, 2021.

\bibitem{tether}
Tether, ``Tether token,'' Website, Accessed: 08-08-2023,
  \url{https://tether.to/}.

\bibitem{DAI}
DAI, ``Dai token,'' Website, Accessed: 08-08-2023,
  \url{https://en.wikipedia.org/wiki/Dai\_(cryptocurrency)}.

\bibitem{Diem}
Diem, ``Diem token,'' Website, Accessed: 08-08-2023,
  \url{https://en.wikipedia.org/wiki/Diem_(digital_currency)}.

\bibitem{cousaert2021sok}
S.~Cousaert, J.~Xu, and T.~Matsui, ``Sok: Yield aggregators in defi,''
  \emph{arXiv preprint arXiv:2105.13891}, 2021.

\bibitem{li2021defi}
J.~Li, ``Defi as an information aggregator,'' in \emph{International Conference
  on Financial Cryptography and Data Security}.\hskip 1em plus 0.5em minus
  0.4em\relax Springer, 2021, pp. 171--176.

\bibitem{1inch}
1inch, ``1inch protocol,'' Website, Accessed: 08-08-2023,
  \url{https://app.1inch.io}.

\bibitem{Matcha}
Matcha, ``Matcha protocol,'' Website, Accessed: 08-08-2023,
  \url{https://matcha.xyz}.

\bibitem{Plasma.Finance}
Plasma.Finance, ``Plasma.finance protocol,'' Website, Accessed: 08-08-2023,
  \url{https://plasma.finance}.

\bibitem{carter2021defi}
N.~Carter and L.~Jeng, ``Defi protocol risks: the paradox of defi,''
  \emph{Regtech, Suptech and Beyond: Innovation and Technology in Financial
  Services” RiskBooks--Forthcoming Q}, vol.~3, 2021.

\bibitem{wang2022speculative}
Z.~Wang, K.~Qin, D.~V. Minh, and A.~Gervais, ``Speculative multipliers on defi:
  Quantifying on-chain leverage risks,'' \emph{Financial Cryptography and Data
  Security}, 2022.

\bibitem{qian2022smart}
P.~Qian, Z.~Liu, Q.~He, B.~Huang, D.~Tian, and X.~Wang, ``Smart contract
  vulnerability detection technique: A survey,'' \emph{arXiv preprint
  arXiv:2209.05872}, 2022.

\bibitem{chu2023survey}
H.~Chu, P.~Zhang, H.~Dong, Y.~Xiao, S.~Ji, and W.~Li, ``A survey on smart
  contract vulnerabilities: Data sources, detection and repair,''
  \emph{Information and Software Technology}, p. 107221, 2023.

\bibitem{ye2020clairvoyance}
J.~Ye, M.~Ma, Y.~Lin, Y.~Sui, and Y.~Xue, ``Clairvoyance: cross-contract static
  analysis for detecting practical reentrancy vulnerabilities in smart
  contracts,'' in \emph{2020 IEEE/ACM 42nd International Conference on Software
  Engineering: Companion Proceedings (ICSE-Companion)}.\hskip 1em plus 0.5em
  minus 0.4em\relax IEEE, 2020, pp. 274--275.

\bibitem{coinCodeCap}
CoinCodeCap, ``Introduction to flash loans: What is a flash loan attack?''
  Website, Accessed: 08-08-2023, \url{https://coincodecap.com/flash-loan}.

\bibitem{lu2021freeswap}
D.~Lu, ``Freeswap exchange protocol,'' 2021.

\bibitem{yuksel2021mitigating}
A.~Y{\"u}ksel, ``Mitigating sandwich attacks in kyber dmm,'' 2021.

\bibitem{adams2021uniswap}
H.~Adams, N.~Zinsmeister, M.~Salem, R.~Keefer, and D.~Robinson, ``Uniswap v3
  core,'' 2021.

\bibitem{weber2017availability}
I.~Weber, V.~Gramoli, A.~Ponomarev, M.~Staples, R.~Holz, A.~B. Tran, and
  P.~Rimba, ``On availability for blockchain-based systems,'' in \emph{2017
  IEEE 36th Symposium on Reliable Distributed Systems (SRDS)}.\hskip 1em plus
  0.5em minus 0.4em\relax IEEE, 2017, pp. 64--73.

\bibitem{zarir2021developing}
A.~A. Zarir, G.~A. Oliva, Z.~M. Jiang, and A.~E. Hassan, ``Developing
  cost-effective blockchain-powered applications: A case study of the gas usage
  of smart contract transactions in the ethereum blockchain platform,''
  \emph{ACM Transactions on Software Engineering and Methodology (TOSEM)},
  vol.~30, no.~3, pp. 1--38, 2021.

\bibitem{xia2021demystifying}
P.~Xia, B.~Gao, W.~Su, Z.~Yu, X.~Luo, C.~Zhang, X.~Xiao, G.~Xu \emph{et~al.},
  ``Demystifying scam tokens on uniswap decentralized exchange,'' \emph{arXiv
  preprint arXiv:2109.00229}, 2021.

\bibitem{scharfman2022decentralized}
J.~Scharfman, ``Decentralized finance (defi) compliance and operations,'' in
  \emph{Cryptocurrency Compliance and Operations}.\hskip 1em plus 0.5em minus
  0.4em\relax Springer, 2022, pp. 171--186.

\bibitem{rugfull}
S.~Malwa, ``Rug pull scams,'' Website, Accessed: 08-08-2023,
  \url{https://www.coindesk.com/markets/2021/12/17/defi-rug-pull-scams-pulled-in-28b-this-year-chainalysis}.

\bibitem{bondly}
Bondly, ``Bondly.finance protocol,'' Website, Accessed: 08-08-2023,
  \url{https://www.bondly.finance/}.

\bibitem{bZx-V1}
K.~J. Kistner, ``Post-mortem,'' Website, Accessed: 08-08-2023,
  \url{https://bzx.network/blog/postmortem-ethdenver}.

\bibitem{makerdao2}
MakerDAO, ``The market collapse of march 12-13, 2020: How it impacted
  makerdao,'' Website, Accessed: 08-08-2023,
  \url{https://blog.makerdao.com/the-market-collapse-of-march-12-2020-how-it-impacted-makerdao/}.

\bibitem{uniswap2}
pNetwork Team, ``Is a new token standard really to blame for the imbtc/uniswap
  and dforce attacks?'' Website, Accessed: 08-08-2023,
  \url{https://medium.com/pnetwork/is-a-new-token-standard-really-to-blame-for-the-imbtc-uniswap-and-dforce-attacks-31c62e2bc799}.

\bibitem{bZx-V2}
K.~J. Kistner, ``itoken duplication incident report,'' Website, Accessed:
  08-08-2023, \url{https://bzx.network/blog/incident}.

\bibitem{eminence}
rekt, ``Eminence - rekt in prod,'' Website, Accessed: 08-08-2023,
  \url{https://rekt.news/eminence-rekt-in-prod/}.

\bibitem{harvest}
H.~Finance, ``Harvest flash loan economic attack post-mortem,'' Website,
  Accessed: 08-08-2023,
  \url{https://medium.com/harvest-finance/harvest-flashloan-economic-attack-post-mortem-3cf900d65217}.

\bibitem{akropolis2}
Akropolis, ``Delphi savings pool exploit,'' Website, Accessed: 08-08-2023,
  \url{https://akropolis.substack.com/p/delphi-savings-pool-exploit?s=r}.

\bibitem{pickle2}
W.~Gottsegen, ``Defi protocol pickle finance hacked for \$20 million,''
  Website, Accessed: 08-08-2023,
  \url{https://decrypt.co/49149/pickle-finance-hack}.

\bibitem{warp}
W.~Finance, ``Warp finance — exploit summary \& recovery of funds,'' Website,
  Accessed: 08-08-2023,
  \url{https://warpfinance.medium.com/warp-finance-exploit-summary-recovery-of-funds-5b8fe4a11898}.

\bibitem{alpha}
C.~Williams, ``Alpha finance exploited in \$37.5 million attack,'' Website,
  Accessed: 08-08-2023,
  \url{https://cryptobriefing.com/alpha-finance-suffers-37-5-million-loss-major-attack/}.

\bibitem{spartan}
SpartanProtocol, ``Post mortem analysis,'' Website, Accessed: 08-08-2023,
  \url{https://spartanprotocol.medium.com/today-spartan-protocol-was-subject-to-an-exploit-targeting-the-liquidity-pools-8589b2069cef}.

\bibitem{valuedefi1}
REKT, ``Value defi (1),'' Website, Accessed: 08-08-2023,
  \url{https://rekt.news/value-defi-rekt/}.

\bibitem{valuedefi2}
------, ``Value defi (2),'' Website, Accessed: 08-08-2023,
  \url{https://rekt.news/value-rekt2/}.

\bibitem{valuedefi3}
------, ``Value defi (3),'' Website, Accessed: 08-08-2023,
  \url{https://rekt.news/value-rekt3/}.

\bibitem{venus}
autofarm.network, ``Venus vaults post-mortem,'' Website, Accessed: 08-08-2023,
  \url{https://medium.com/autofarm-network/21-april-2021-venus-vaults-post-mortem-1518ae7399c6}.

\bibitem{pancakeBunny}
B.~Finance, ``Post mortem analysis,'' Website, Accessed: 08-08-2023,
  \url{https://pancakebunny.medium.com/hello-bunny-fam-a7bf0c7a07ba}.

\bibitem{pancakeBunny2}
------, ``Polybunny post-mortem \& compensation,'' Website, Accessed:
  08-08-2023,
  \url{https://pancakebunny.medium.com/polybunny-post-mortem-compensation-42b5c35ce957}.

\bibitem{Burgerswap}
E.~Genç, ``Burgerswap explains \$7.2 million flash loan attack in
  post-mortem,'' Website, Accessed: 08-08-2023,
  \url{https://decrypt.co/72194/burgerswap-explains-7-2-million-flash-loan-attack-in-post-mortem}.

\bibitem{julswap}
JustLiquidity, ``Flash loan farming / julb / bnb,'' Website, Accessed:
  08-08-2023,
  \url{https://justliquidity.medium.com/flash-loan-farming-julb-bnb-14c6c128f5dd}.

\bibitem{belt2}
B.~Finance, ``May 29 incident report,'' Website, Accessed: 08-08-2023,
  \url{https://medium.com/belt-finance/may-29-incident-report-865d20cc46ca}.

\bibitem{sushiSwap}
SlowMist, ``A brief analysis of the story of the sushi swap attack,'' Website,
  Accessed: 08-08-2023,
  \url{https://slowmist.medium.com/slowmist-a-brief-analysis-of-the-story-of-the-sushi-swap-attack-c7bc6709adea}.

\bibitem{THORChain1}
THORChain, ``Certik completes thorchain audit,'' Website, Accessed: 08-08-2023,
  \url{https://medium.com/thorchain/certik-completes-thorchain-audit-c6d88fad3613}.

\bibitem{vee}
V.~Finance, ``Vee.finance incident announcement,'' Website, Accessed:
  08-08-2023,
  \url{https://veefi.medium.com/vee-finance-accident-announcement-5e75ff197da6}.

\bibitem{indexed}
I.~Finance, ``Indexed attack post-mortem,'' Website, Accessed: 08-08-2023,
  \url{https://ndxfi.medium.com/indexed-attack-post-mortem-b006094f0bdc}.

\bibitem{cream.finance2}
A.~Thurman, ``Cream finance suffers third hack, loses over \$130 million,''
  Website, Accessed: 08-08-2023,
  \url{https://decrypt.co/84590/cream-finance-suffers-third-hack-losing-over-130-million}.

\bibitem{nerve}
BlockSec, ``The analysis of nerve bridge security incident,'' Website,
  Accessed: 08-08-2023,
  \url{https://blocksecteam.medium.com/the-analysis-of-nerve-bridge-security-incident-ead361a21025}.

\bibitem{beanstalk}
J.~Benson, ``Ethereum defi protocol beanstalk hacked for \$182 million—what
  you need to know,'' Website, Accessed: 08-08-2023,
  \url{https://decrypt.co/98118/ethereum-defi-protocol-beanstalk-hacked-182-million-what-you-need-know}.

\bibitem{saddle}
P.~Boyle, ``Saddle finance loses over \$10 million in hack,'' Website,
  Accessed: 08-08-2023,
  \url{https://medium.com/coinmonks/saddle-finance-loses-over-10-million-in-hack-derev-blog-2ba4b5d66527}.

\bibitem{balancer2}
M.~McDonald, ``Incident with non-standard erc20 deflationary tokens,'' Website,
  Accessed: 08-08-2023,
  \url{https://medium.com/balancer-protocol/incident-with-non-standard-erc20-deflationary-tokens-95a0f6d46dea}.

\bibitem{sanshu2}
BlockSec, ``The analysis of the sanshu inu security incident,'' Website,
  Accessed: 08-08-2023,
  \url{https://blocksecteam.medium.com/the-analysis-of-the-sanshu-inu-security-incident-28c0c7c0e783}.

\bibitem{Ankr}
M.~Maiboroda, ``Another huge defi-exploit, ankr protocol attacked for 10
  trillion abnbc tokens,'' Website, Accessed: 08-08-2023,
  \url{https://boxmining.com/attack-on-ankr-abnbc/}.

\bibitem{lendf.me}
PeckShield, ``Uniswap/lendf.me hacks: Root cause and loss analysis,'' Website,
  Accessed: 08-08-2023,
  \url{https://peckshield.medium.com/uniswap-lendf-me-hacks-root-cause-and-loss-analysis-50f3263dcc09}.

\bibitem{origin}
M.~Liu, ``Urgent: Ousd was hacked and there has been a loss of funds,''
  Website, Accessed: 08-08-2023,
  \url{https://blog.originprotocol.com/urgent-ousd-has-hacked-and-there-has-been-a-loss-of-funds-7b8c4a7d534c}.

\bibitem{grim}
J.~Benson, ``Grim finance hacked for \$30 million in fantom tokens,'' Website,
  Accessed: 08-08-2023,
  \url{https://decrypt.co/88727/grim-finance-hacked-30-million-fantom-tokens}.

\bibitem{rari2}
D.~Lucid, ``Rari capital ethereum pool — post-mortem,'' Website, Accessed:
  08-08-2023,
  \url{https://medium.com/rari-capital/5-8-2021-rari-ethereum-pool-post-mortem-60aab6a6f8f9}.

\bibitem{rari3}
C.~Williams, ``\$80m lost in attack on rari capital,'' Website, Accessed:
  08-08-2023,
  \url{https://cryptobriefing.com/80m-lost-in-attack-on-rari-capital/}.

\bibitem{Fei}
CertiK, ``Fei protocol incident analysis,'' Website, Accessed: 08-08-2023,
  \url{https://certik.medium.com/fei-protocol-incident-analysis-8527440696cc}.

\bibitem{Mango}
S.~Malwa, ``How market manipulation led to \$100m exploit on solana defi
  exchange mango,'' Website, Accessed: 08-08-2023,
  \url{https://www.coindesk.com/markets/2022/10/12/how-market-manipulation-led-to-a-100m-exploit-on-solana-defi-exchange-mango/}.

\bibitem{BONDLY2}
M.~DAO, ``Bondly exploit — how it unfolded on zenterest postmortem,''
  Website, Accessed: 08-08-2023,
  \url{https://medium.com/mantra-dao/bondly-exploit-how-it-unfolded-on-zenterest-postmortem-d8120d8d784b}.

\bibitem{bancor}
J.~Russell, ``The crypto world’s latest hack sees bancor lose \$23.5m,''
  Website, Accessed: 08-08-2023,
  \url{https://techcrunch.com/2018/07/10/bancor-loses-23-5m/}.

\bibitem{dodo}
DODO, ``Dodo pool incident postmortem: With a little help from our friends,''
  Website, Accessed: 08-08-2023,
  \url{https://blog.dodoex.io/dodo-pool-incident-postmortem-with-a-little-help-from-our-friends-327e66872d42}.

\bibitem{Uranium}
S.~Foundation, ``Uranium finance exploit analysis,'' Website, Accessed:
  08-08-2023,
  \url{https://medium.com/shentu-foundation/uranium-finance-exploit-analysis-d135055d6a6a}.

\bibitem{ChainSwap1}
ChainSwap, ``Chainswap post-mortem and compensation plan,'' Website, Accessed:
  08-08-2023,
  \url{https://chain-swap.medium.com/chainswap-post-mortem-and-compensation-plan-90cad50898ab}.

\bibitem{ChainSwap2}
------, ``Chainswap exploit 11 july 2021 post-mortem,'' Website, Accessed:
  08-08-2023,
  \url{https://chain-swap.medium.com/chainswap-exploit-11-july-2021-post-mortem-6e4e346e5a32}.

\bibitem{popsicle2}
Popsicle.Finance, ``Popsicle finance post mortem- after fragola hack,''
  Website, Accessed: 08-08-2023,
  \url{https://popsiclefinance.medium.com/popsicle-finance-post-mortem-after-fragola-hack-f45b302362e0}.

\bibitem{poly2}
T.~Gagliardoni, ``The poly network hack explained,'' Website, Accessed:
  08-08-2023,
  \url{https://research.kudelskisecurity.com/2021/08/12/the-poly-network-hack-explained/}.

\bibitem{compound2}
J.~Gogo, ``\$160m at risk due to bug in defi lending protocol compound,''
  Website, Accessed: 08-08-2023,
  \url{https://decrypt.co/82499/compound-exploit-drains-21m-from-lending-protocol}.

\bibitem{PAID2}
P.~NETWORK, ``Paid network attack postmortem,'' Website, Accessed: 08-08-2023,
  \url{https://paidnetwork.medium.com/paid-network-attack-postmortem-march-7-2021-9e4c0fef0e07}.

\bibitem{oypn}
PeckShield, ``Opyn hacks: Root cause analysis,'' Website, Accessed: 08-08-2023,
  \url{https://peckshield.medium.com/opyn-hacks-root-cause-analysis-c65f3fe249db}.

\bibitem{Chainlink}
M.~Dalton, ``Chainlink endures spam attack: Congestion, high fees,'' Website,
  Accessed: 08-08-2023,
  \url{https://cryptobriefing.com/chainlink-endures-spam-attack-congestion-high-fees/}.

\bibitem{cover}
C.~Protocol, ``12/28 post-mortem,'' Website, Accessed: 08-08-2023,
  \url{https://coverprotocol.medium.com/12-28-post-mortem-34c5f9f718d4}.

\bibitem{ycredit2}
BlockSec, ``Deposit less, get more: ycredit attack details,'' Website,
  Accessed: 08-08-2023,
  \url{https://blocksecteam.medium.com/deposit-less-get-more-ycredit-attack-details-f589f71674c3}.

\bibitem{yearn2}
D.~H. Staff, ``yearn.finance reveals details on hack that triggered \$11
  million loss,'' Website, Accessed: 08-08-2023,
  \url{https://dailyhodl.com/2021/02/07/yearn-finance-reveals-details-on-hack-that-triggered-11-million-loss/}.

\bibitem{furucombo2}
FURUCOMBO, ``Furucombo post-mortem march 2021,'' Website, Accessed: 08-08-2023,
  \url{https://medium.com/furucombo/furucombo-post-mortem-march-2021-ad19afd415e}.

\bibitem{easyfi2}
EasyFi, ``Easyfi security incident. pre-post mortem,'' Website, Accessed:
  08-08-2023,
  \url{http://web.archive.org/web/20210420195832/https://medium.com/easify-network/easyfi-security-incident-pre-post-mortem-33f2942016e9}.

\bibitem{vaultSX}
VaultSX, ``Vaultsx hack: Lessons learned and other thoughts,'' Website,
  Accessed: 08-08-2023,
  \url{https://www.eosgo.io/news/vaultsx-hack-lessnos-learned-and-thoughts}.

\bibitem{anyswap}
M.~P. Anyswap), ``Anyswap multichain router v3 exploit statement,'' Website,
  Accessed: 08-08-2023,
  \url{https://medium.com/multichainorg/anyswap-multichain-router-v3-exploit-statement-6833f1b7e6fb}.

\bibitem{safeDollar}
S.~Dollar, ``Saddle finance loses over \$10 million in hack,'' Website,
  Accessed: 08-08-2023,
  \url{https://safedollar.medium.com/safedollar-post-mortem-analysis-cb2769fe059}.

\bibitem{badgerdao}
Halborn, ``Explained: The badgerdao hack,'' Website, Accessed: 08-08-2023,
  \url{https://halborn.com/explained-the-badgerdao-hack-december-2021/}.

\bibitem{Wormhole}
Wormhole, ``Wormhole incident report,'' Website, Accessed: 08-08-2023,
  \url{https://wormholecrypto.medium.com/wormhole-incident-report-02-02-22-ad9b8f21eec6}.

\bibitem{Nomad}
B.~B. Sam~Kessler, ``Crypto bridge nomad drained of nearly \$200m in exploit,''
  Website, Accessed: 08-08-2023,
  \url{https://www.coindesk.com/tech/2022/08/02/nomad-bridge-drained-of-nearly-200-million-in-exploit/}.

\bibitem{quadriga}
Q.~Initiative, ``Cryptocurrency exchange hacks, frauds, and scams,'' Website,
  Accessed: 08-08-2023, \url{https://www.quadrigainitiative.com/education.php}.

\bibitem{securify}
P.~Tsankov, A.~Dan, D.~Drachsler-Cohen, A.~Gervais, F.~Buenzli, and M.~Vechev,
  ``Securify: Practical security analysis of smart contracts,'' in
  \emph{Proceedings of the 2018 ACM SIGSAC Conference on Computer and
  Communications Security}.\hskip 1em plus 0.5em minus 0.4em\relax ACM, 2018,
  pp. 67--82.

\bibitem{securify2}
------, ``Securify2,'' Website, Accessed: 08-08-2023,
  \url{https://github.com/eth-sri/securify2}.

\bibitem{so2020verismart}
S.~So, M.~Lee, J.~Park, H.~Lee, and H.~Oh, ``Verismart: A highly precise safety
  verifier for ethereum smart contracts,'' in \emph{2020 IEEE Symposium on
  Security and Privacy (SP)}.\hskip 1em plus 0.5em minus 0.4em\relax IEEE,
  2020, pp. 1678--1694.

\bibitem{wang2018formal}
Y.~Wang, S.~K. Lahiri, S.~Chen, R.~Pan, I.~Dillig, C.~Born, and I.~Naseer,
  ``Formal specification and verification of smart contracts for azure
  blockchain,'' \emph{arXiv preprint arXiv:1812.08829}, 2018.

\bibitem{kalra2018zeus}
S.~Kalra, S.~Goel, M.~Dhawan, and S.~Sharma, ``Zeus: analyzing safety of smart
  contracts.'' in \emph{Ndss}, 2018, pp. 1--12.

\bibitem{chen2021defectchecker}
J.~Chen, X.~Xia, D.~Lo, J.~Grundy, X.~Luo, and T.~Chen, ``Defectchecker:
  Automated smart contract defect detection by analyzing evm bytecode,''
  \emph{IEEE Transactions on Software Engineering}, vol.~48, no.~07, pp.
  2189--2207, 2022.

\bibitem{torres2019art}
C.~F. Torres, M.~Steichen \emph{et~al.}, ``The art of the scam: Demystifying
  honeypots in ethereum smart contracts,'' in \emph{28th USENIX Security
  Symposium (USENIX Security 19)}, 2019, pp. 1591--1607.

\bibitem{nikolic2018finding}
I.~Nikoli{\'c}, A.~Kolluri, I.~Sergey, P.~Saxena, and A.~Hobor, ``Finding the
  greedy, prodigal, and suicidal contracts at scale,'' in \emph{Proceedings of
  the 34th annual computer security applications conference}, 2018, pp.
  653--663.

\bibitem{mossberg2019manticore}
M.~Mossberg, F.~Manzano, E.~Hennenfent, A.~Groce, G.~Grieco, J.~Feist,
  T.~Brunson, and A.~Dinaburg, ``Manticore: A user-friendly symbolic execution
  framework for binaries and smart contracts,'' in \emph{2019 34th IEEE/ACM
  International Conference on Automated Software Engineering (ASE)}.\hskip 1em
  plus 0.5em minus 0.4em\relax IEEE, 2019, pp. 1186--1189.

\bibitem{mythril}
B.~Mueller, ``A framework for bug hunting on the ethereum blockchain,''
  Website, Accessed: 08-08-2023, \url{https://github.com/ConsenSys/mythril}.

\bibitem{oyente}
L.~Luu, D.-H. Chu, H.~Olickel, P.~Saxena, and A.~Hobor, ``Making smart
  contracts smarter,'' in \emph{Proceedings of the 2016 ACM SIGSAC conference
  on computer and communications security}.\hskip 1em plus 0.5em minus
  0.4em\relax ACM, 2016, pp. 254--269.

\bibitem{torres2018osiris}
C.~F. Torres, J.~Sch{\"u}tte, and R.~State, ``Osiris: Hunting for integer bugs
  in ethereum smart contracts,'' in \emph{Proceedings of the 34th Annual
  Computer Security Applications Conference}, 2018, pp. 664--676.

\bibitem{rodler2018sereum}
M.~Rodler, W.~Li, G.~O. Karame, and L.~Davi, ``Sereum: Protecting existing
  smart contracts against re-entrancy attacks,'' \emph{arXiv preprint
  arXiv:1812.05934}, 2018.

\bibitem{krupp2018teether}
J.~Krupp and C.~Rossow, ``$\{$teEther$\}$: Gnawing at ethereum to automatically
  exploit smart contracts,'' in \emph{27th USENIX Security Symposium (USENIX
  Security 18)}, 2018, pp. 1317--1333.

\bibitem{permenev2020verx}
A.~Permenev, D.~Dimitrov, P.~Tsankov, D.~Drachsler-Cohen, and M.~Vechev,
  ``Verx: Safety verification of smart contracts,'' in \emph{2020 IEEE
  symposium on security and privacy (SP)}.\hskip 1em plus 0.5em minus
  0.4em\relax IEEE, 2020, pp. 1661--1677.

\bibitem{albert2018ethir}
E.~Albert, P.~Gordillo, B.~Livshits, A.~Rubio, and I.~Sergey, ``Ethir: A
  framework for high-level analysis of ethereum bytecode,'' in
  \emph{International symposium on automated technology for verification and
  analysis}.\hskip 1em plus 0.5em minus 0.4em\relax Springer, 2018, pp.
  513--520.

\bibitem{tikhomirov2018smartcheck}
S.~Tikhomirov, E.~Voskresenskaya, I.~Ivanitskiy, R.~Takhaviev, E.~Marchenko,
  and Y.~Alexandrov, ``Smartcheck: Static analysis of ethereum smart
  contracts,'' in \emph{Proceedings of the 1st International Workshop on
  Emerging Trends in Software Engineering for Blockchain}, 2018, pp. 9--16.

\bibitem{brent2018vandal}
L.~Brent, A.~Jurisevic, M.~Kong, E.~Liu, F.~Gauthier, V.~Gramoli, R.~Holz, and
  B.~Scholz, ``Vandal: A scalable security analysis framework for smart
  contracts,'' \emph{arXiv preprint arXiv:1809.03981}, 2018.

\bibitem{wang2020oracle}
H.~Wang, Y.~Liu, Y.~Li, S.-W. Lin, C.~Artho, L.~Ma, and Y.~Liu,
  ``Oracle-supported dynamic exploit generation for smart contracts,''
  \emph{IEEE Transactions on Dependable and Secure Computing}, 2020.

\bibitem{wang2019vultron}
H.~Wang, Y.~Li, S.-W. Lin, L.~Ma, and Y.~Liu, ``Vultron: catching vulnerable
  smart contracts once and for all,'' in \emph{2019 IEEE/ACM 41st International
  Conference on Software Engineering: New Ideas and Emerging Results
  (ICSE-NIER)}.\hskip 1em plus 0.5em minus 0.4em\relax IEEE, 2019, pp. 1--4.

\bibitem{liu2018reguard}
C.~Liu, H.~Liu, Z.~Cao, Z.~Chen, B.~Chen, and B.~Roscoe, ``Reguard: finding
  reentrancy bugs in smart contracts,'' in \emph{2018 IEEE/ACM 40th
  International Conference on Software Engineering: Companion
  (ICSE-Companion)}.\hskip 1em plus 0.5em minus 0.4em\relax IEEE, 2018, pp.
  65--68.

\bibitem{he2019learning}
J.~He, M.~Balunovi{\'c}, N.~Ambroladze, P.~Tsankov, and M.~Vechev, ``Learning
  to fuzz from symbolic execution with application to smart contracts,'' in
  \emph{Proceedings of the 2019 ACM SIGSAC Conference on Computer and
  Communications Security}, 2019, pp. 531--548.

\bibitem{wustholz2020harvey}
V.~W{\"u}stholz and M.~Christakis, ``Harvey: A greybox fuzzer for smart
  contracts,'' in \emph{Proceedings of the 28th ACM Joint Meeting on European
  Software Engineering Conference and Symposium on the Foundations of Software
  Engineering}, 2020, pp. 1398--1409.

\bibitem{torres2021confuzzius}
C.~F. Torres, A.~K. Iannillo, A.~Gervais, and R.~State, ``Confuzzius: A data
  dependency-aware hybrid fuzzer for smart contracts,'' in \emph{2021 IEEE
  European Symposium on Security and Privacy (EuroS\&P)}.\hskip 1em plus 0.5em
  minus 0.4em\relax IEEE, 2021, pp. 103--119.

\bibitem{nguyen2020sfuzz}
T.~D. Nguyen, L.~H. Pham, J.~Sun, Y.~Lin, and Q.~T. Minh, ``sfuzz: An efficient
  adaptive fuzzer for solidity smart contracts,'' in \emph{Proceedings of the
  ACM/IEEE 42nd International Conference on Software Engineering}, 2020, pp.
  778--788.

\bibitem{9795233}
Y.~Xue, J.~Ye, W.~Zhang, J.~Sun, L.~Ma, H.~Wang, and J.~Zhao, ``xfuzz: Machine
  learning guided cross-contract fuzzing,'' \emph{IEEE Transactions on
  Dependable and Secure Computing}, pp. 1--14, 2022.

\bibitem{choi2021smartian}
J.~Choi, D.~Kim, S.~Kim, G.~Grieco, A.~Groce, and S.~K. Cha, ``Smartian:
  Enhancing smart contract fuzzing with static and dynamic data-flow
  analyses,'' in \emph{2021 36th IEEE/ACM International Conference on Automated
  Software Engineering (ASE)}.\hskip 1em plus 0.5em minus 0.4em\relax IEEE,
  2021, pp. 227--239.

\bibitem{rlf}
J.~Su, H.-N. Dai, L.~Zhao, Z.~Zheng, and X.~Luo, ``Effectively generating
  vulnerable transaction sequences in smart contracts with reinforcement
  learning-guided fuzzing,'' in \emph{Proceedings of the 37th IEEE/ACM
  International Conference on Automated Software Engineering}, ser. ASE '22,
  New York, NY, USA, 2023.

\bibitem{10018241}
Z.~Liu, P.~Qian, J.~Yang, L.~Liu, X.~Xu, Q.~He, and X.~Zhang, ``Rethinking
  smart contract fuzzing: Fuzzing with invocation ordering and important branch
  revisiting,'' \emph{IEEE Transactions on Information Forensics and Security},
  vol.~18, pp. 1237--1251, 2023.

\bibitem{ItyFuzz}
C.~Shou, S.~Tan, and K.~Sen, ``Ityfuzz: Snapshot-based fuzzer for smart
  contract,'' in \emph{Proceedings of the 32nd ACM SIGSOFT International
  Symposium on Software Testing and Analysis}, ser. ISSTA 2023, New York, NY,
  USA, 2023, p. 322–333.

\bibitem{tann2018towards}
W.~J.-W. Tann, X.~J. Han, S.~S. Gupta, and Y.-S. Ong, ``Towards safer smart
  contracts: A sequence learning approach to detecting security threats,''
  \emph{arXiv preprint arXiv:1811.06632}, 2018.

\bibitem{qian2020towards}
P.~Qian, Z.~Liu, Q.~He, R.~Zimmermann, and X.~Wang, ``Towards automated
  reentrancy detection for smart contracts based on sequential models,''
  \emph{IEEE Access}, vol.~8, pp. 19\,685--19\,695, 2020.

\bibitem{wang2020contractward}
W.~Wang, J.~Song, G.~Xu, Y.~Li, H.~Wang, and C.~Su, ``Contractward: Automated
  vulnerability detection models for ethereum smart contracts,'' \emph{IEEE
  Transactions on Network Science and Engineering}, vol.~8, no.~2, pp.
  1133--1144, 2020.

\bibitem{liu2018s}
H.~Liu, C.~Liu, W.~Zhao, Y.~Jiang, and J.~Sun, ``S-gram: towards semantic-aware
  security auditing for ethereum smart contracts,'' in \emph{2018 33rd IEEE/ACM
  International Conference on Automated Software Engineering (ASE)}.\hskip 1em
  plus 0.5em minus 0.4em\relax IEEE, 2018, pp. 814--819.

\bibitem{zhuangsmart}
Y.~Zhuang, Z.~Liu, P.~Qian, Q.~Liu, X.~Wang, and Q.~He, ``Smart contract
  vulnerability detection using graph neural network,'' in \emph{IJCAI}, 2020,
  pp. 3283--3290.

\bibitem{9534324}
X.~Yu, H.~Zhao, B.~Hou, Z.~Ying, and B.~Wu, ``Deescvhunter: A deep
  learning-based framework for smart contract vulnerability detection,'' in
  \emph{2021 International Joint Conference on Neural Networks (IJCNN)}, 2021,
  pp. 1--8.

\bibitem{9740682}
S.-J. Hwang, S.-H. Choi, J.~Shin, and Y.-H. Choi, ``Codenet: Code-targeted
  convolutional neural network architecture for smart contract vulnerability
  detection,'' \emph{IEEE Access}, vol.~10, pp. 32\,595--32\,607, 2022.

\bibitem{s23167246}
W.~Deng, H.~Wei, T.~Huang, C.~Cao, Y.~Peng, and X.~Hu, ``Smart contract
  vulnerability detection based on deep learning and multimodal decision
  fusion,'' \emph{Sensors}, vol.~23, no.~16, 2023.

\bibitem{durieux2020empirical}
T.~Durieux, J.~F. Ferreira, R.~Abreu, and P.~Cruz, ``Empirical review of
  automated analysis tools on 47,587 ethereum smart contracts,'' in
  \emph{Proceedings of the ACM/IEEE 42nd International conference on software
  engineering}, 2020, pp. 530--541.

\bibitem{10.1145/3460319.3464837}
M.~Ren, Z.~Yin, F.~Ma, Z.~Xu, Y.~Jiang, C.~Sun, H.~Li, and Y.~Cai, ``Empirical
  evaluation of smart contract testing: What is the best choice?'' in
  \emph{Proceedings of the 30th ACM SIGSOFT International Symposium on Software
  Testing and Analysis}, ser. ISSTA 2021, 2021, p. 566–579.

\bibitem{Bhargavan}
K.~Bhargavan, A.~Delignat-Lavaud, C.~Fournet, A.~Gollamudi, G.~Gonthier,
  N.~Kobeissi, N.~Kulatova, A.~Rastogi, T.~Sibut-Pinote, N.~Swamy
  \emph{et~al.}, ``Formal verification of smart contracts: Short paper,'' in
  \emph{Proceedings of the 2016 ACM Workshop on Programming Languages and
  Analysis for Security}.\hskip 1em plus 0.5em minus 0.4em\relax ACM, 2016, pp.
  91--96.

\bibitem{yang2020seraph}
Z.~Yang, H.~Liu, Y.~Li, H.~Zheng, L.~Wang, and B.~Chen, ``Seraph: enabling
  cross-platform security analysis for evm and wasm smart contracts,'' in
  \emph{2020 IEEE/ACM 42nd International Conference on Software Engineering:
  Companion Proceedings (ICSE-Companion)}.\hskip 1em plus 0.5em minus
  0.4em\relax IEEE, 2020, pp. 21--24.

\bibitem{reis2020tezla}
J.~S. Reis, P.~Crocker, and S.~M. de~Sousa, ``Tezla, an intermediate
  representation for static analysis of michelson smart contracts,'' in
  \emph{2nd Workshop on Formal Methods for Blockchains}, 2020.

\bibitem{murray2019survey}
Y.~Murray and D.~A. Anisi, ``Survey of formal verification methods for smart
  contracts on blockchain,'' in \emph{2019 10th IFIP International Conference
  on New Technologies, Mobility and Security (NTMS)}.\hskip 1em plus 0.5em
  minus 0.4em\relax IEEE, 2019, pp. 1--6.

\bibitem{gu2018formal}
R.~Gu, R.~Marinescu, C.~Seceleanu, and K.~Lundqvist, ``Formal verification of
  an autonomous wheel loader by model checking,'' in \emph{Proceedings of the
  6th Conference on Formal Methods in Software Engineering}, 2018, pp. 74--83.

\bibitem{liu2019formal}
J.~Liu, B.~Zhan, S.~Wang, S.~Ying, T.~Liu, Y.~Li, M.~Ying, and N.~Zhan,
  ``Formal verification of quantum algorithms using quantum hoare logic,'' in
  \emph{International conference on computer aided verification}.\hskip 1em
  plus 0.5em minus 0.4em\relax Springer, 2019, pp. 187--207.

\bibitem{baldoni2018survey}
R.~Baldoni, E.~Coppa, D.~C. D’elia, C.~Demetrescu, and I.~Finocchi, ``A
  survey of symbolic execution techniques,'' \emph{ACM Computing Surveys
  (CSUR)}, vol.~51, no.~3, pp. 1--39, 2018.

\bibitem{consensys}
ConsenSys, ``Consensys: a market-leading blockchain technology company.''
  Website, Accessed: 08-08-2023, \url{https://consensys.net}.

\bibitem{pierro2021analysis}
G.~A. Pierro and R.~Tonelli, ``Analysis of source code duplication in ethreum
  smart contracts,'' in \emph{2021 IEEE International Conference on Software
  Analysis, Evolution and Reengineering (SANER)}.\hskip 1em plus 0.5em minus
  0.4em\relax IEEE, 2021, pp. 701--707.

\bibitem{li2020stan}
X.~Li, T.~Chen, X.~Luo, T.~Zhang, L.~Yu, and Z.~Xu, ``Stan: Towards describing
  bytecodes of smart contract,'' in \emph{2020 IEEE 20th International
  Conference on Software Quality, Reliability and Security (QRS)}.\hskip 1em
  plus 0.5em minus 0.4em\relax IEEE, 2020, pp. 273--284.

\bibitem{bistarelli2020ethereum}
S.~Bistarelli, G.~Mazzante, M.~Micheletti, L.~Mostarda, D.~Sestili, and
  F.~Tiezzi, ``Ethereum smart contracts: Analysis and statistics of their
  source code and opcodes,'' \emph{Internet of Things}, vol.~11, p. 100198,
  2020.

\bibitem{lu2021neucheck}
N.~Lu, B.~Wang, Y.~Zhang, W.~Shi, and C.~Esposito, ``Neucheck: A more practical
  ethereum smart contract security analysis tool,'' \emph{Software: Practice
  and Experience}, vol.~51, no.~10, pp. 2065--2084, 2021.

\bibitem{grech2019gigahorse}
N.~Grech, L.~Brent, B.~Scholz, and Y.~Smaragdakis, ``Gigahorse: thorough,
  declarative decompilation of smart contracts,'' in \emph{2019 IEEE/ACM 41st
  International Conference on Software Engineering (ICSE)}.\hskip 1em plus
  0.5em minus 0.4em\relax IEEE, 2019, pp. 1176--1186.

\bibitem{perez2020cost}
V.~P{\'e}rez, M.~Klemen, P.~L{\'o}pez-Garc{\'\i}a, J.~F. Morales, and
  M.~Hermenegildo, ``Cost analysis of smart contracts via parametric resource
  analysis,'' in \emph{International Static Analysis Symposium}.\hskip 1em plus
  0.5em minus 0.4em\relax Springer, 2020, pp. 7--31.

\bibitem{manes2019art}
V.~J. Man{\`e}s, H.~Han, C.~Han, S.~K. Cha, M.~Egele, E.~J. Schwartz, and
  M.~Woo, ``The art, science, and engineering of fuzzing: A survey,''
  \emph{IEEE Transactions on Software Engineering}, vol.~47, no.~11, pp.
  2312--2331, 2019.

\bibitem{zheng2019firm}
Y.~Zheng, A.~Davanian, H.~Yin, C.~Song, H.~Zhu, and L.~Sun, ``Firm-afl:
  high-throughput greybox fuzzing of iot firmware via augmented process
  emulation,'' in \emph{28th $\{$USENIX$\}$ Security Symposium ($\{$USENIX$\}$
  Security 19)}, 2019, pp. 1099--1114.

\bibitem{bohme2017directed}
M.~B{\"o}hme, V.-T. Pham, M.-D. Nguyen, and A.~Roychoudhury, ``Directed greybox
  fuzzing,'' in \emph{Proceedings of the 2017 ACM SIGSAC Conference on Computer
  and Communications Security}, 2017, pp. 2329--2344.

\bibitem{xu2019fuzzing}
W.~Xu, H.~Moon, S.~Kashyap, P.-N. Tseng, and T.~Kim, ``Fuzzing file systems via
  two-dimensional input space exploration,'' in \emph{2019 IEEE Symposium on
  Security and Privacy (SP)}.\hskip 1em plus 0.5em minus 0.4em\relax IEEE,
  2019, pp. 818--834.

\bibitem{li2018vuldeepecker}
Z.~Li, D.~Zou, S.~Xu, X.~Ou, H.~Jin, S.~Wang, Z.~Deng, and Y.~Zhong,
  ``Vuldeepecker: A deep learning-based system for vulnerability detection,''
  \emph{arXiv preprint arXiv:1801.01681}, 2018.

\bibitem{lin2020software}
G.~Lin, S.~Wen, Q.-L. Han, J.~Zhang, and Y.~Xiang, ``Software vulnerability
  detection using deep neural networks: a survey,'' \emph{Proceedings of the
  IEEE}, vol. 108, no.~10, pp. 1825--1848, 2020.

\bibitem{chakraborty2021deep}
S.~Chakraborty, R.~Krishna, Y.~Ding, and B.~Ray, ``Deep learning based
  vulnerability detection: Are we there yet?'' \emph{IEEE Transactions on
  Software Engineering}, vol.~48, no.~09, pp. 3280--3296, 2022.

\bibitem{zou2019mu}
D.~Zou, S.~Wang, S.~Xu, Z.~Li, and H.~Jin, ``Vuldeepecker: A deep
  learning-based system for multiclass vulnerability detection,'' \emph{IEEE
  Transactions on Dependable and Secure Computing}, vol.~18, no.~5, pp.
  2224--2236, 2019.

\bibitem{10.1145/3402450}
X.~L. Yu, O.~Al-Bataineh, D.~Lo, and A.~Roychoudhury, ``Smart contract
  repair,'' \emph{ACM Trans. Softw. Eng. Methodol.}, vol.~29, no.~4, 2020.

\bibitem{zhang2020smartshield}
Y.~Zhang, S.~Ma, J.~Li, K.~Li, S.~Nepal, and D.~Gu, ``Smartshield: Automatic
  smart contract protection made easy,'' in \emph{2020 IEEE 27th International
  Conference on Software Analysis, Evolution and Reengineering (SANER)}.\hskip
  1em plus 0.5em minus 0.4em\relax IEEE, 2020, pp. 23--34.

\bibitem{nguyen2021sguard}
T.~D. Nguyen, L.~H. Pham, and J.~Sun, ``Sguard: towards fixing vulnerable smart
  contracts automatically,'' in \emph{2021 IEEE Symposium on Security and
  Privacy (SP)}.\hskip 1em plus 0.5em minus 0.4em\relax IEEE, 2021, pp.
  1215--1229.

\bibitem{rodler2021evmpatch}
M.~Rodler, W.~Li, G.~O. Karame, and L.~Davi, ``$\{$EVMPatch$\}$: Timely and
  automated patching of ethereum smart contracts,'' in \emph{30th USENIX
  Security Symposium (USENIX Security 21)}, 2021, pp. 1289--1306.

\bibitem{jin2021aroc}
H.~Jin, Z.~Wang, M.~Wen, W.~Dai, Y.~Zhu, and D.~Zou, ``Aroc: An automatic
  repair framework for on-chain smart contracts,'' \emph{IEEE Transactions on
  Software Engineering}, vol.~48, no.~11, pp. 4611--4629, 2021.

\bibitem{giesen2022practical}
J.-R. Giesen, S.~Andreina, M.~Rodler, G.~O. Karame, and L.~Davi, ``Practical
  mitigation of smart contract bugs,'' \emph{arXiv preprint arXiv:2203.00364},
  2022.

\bibitem{ferreira2022elysium}
C.~Ferreira~Torres, H.~Jonker, and R.~State, ``Elysium: Context-aware
  bytecode-level patching to automatically heal vulnerable smart contracts,''
  in \emph{Proceedings of the 25th International Symposium on Research in
  Attacks, Intrusions and Defenses}, 2022, pp. 115--128.

\bibitem{tolmach2022property}
P.~Tolmach, Y.~Li, and S.-W. Lin, ``Property-based automated repair of defi
  protocols,'' in \emph{Proceedings of the 37th IEEE/ACM International
  Conference on Automated Software Engineering}, 2022, pp. 1--5.

\bibitem{eskandari2019sok}
S.~Eskandari, S.~Moosavi, and J.~Clark, ``Sok: Transparent dishonesty:
  front-running attacks on blockchain,'' in \emph{International Conference on
  Financial Cryptography and Data Security}.\hskip 1em plus 0.5em minus
  0.4em\relax Springer, 2019, pp. 170--189.

\bibitem{babel2021clockwork}
K.~Babel, P.~Daian, M.~Kelkar, and A.~Juels, ``Clockwork finance: Automated
  analysis of economic security in smart contracts,'' \emph{arXiv preprint
  arXiv:2109.04347}, 2021.

\bibitem{boussard2019steward}
M.~Boussard, S.~Papillon, P.~Peloso, M.~Signorini, and E.~Waisbard, ``Steward:
  Sdn and blockchain-based trust evaluation for automated risk management on
  iot devices,'' in \emph{IEEE INFOCOM 2019-IEEE Conference on Computer
  Communications Workshops (INFOCOM WKSHPS)}.\hskip 1em plus 0.5em minus
  0.4em\relax IEEE, 2019, pp. 841--846.

\bibitem{defiscore}
L.~Jordan and C.~Jack, ``Defi score: An open framework for evaluating defi
  protocols,'' Website, Accessed: 08-08-2023,
  \url{https://github.com/ConsenSys/defi-score}.

\bibitem{defisafety}
H.~Rex, H.~Lucas, A.~Mohamed, D.~David, S.~Nick, V.~Nic, and C.~Carter,
  ``Defi-safety is an independent ratings organization that evaluates
  decentralized finance products,'' Website, Accessed: 08-08-2023,
  \url{https://www.defisafety.com/}.

\bibitem{yCredit}
BlockSec, ``Deposit less, get more: ycredit attack details,'' Website,
  Accessed: 08-08-2023,
  \url{https://blocksecteam.medium.com/deposit-less-get-more-ycredit-attack-details-f589f71674c3}.

\bibitem{opyn}
Opyn, ``Opyn eth put exploit post mortem,'' Website, Accessed: 08-08-2023,
  \url{https://medium.com/opyn/opyn-eth-put-exploit-post-mortem-1a009e3347a8}.

\bibitem{chaliasos2023smart}
S.~Chaliasos, M.~A. Charalambous, L.~Zhou, R.~Galanopoulou, A.~Gervais,
  D.~Mitropoulos, and B.~Livshits, ``Smart contract and defi security: Insights
  from tool evaluations and practitioner surveys,'' 2023.

\bibitem{heilman2015eclipse}
E.~Heilman, A.~Kendler, A.~Zohar, and S.~Goldberg, ``Eclipse attacks on
  $\{$Bitcoin’s$\}$$\{$Peer-to-Peer$\}$ network,'' in \emph{24th USENIX
  Security Symposium (USENIX Security 15)}, 2015, pp. 129--144.

\bibitem{gervais2016security}
A.~Gervais, G.~O. Karame, K.~W{\"u}st, V.~Glykantzis, H.~Ritzdorf, and
  S.~Capkun, ``On the security and performance of proof of work blockchains,''
  in \emph{Proceedings of the 2016 ACM SIGSAC conference on computer and
  communications security}, 2016, pp. 3--16.

\bibitem{zhang2019lay}
R.~Zhang and B.~Preneel, ``Lay down the common metrics: Evaluating
  proof-of-work consensus protocols' security,'' in \emph{2019 IEEE Symposium
  on Security and Privacy (SP)}.\hskip 1em plus 0.5em minus 0.4em\relax IEEE,
  2019, pp. 175--192.

\bibitem{hou2019squirrl}
C.~Hou, M.~Zhou, Y.~Ji, P.~Daian, F.~Tramer, G.~Fanti, and A.~Juels, ``Squirrl:
  Automating attack analysis on blockchain incentive mechanisms with deep
  reinforcement learning,'' \emph{arXiv preprint arXiv:1912.01798}, 2019.

\bibitem{ahn2020packet}
S.~Ahn, T.~Kim, Y.~Kwon, and S.~Cho, ``Packet aggregation scheme to mitigate
  the network congestion in blockchain networks,'' in \emph{2020 International
  Conference on Electronics, Information, and Communication (ICEIC)}.\hskip 1em
  plus 0.5em minus 0.4em\relax IEEE, 2020, pp. 1--3.

\bibitem{zhou2021a2mm}
L.~Zhou, K.~Qin, and A.~Gervais, ``A2mm: Mitigating frontrunning, transaction
  reordering and consensus instability in decentralized exchanges,''
  \emph{arXiv preprint arXiv:2106.07371}, 2021.

\bibitem{su2021evil}
L.~Su, X.~Shen, X.~Du, X.~Liao, X.~Wang, L.~Xing, and B.~Liu, ``Evil under the
  sun: Understanding and discovering attacks on ethereum decentralized
  applications,'' in \emph{30th USENIX Security Symposium (USENIX Security
  21)}, 2021, pp. 1307--1324.

\bibitem{flashbot1}
Flashbots, ``Everything there is to know about flashbots,'' Website, Accessed:
  08-08-2023, \url{https://github.com/flashbots/pm}.

\bibitem{flashbot2}
A.~Obadia, ``Flashbots: Frontrunning the mev crisis,'' Website, Accessed:
  08-08-2023,
  \url{https://medium.com/flashbots/frontrunning-the-mev-crisis-40629a613752}.

\bibitem{Thurman}
A.~Thurman, ``Value defi protocol suffers \$6 million flash loan exploit
  (2020),'' Website, Accessed: 08-08-2023,
  \url{https://cointelegraph.com/news/value-defi-protocol-suffers-6-million-flash-loan-exploit}.

\bibitem{gu2020empirical}
W.~C. Gu, A.~Raghuvanshi, and D.~Boneh, ``Empirical measurements on pricing
  oracles and decentralized governance for stablecoins,'' \emph{Available at
  SSRN 3611231}, 2020.

\bibitem{xu2017enabling}
L.~Xu, N.~Shah, L.~Chen, N.~Diallo, Z.~Gao, Y.~Lu, and W.~Shi, ``Enabling the
  sharing economy: Privacy respecting contract based on public blockchain,'' in
  \emph{Proceedings of the ACM Workshop on Blockchain, Cryptocurrencies and
  Contracts}, 2017, pp. 15--21.

\bibitem{chu2021manta}
S.~Chu, Y.~Xia, and Z.~Zhang, ``Manta: a plug and play private defi stack,''
  \emph{Cryptology ePrint Archive}, 2021.

\bibitem{dai2021flexible}
W.~Dai, ``Flexible anonymous transactions (flax): Towards privacy-preserving
  and composable decentralized finance,'' \emph{Cryptology ePrint Archive},
  2021.

\bibitem{xie2019libra}
T.~Xie, J.~Zhang, Y.~Zhang, C.~Papamanthou, and D.~Song, ``Libra: Succinct
  zero-knowledge proofs with optimal prover computation,'' in \emph{Annual
  International Cryptology Conference}.\hskip 1em plus 0.5em minus 0.4em\relax
  Springer, 2019, pp. 733--764.

\bibitem{boyle2019practical}
E.~Boyle, N.~Gilboa, Y.~Ishai, and A.~Nof, ``Practical fully secure three-party
  computation via sublinear distributed zero-knowledge proofs,'' in
  \emph{Proceedings of the 2019 ACM SIGSAC Conference on Computer and
  Communications Security}, 2019, pp. 869--886.

\bibitem{reich2019privacy}
D.~Reich, A.~Todoki, R.~Dowsley, M.~De~Cock \emph{et~al.}, ``Privacy-preserving
  classification of personal text messages with secure multi-party
  computation,'' \emph{Advances in Neural Information Processing Systems},
  vol.~32, 2019.

\bibitem{bayatbabolghani2018secure}
F.~Bayatbabolghani and M.~Blanton, ``Secure multi-party computation,'' in
  \emph{Proceedings of the 2018 ACM SIGSAC Conference on Computer and
  Communications Security}, 2018, pp. 2157--2159.

\end{thebibliography}

\end{sloppypar}

\end{document}